\documentclass[preprint2]{aastex} 

\newcommand{\kms}{$\rm km\;s^{-1}$}    
    
\newcommand{\MgI}{Mg~{\small I}}    
 
\newcommand{\NII}{[N~{\small II}]} 
\newcommand{\SII}{[S~{\small II}]} 
  
\newcommand{\Ha}{H$\alpha$}   
\newcommand{\Hb}{H$\beta$}   
\newcommand{\Fe}{$\langle$Fe$\rangle$}  
\newcommand{\Mgb}{Mg$b$}  
\newcommand{\Mgd}{Mg$_2$}   
 
\def\fdg{\hbox{$.\!\!^\circ$}}

\def\kms{\mathrm{km\;s^{-1}}} 
\def\kmsmpc{\mathrm{km\;s^{-1}\;Mpc^{-1}}} 
\def\lsun{L$_{\odot}$} 
\def\msun{M$_{\odot}$}

\def\kroupa{\Upsilon_\mathrm{Krou}} 
 
\def\ratml{\mldyn/\kroupa} 
\def\sigeff{\sigma_\mathrm{eff}} 
\def\reff{r_\mathrm{eff}} 
\def\avdmfrac{\langle \dmfrac \rangle}
\def\mueff{\langle \mu \rangle_\mathrm{eff}} 
\def\pa{\mathrm{P.A.}} 
\def\pastars{\mathrm{P.A.}_\mathrm{stars}} 
\def\pagas{\mathrm{P.A.}_\mathrm{gas}} 
\def\istars{i_\mathrm{stars}} 
\def\igas{i_\mathrm{gas}} 
\def\kpc{\mathrm{kpc}}
\def\zdm{z_\mathrm{DM}}
\def\zstars{z_\ast}

\def\stellar{\mathrm{\ast}}
\def\dm{\mathrm{halo}}
\def\lum{\mathrm{\ast}}

\def\mstars{M_\stellar}
\def\mldyn{\Upsilon_\stellar} 
\def\mllum{\Upsilon_\lum} 
\def\mltrue{\Upsilon_\mathrm{stars}} 
\def\rhodm{\rho_\dm} 
\def\rholum{\rho_\lum}
\def\dmfrac{f_\dm} 

\shorttitle{Early-type galaxies in Abell~262 cluster}  
\shortauthors{Wegner et al.}  
  
\begin{document}  
 
\title{Further evidence for large central mass-to-light ratios in early-type galaxies:
the case of ellipticals and lenticulars in the Abell~262 cluster\footnote{Based on data
    collected with the 2.4-m Hiltner Telescope.}}

\slugcomment{To be published in AJ. \today}  
  
\author{G. A. Wegner\altaffilmark{2}}  
\affil{Department of Physics and Astronomy, 6127 Wilder Laboratory,   
    Dartmouth College, Hanover, NH 03755-3528, USA}  
 
\author{E. M. Corsini}  
\affil{Dipartimento di Fisica e Astronomia `G. Galilei', Universit\`a di Padova, 
  vicolo dell'Osservatorio 3, 35122 Padova, Italy} 
\affil{INAF--Osservatorio Astronomico di Padova, 
  vicolo dell'Osservatorio 2, 35122 Padova, Italy} 
 
\author{J. Thomas}  
\affil{Max-Planck-Institut f\"ur extraterrestrische Physik,  
    Giessenbachstra{\ss}e, D-85748 Garching, Germany}  
 
\author{R. P. Saglia}  
\affil{Max-Planck-Institut f\"ur extraterrestrische Physik,  
    Giessenbachstra{\ss}e, D-85748 Garching, Germany}  
 
\author{R. Bender}  
\affil{Max-Planck-Institut f\"ur extraterrestrische Physik,  
    Giessenbachstra{\ss}e, D-85748 Garching, Germany}  
\affil{Universit\"ats-Sternwarte M\"unchen, Scheinerstra{\ss}e~1,  
    D-81679 M\"unchen, Germany}  
 
\author{S. B. Pu}  
\affil{The Beijing No. 12 High School, No. 15, YiZe Road, FengTai District, 
  100071 Beijing, China}  

\altaffiltext{2}{Visiting Astronomer, MDM Observatory, Kitt Peak, 
  Arizona, operated by a consortium of Dartmouth College, the University 
  of Michigan, Columbia University, the Ohio State University, and Ohio 
  University.} 
 
\begin{abstract}   
  We present new radially resolved spectroscopy of 8 early-type
  galaxies in the Abell~262 cluster. The measurements include stellar
  rotation, velocity dispersion, $H_3$ and $H_4$ coefficients of
  the line-of-sight velocity distribution along the major and minor
  axes and an intermediate axis as well as line-strength index
  profiles of Mg, Fe and H$\beta$. The ionized-gas velocity and
  velocity dispersion is measured for 6 sample galaxies along
  different axes.  We derive dynamical mass-to-light ratios and dark
  matter densities from orbit-based dynamical models, complemented by
  the galaxies' ages, metallicities, and $\alpha$-elements abundances
  from single stellar population models. The ionized-gas kinematics
  gives a valuable consistency check for the model assumptions about
  orientation and intrinsic shape of the galaxies. Four galaxies have
  a significant detection of dark matter and their halos are about 10
  times denser than in spirals of the same stellar mass. Calibrating
  dark matter densities to cosmological simulations we find assembly
  redshifts $\zdm\approx 1-3$, as previously reported for Coma. {
    The dynamical mass that follows the light is} larger than expected
  for a Kroupa stellar initial mass function (IMF), especially in
  galaxies with high velocity dispersion $\sigeff$ inside the
  effective radius $\reff$. This could indicate a `massive' IMF in
  massive galaxies. Alternatively, some of the dark matter in massive
  galaxies could follow the light very closely. In combination with
  our comparison sample of Coma early-type galaxies, we now have 5 of
  24 galaxies where (1) mass follows light to $1-3\,\reff$, (2) the
  dynamical mass-to-light ratio { of all the mass that follows the
    light is large ($\approx\,8-10$ in the Kron-Cousins $R$ band)},
  (3) the dark matter fraction is negligible to $1-3\,\reff$. Unless
  the IMF in these galaxies is particularly `massive' and somehow
  coupled to the dark matter content, there seems to be a significant
  degeneracy between luminous and dark matter in at least some
  early-type galaxies. { The role of violent relaxation is briefly
  discussed.}
\end{abstract}  
  

\section{Introduction} 
\label{sec:introduction} 

The distribution of dark matter surrounding elliptical galaxies as
well as the dynamical structure of these systems is not known in
general. While the dynamical structure contains information about the
assembly mechanism of elliptical galaxies, the mass distribution can
be used to constrain the assembly epoch. For example, the simple
spherical collapse model predicts that the central dark matter density
scales with $(1+z_{\rm DM})^3$, where $z_{\rm DM}$ is the assembly
redshift of the halo. Numerical simulations likewise predict a close
relationship between halo concentration and assembly epoch
\citep{Wechsler2002}.

Dynamical modeling of stellar kinematics allows the reconstruction of
both the orbital structure and the mass distribution in elliptical
galaxies, and, thus offers crucial information about when and how
elliptical galaxies formed.  In the past years we published a series
of papers \citep{Thomas2004, Thomas2005, Thomas2007a, Thomas2007b,
  Thomas2009a, Thomas2009b, Thomas2011} studying the distribution of
dark matter and the dynamical structure of the elliptical galaxies of
the Coma cluster. For the dynamical analysis we employed a refined
three-integral axisymmetric orbit superposition code to model the
systems of stars in order to determine their mass compositions and
orbital structures. For each trial potential on a grid systematically
probing { a mass component that follows the light (with
  mass-to-light ratio $\mldyn$)} and different dark matter profiles,
we determined an orbit superposition that best matched the kinematic
data published by \citet{Mehlert2000}, \citet{Wegner2002}, and
\citet{Corsini2008} with the additional requirement of maximizing the
model's entropy. The optimal entropy maximalization and fit to the
data were calibrated using Monte-Carlo simulations of observationally
motivated galaxy models. These simulations revealed that from typical
Coma data, the mass distribution can be reconstructed with an accuracy
of $\approx\,15\%$. The comparison with the masses derived through
strong gravitational lensing for elliptical galaxies with similar
velocity dispersion confirmed this result.

We compared $\mldyn$ to the mass-to-light ratio $\Upsilon_{\rm SSP}$
predicted by the best-fit single stellar population (SSP) models to
the line-strength indices as in \citet{Mehlert2003}. We found that at
face value a Salpeter initial mass function (IMF,
\citealt{Salpeter1955}) gives on average the best agreement between
$\mldyn$ and $\Upsilon_{\rm SSP}$. However, several arguments favor a
Kroupa IMF \citep{Kroupa2001} for the underlying true stellar mass
distribution.

We derived dark matter densities for the Coma ellipticals that are on
average at least $13$ times higher than in spirals of the same stellar
mass, consistent with the findings of \citet{Gerhard2001} for round
galaxies.  We concluded that elliptical galaxy halos have assembled
earlier, around $z_{\rm DM}\,\approx\,1-3$. This agrees with the
predictions of the semi-analytic galaxy formation models (SAMs) of
\citet{DeLucia2006}, with simulated ellipticals in a Coma-like cluster
environment. Halos form hierarchically -- less massive halos first. In
contrast, studies on the star formation histories based on line
indices \citep{Nelan2005, ThomasD2005} show that stars form
anti-hierarchically -- the least massive galaxies have the youngest
stellar populations.  { This would result from star formation
  following the hierarchical mass assembly to the time that the
  available gas is consumed or expelled; early star formation would
  occur in low-mass systems that subsequently merge forming massive
  systems and today, low-mass galaxies are generally late-assembling
  galaxies with younger stellar populations.}  We found that in a
third of our Coma galaxies star-formation and halo assembly redshifts
match. The youngest galaxies must have experienced some star-formation
after the main halo-assembly epoch -- in line with their orbital
structure being inconsistent with collisionless $N$-body mergers with
no star formation of disk galaxies \citep{Thomas2009b}.
 
The Coma cluster represents a high density local cluster sample.  In
lower-density local environments, the stellar populations of
ellipticals are formed at $z_\ast\,\le\,1$ \citep{Wegner2008} { and
  have a greater spread of ages -- the oldest are about as old as
  cluster ellipticals, but the youngest are much younger
  \citep{Collobert2006}.}  Dark matter densities of ellipticals there
are predicted to be lower, because they assemble later. These
predictions need now to be verified by models of real galaxies.
 
Here we present the new data set we collected for a sample of
early-type galaxies of the nearby poor cluster Abell~262. The results
of the dynamical analysis we performed are compared to the Coma
cluster case.
Abell~262 is one of the most conspicuous condensations in the
Pisces-Perseus supercluster. The center of the cluster coincides with
the position of the cD galaxy NGC~708. Abell~262 is far less densely
populated than the Coma cluster and is comparable to the Virgo
cluster.  Indeed, it falls into richness class 0 \citep{Abell1989}.
The cluster redshift of $z\,=\,0.0163$ \citep{Struble1999} results
in a distance of 70 Mpc, assuming $H_0\,=\,70\;\kmsmpc$ . At this
redshift the angular distance $1''$ corresponds to 339 pc. The cluster
mass within the virial radius ($R_{\rm vir}\,=\,1\fdg52$ corresponding
to 1.9 Mpc) and calculated from the line-of-sight velocity dispersion
of the cluster galaxies ($\sigma\,=\,548\pm35\;\kms$) is $M_{\rm
  vir}\,=\,(2.5\pm0.5)\times10^{14}$ \msun . It is in agreement with
the mass estimated from X-ray data assuming hydrostatic equilibrium
\citep{Neill2001}.
The low X-ray luminosity \citep{David1993}, low temperature of the hot
intracluster medium \citep{Sato2009} with a complex structure
\citep{Clarke2009}, and a central cooling flow \citep{Blanton2004} are
the typical signatures of a less evolved, dynamically young cluster.
 
The structure of the paper is as follows. In Sec.~\ref{sec:sample} we
present the galaxy sample. We discuss the observations, data reduction
and analysis of the imaging dataset in Sec.~\ref{sec:imaging} and of
the spectroscopic dataset and the SSP analysis of the line-strength
indices in Sec.~\ref{sec:spectroscopy}.  A summary of the dynamical
modeling technique is given in Sec.~\ref{sec:dynamics} and a
comparison with the gas kinematics and strong gravitational lensing
masses follows in Sec.~\ref{sec:comparison}. The results are presented
in Sec.~\ref{sec:results} and the implications discussed in
Sec.~\ref{sec:discussion}. A summary of the paper follows in
Sec.~\ref{sec:conclusions}.

\section{Galaxy sample} 
\label{sec:sample} 
 
All the sample galaxies in Abell 262 were studied within the EFAR
(Ellipticals FAR away) project. The latter was aimed at constructing
an accurate and homogeneous photometric and spectroscopy database for
a large sample of early-type galaxies at redshift between 6000 and
$15000\;\kms$ and distributed in 85 clusters in the Hercules-Corona
Borealis and Perseus-Pisces-Cetus regions.
Details about the selection of the EFAR galaxy and cluster sample are
given by \citet{Wegner1996}. \citet{Saglia1997a} and
\citet{Wegner1999} presented the photometric and spectroscopic
databases, respectively. Results addressed the galaxy structural
parameters \citep{Saglia1997b}, the properties of the stellar
populations \citep{Colless1999}, the fundamental plane, peculiar
velocities, and bulk motions \citep{Saglia2001, Colless2001}.
 
The galaxies studied here were selected to be round and classified as
early-type (cD, E, and E/S0) in \citet{Wegner1996}. Their basic
properties which include morphological type, structural parameters,
total magnitude, velocity dispersion, and redshift can be found in
Table~\ref{tab:properties}.

\begin{table*}  
\caption{Properties of the sample galaxies \label{tab:properties}}     
\begin{small} 
\begin{tabular}{lllrccrccr}    
\tableline  
\tableline   
\multicolumn{1}{c}{Object} & \multicolumn{1}{c}{Alt. Name} &  
\multicolumn{1}{c}{Type} & \multicolumn{1}{c}{$\reff$} &   
\multicolumn{1}{c}{$\mueff$} &  
\multicolumn{1}{c}{$R_{\rm T}$} & \multicolumn{1}{c}{P.A.} &  
\multicolumn{1}{c}{$cz$} & \multicolumn{1}{c}{$\sigma$} & 
\multicolumn{1}{c}{$d$} \\   
\multicolumn{1}{c}{} & \multicolumn{1}{c}{} & 
\multicolumn{1}{c}{} & \multicolumn{1}{c}{(arcsec)} & 
\multicolumn{1}{c}{(mag arcsec$^{-1}$)} &  
\multicolumn{1}{c}{(mag)} & \multicolumn{1}{c}{($^\circ$)} &  
\multicolumn{1}{c}{($\kms$)} & \multicolumn{1}{c}{($\kms$)} & 
\multicolumn{1}{c}{(Mpc)}\\ 
\multicolumn{1}{c}{(1)} & \multicolumn{1}{c}{(2)} &   
\multicolumn{1}{c}{(3)} & \multicolumn{1}{c}{(4)} &   
\multicolumn{1}{c}{(5)} & \multicolumn{1}{c}{(6)} &   
\multicolumn{1}{c}{(7)} & \multicolumn{1}{c}{(8)} & 
\multicolumn{1}{c}{(9)} & \multicolumn{1}{c}{(10)} \\   
\tableline    
\object{IC   171} & \object {Abell~262 E} & E    & 30.31 & 20.47 & 11.16 &     108 & 5346 & 178 & 1.22\\ 
\object{NGC  679} & \object {Abell~262 D} & E    & 10.37 & 18.80 & 11.79 &      85 & 5036 & 229 & 0.88\\ 
\object{NGC  687} & \object {Abell~262 C} & E/S0 & 12.79 & 19.19 & 11.74 & \nodata & 5086 & 224 & 0.61\\ 
\object{NGC  703} & \object {Abell~262 I} & E/S0 &  9.41 & 19.45 & 12.65 &      47 & 5584 & 180 & 0.04\\ 
\object{NGC  708} & \object {Abell~262 A} & cD   & 33.03 & 21.00 & 11.48 &      39 & 4845 & 194 & 0\\ 
\object{NGC  712} & \object {Abell~262 F} & E/S0 & 10.31 & 19.18 & 12.19 &      83 & 5330 & 221 & 0.82\\ 
\object{NGC  759} & \object {Abell~262 G} & E/S0 & 16.24 & 19.72 & 11.74 & \nodata & 4644 & 241 & 1.27\\ 
\object{UGC 1308} & \object {Abell~262 B} & E    & 23.88 & 19.44 & 10.63 &     140 & 5227 & 216 & 0.50\\ 
\tableline    
\end{tabular}     
\end{small} 
\tablecomments{    
Col. 1: Name. 
Col. 2: Alternative name from \citet{Wegner1996}. 
Col. 3: Morphological type from \citet{Wegner1996}. 
        { In spite of the cD classification, the properties of
          NGC~708 are not particularly extreme with respect to the
          other Abell~262 galaxies listed here.}
Col. 4: Effective radius obtained by \citet{Saglia1997a}  
        in the Kron-Cousins $R$ band. 
Col. 5: Average surface brightness within the effective radius  
        obtained by \citet{Saglia1997a} in the Kron-Cousins $R$ band 
        after applying the Galactic absorption, $K_R$, and cosmological 
        dimming corrections. 
Col. 6: Total magnitude measured by \citet{Saglia1997a} in the  
        Kron-Cousins $R$ band after applying 
        the Galactic absorption and $K_R$ corrections. 
Col. 7: Major-axis position angle measured North through East  
        from HyperLeda catalog \citep{Paturel2003}.         
Col. 8: Heliocentric systemic velocity from \citet{Wegner1999}. 
Col. 9: Velocity dispersion measured by by \citet{Wegner1999} after applying  
        the correction to the standard metric aperture of 0.54 $h^{-1}$ kpc  
        in radius.  
Col. 10: Projected distance from NGC~708.} 
\end{table*}    

\section{Observations, data reduction, and analysis: imaging} 
\label{sec:imaging} 
 
\subsection{Observations and data reduction} 
 
As part of the HST Snapshot Proposal 10884 (P.I. G. A. Wegner), the
galaxies NGC~679, NGC~687, NGC~708, NGC~759, and UGC~1308 were
observed with Wide Field Planetary Camera 2 (WFPC2) on board the HST
between 2007 July 14 and 2008 September 27. For each galaxy two 300-s
exposures were taken with the filter F622W. All exposures were
performed with the telescope guiding in fine lock, which typically
gave an rms tracking error of $0\farcs003$.
The centers of the galaxies were positioned on the Planetary Camera
chip (PC) in order to get the best possible spatial resolution. This
consists of $800\,\times\,800$ pixels of
$0\farcs0455\,\times\,0\farcs0455$ each, yielding a field of view of
about $36''\,\times\,36''$. The two pointings were shifted along the
pixel diagonal by $0\farcs3535$. This pattern corresponds to a shift
of $5.5\,\times\,5.5$ pixels on the PC chip.
  
In the following our photometric analysis is limited to the PC chip,
since the nuclear surface brightness profiles were matched to
available radially extended ground-based photometry. Indeed, all the
sample galaxies were observed in the $R$ band of the Kron-Cousins
system as a part of the EFAR project \citep{Colless1993, Saglia1997a}.
  
The WFPC2 images were reduced using the CalWFPC reduction pipeline in
IRAF\footnote{Imaging Reduction and Analysis Facilities (IRAF) is
  distributed by the National Optical Astronomy Observatories which
  are operated by the Association of Universities for Research in
  Astronomy (AURA) under cooperative agreement with the National
  Science Foundation.} maintained by the Space Telescope Science
Institute. Reduction steps include bias subtraction, dark current
subtraction, and flat-fielding, as described in detail in the WFPC2
instrument and data handbooks \citep{Baggett2002, McMaster2008}.
Subsequent analysis was performed using IRAF standard tasks. The bad
pixels were corrected by means of a linear one-dimensional
interpolation using the data quality files and the WFIXUP task. For
each galaxy, the alignment of the images was checked by comparing the
centroids of stars in the field of view. The images were aligned to an
accuracy of a few hundredths of a pixel using IMSHIFT and knowledge of
the offsets. They were then combined with IMCOMBINE. A check was done
to verify that the alignment and combination did not introduce a
significant blurring of the data. To this aim, the full-width at half
maximum (FWHM) of the Gaussian fitting to the field stars was measured
in the original and combined frames. It was found that they did not
change to within a few percent. In combining the images, pixels
deviating by more than three times the local standard deviation
(calculated from the combined effect of Poisson and read-out noise)
were flagged as cosmic rays and rejected. The residual cosmic rays and
bad pixels were corrected by manually editing the resulting image with
IMEDIT.

\begin{figure*}[t!]
\centering
\includegraphics[width=0.8\textwidth]{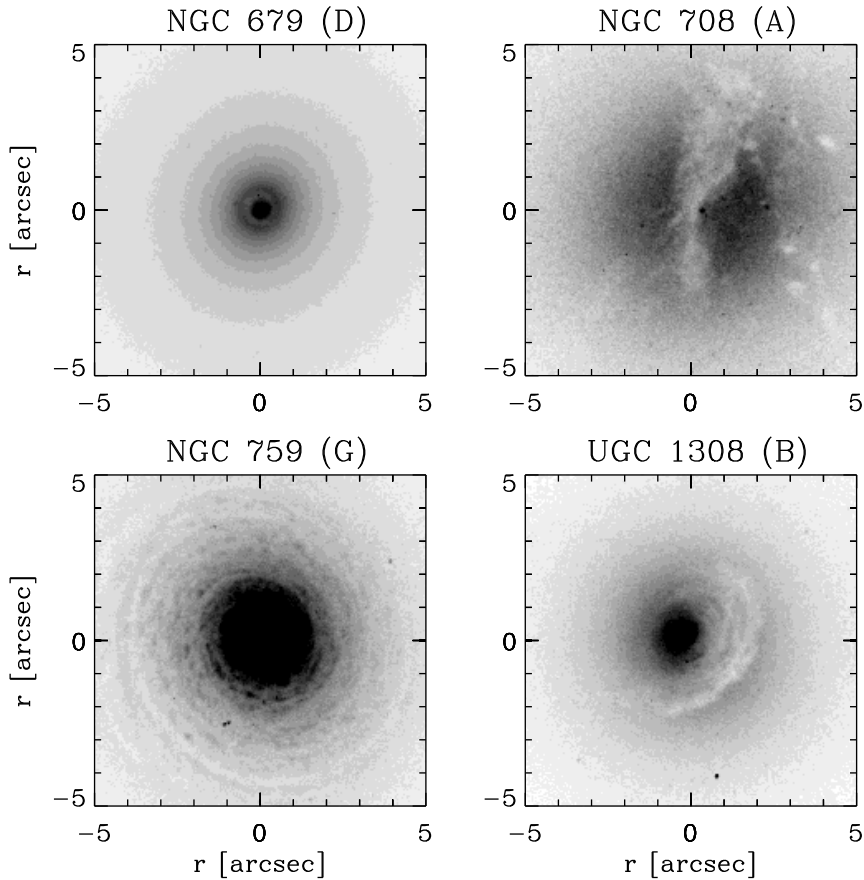}
\caption{\footnotesize Central portions of HST WFPC2/F622W images of
  the galaxies showing dust features. Each frame uses a negative
  intensity scale and North is at the top with East to the left.}
\label{fig:dustdisks}
\end{figure*}

\subsection{Isophotal analysis} 
 
The reduced EFAR images were available to one of us (RPS). To
calibrate the images the photometric zero-points in
\citet{Saglia1997a} were adopted and their coefficients were applied
to correct the measured magnitudes for Galactic absorption and $K_R$
correction and to correct the measured surface brightnesses for
Galactic absorption, $K_R$ correction, and cosmological dimming. The
EFAR images were used to determine the zero-points and calibrate the
WFPC2 images.
 
To this aim, the isophotal profiles of the galaxies were measured by
fitting the isophotes in the WFPC2 and EFAR images with ellipses using
the isophotal shape algorithm described by \citet{Bender1987}. For the
fitted isophotes the algorithm provides the azimuthally-averaged
surface brightness ($\mu$), ellipticity ($\epsilon$), position angle
(P.A.), center coordinates ($x_0,y_0$) and third, fourth, and sixth
cosine ($a_3$, $a_4$, and $a_6$) and sine ($b_3$, $b_4$, and $b_6$)
Fourier coefficients describing the deviation of isophotal shape from
perfect ellipse.
Foreground stars, residual cosmic rays, and bad pixels were masked
before fitting. Moreover, the centers of ellipses were allowed to
vary. As a first step, the sky background was measured in regions free
of sources at the edges of the field of view and then subtracted. Its
final value in the WFPC2 images was determined through the matching of
the ground-based photometry.
 
The surface-brightness radial profiles measured in the WFPC2 images
were matched to the EFAR ones typically between $3''$ and $10''$ by
determining the zero-point and sky value that minimize the
surface-brightness flux differences. The galaxies were assumed to have
no $R - m_{\rm F622W}$ color gradient in the matched radial
region. The resulting distribution of the differences $\Delta m = R -
m_{\rm F622W}$ after minimization gives a standard deviation about the
mean $\sigma_{\Delta m} = 0.035$ mag for NGC~679, $\sigma_{\Delta m} =
0.017$ mag for NGC~687, and $\sigma_{\Delta m} = 0.050$ mag for
UGC~1308. Therefore, the scatter $\sigma_{\Delta m}$ is the dominant
source of error in the zero-point of the WFPC2 images, because the
EFAR final data set has a common zero-point to better than 0.01 mag
\citep{Saglia1997a}.

The prominent dust lanes crossing the nucleus of NGC~708
(Fig.~\ref{fig:dustdisks}) did not allow us to measure reliable
isophotal parameters within a few arcsecs from the galaxy center in
both the WFPC2 and EFAR images. To minimize the impact of dust on the
measurement of the isophote shape, we analyzed the $H$-band image
obtained by \citet{Gavazzi1996}, retrieved from the GOLDMine Archive
\citep{Gavazzi2003}.  { The near-infrared data were taken on 1995
  February 13 at Calar Alto Observatory, Spain. The 2.2-m telescope
  mounted the Max-Planck-Institut f\"ur Astronomie General-purpose
  Infrared Camera (MAGIC) with a Rockwell $256\,\times\,256$ NICMOS3
  HgCdTe detector array. The spatial scale was $1\farcs61$
  pixel$^{-1}$ yielding a field of view of
  $6\farcm8\,\times\,6\farcm8$. The total exposure times was 192 s. A
  two-dimensional fit with a circular Gaussian to the field stars in
  the resulting image yielded a $\rm FWHM\,=\,2\farcs2$.}  The
surface-brightness radial profiles measured in the GOLDMine and EFAR
images were matched between $5''$ and $26''$ with $\langle \Delta m
\rangle = +0.006$ mag and $\sigma_{\Delta m} = 0.026$ mag.
 
For NGC~679, NGC~687, NGC~759, and UGC~1308, the position angles
measured in the available HST and ground-based available images match
within 2$^\circ$, ellipticities within less than 0.05, cosine and sine
Fourier coefficients within less than $1\%$, and center coordinates of
the fitting ellipses within $0\farcs2$.
 
The radial profiles of the azimuthally averaged surface brightness,
ellipticity, position angle, center coordinates and third, fourth, and
sixth cosine and sine Fourier coefficients of the sample galaxies are
presented in Fig.~\ref{fig:photometry} and
Table~\ref{tab:photometry}. For NGC~679, NGC~687, NGC~759, and
UGC~1308 they are the combination of HST and EFAR data.  For NGC~708
they are measured on the GOLDMine image. For IC~171, NGC~703, and
NGC~712 they are obtained from the EFAR images.

\subsection{Morphologies of the galaxies with dust} 
 
Classed as early-type galaxies on ground-based data
(\citealt{Wegner1996}, but see also \citealt{RC3}), the HST images
show that four of the galaxies in this investigation, NGC~679, NGC
708, UGC 1308, and NGC 759, have absorbing material in their central
regions.  The images are shown in Fig.~\ref{fig:dustdisks}. These
resemble the dust features described elsewhere
\citep[e.g.,][]{Martini2003, Lauer2005} and found in two of our Coma
cluster galaxies \citep{Corsini2008}. Only NGC~687 shows no dust
features in its nucleus.
  
\paragraph{NGC~679.} 
The nucleus of the galaxy hosts a nearly face-on ring of dust. It has
a radius of $4\farcs5$ (1.5 kpc).

\paragraph{NGC~708.} 
The strong dark stripe of NGC~708 appears to be consistent with a dust
disk seen nearly edge-on as it runs roughly North-South through the
center of the galaxy image. It extends out to $3\farcs2$ (1.1 kpc) on
its South side and $5\farcs8$ (2.0 kpc) on the North side.  Individual
dark blobs can be seen along the edge of the band, particularly to the
West side.
 
\paragraph{NGC~759.} 
The galaxy has a nuclear dusty disk which is close to being
face-on. It has a radius of $5\farcs0$ (1.7 kpc) and it is
characterized by tightly wound multiple spiral arms.

\paragraph{UGC~1308.} 
The galaxy nucleus appears to have an inclined dust disk. The
projected elliptical shape has a major axis of $3\farcs2$ (1.1 kpc)
across in the North-South direction and minor axis $2\farcs6$ (0.9
kpc). It is asymmetric in that it is stronger on the West side and is
not present on the East side.

\begin{figure*}[t!] 
\includegraphics[width=0.5\textwidth]{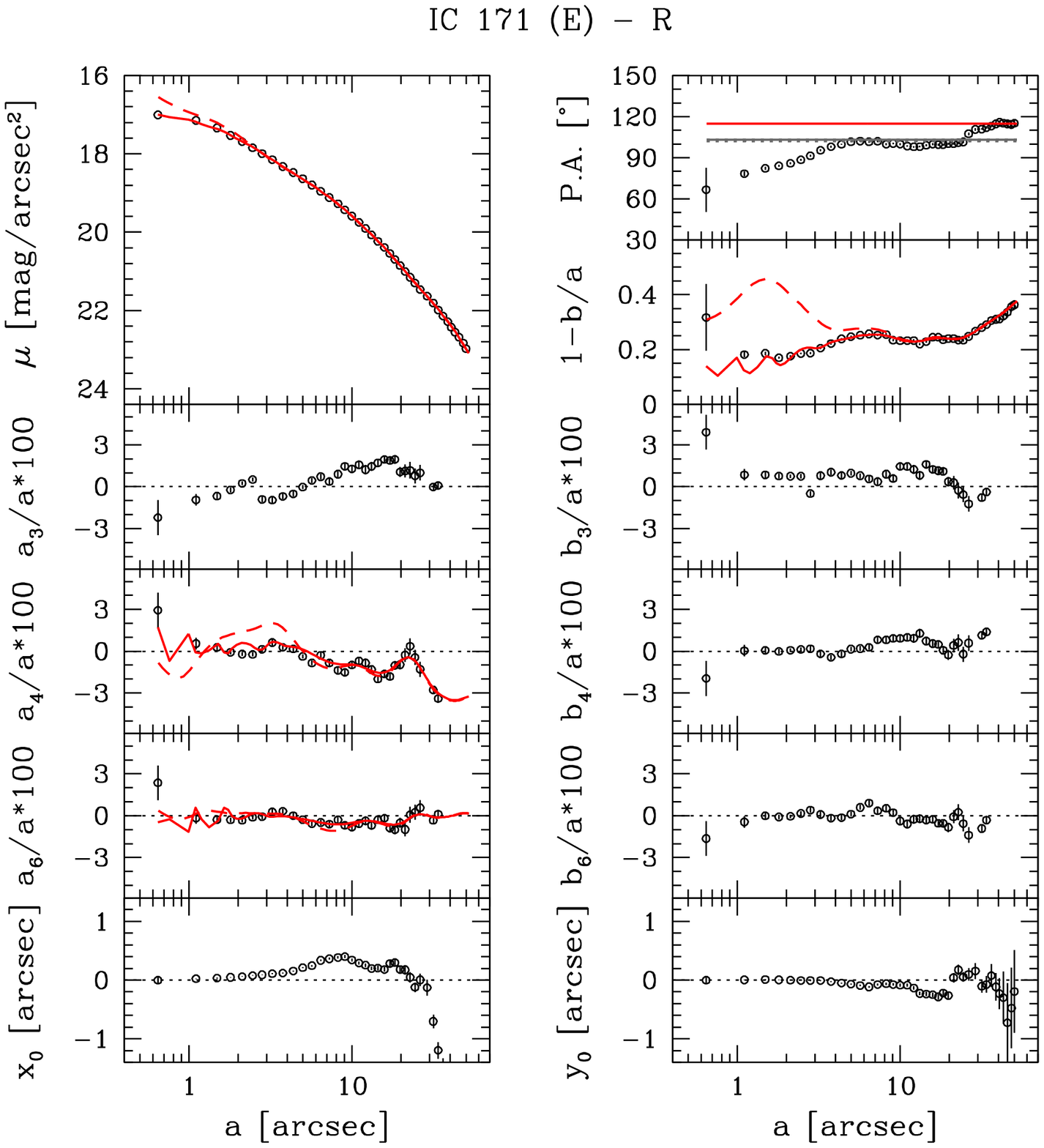} 
\includegraphics[width=0.5\textwidth]{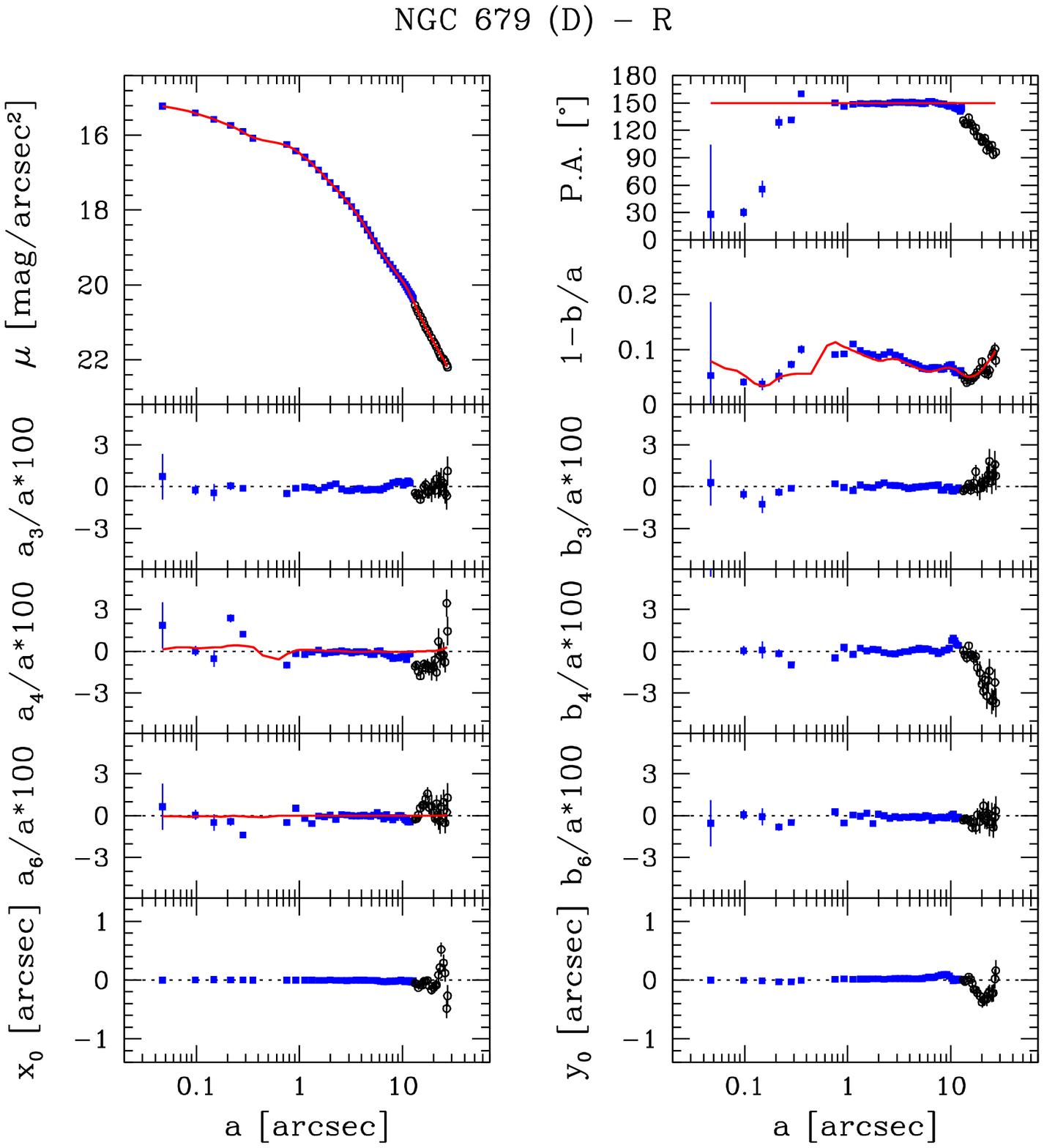}  

\caption{Isophotal parameters of the galaxies from ground-based (black 
  open circles) and HST/WFPC2 images (blue filled squares) as a 
  function of the logarithm of the semi-major axis distance in arcsec. 
  The radial profiles of Kron-Cousins $R$-band surface brightness
  ($\mu$), third, fourth, and sixth cosine Fourier coefficients
  ($a_3$, $a_4$, and $a_6$), and $x-$coordinate of the center ($x_0$)
  are plotted in the left panels (from top to bottom).
  The radial profiles of position angle (P.A.), ellipticity ($1-b/a$),
  third, fourth, and sixth sine Fourier coefficients ($b_3$, $b_4$,
  and $b_6$), and $y-$coordinate of the center ($y_0$), are plotted in
  the right panels (from top to bottom). 
  For galaxies with only ground-based photometry, the solid and dashed
  red lines correspond to the surface brightness, ellipticity, $a_4$,
  and $a_6$ profiles obtained by projecting the deprojected stellar
  luminosity density with and without the correction for seeing
  convolution, respectively. For the remaining galaxies, the solid red
  lines are obtained by projecting the stellar luminosity density
  without the correction for seeing convolution. The solid grey line
  corresponds to the galaxy position angle ($\pagas$) derived by
  minimizing the difference between the rotation velocity of the
  ionized gas and the circular velocity from the dynamical model (see
  Sec. \ref{sec:gasresults} for details). The grey dotted lines
  delimit the $68\%$ confidence region for $\pagas$.}
\label{fig:photometry}  
\end{figure*}  
 
\addtocounter{figure}{-1} 
\begin{figure*}[t!]  
\epsscale{1.0}  
\includegraphics[width=0.5\textwidth]{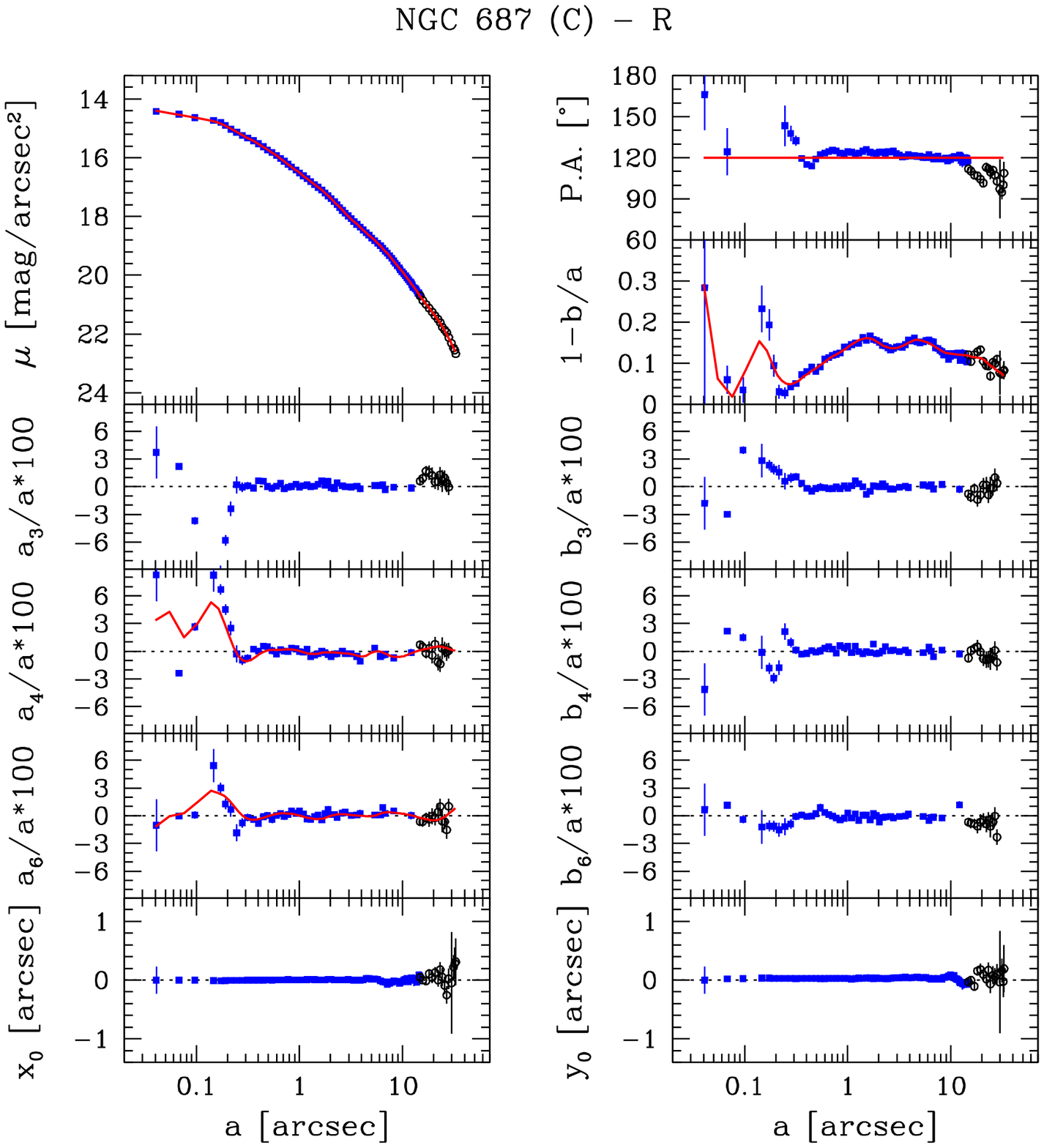} 
\includegraphics[width=0.5\textwidth]{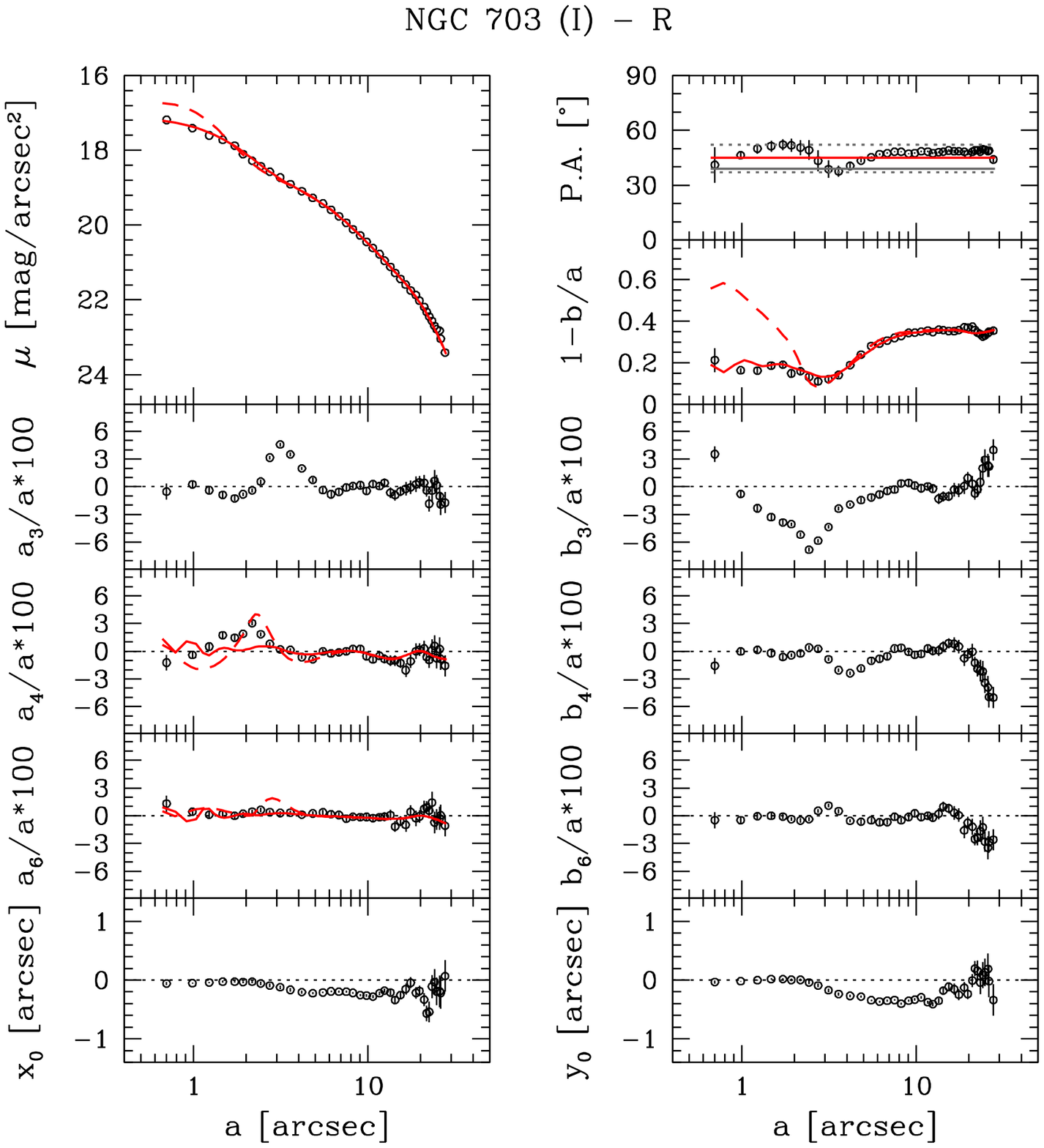} 
\includegraphics[width=0.5\textwidth]{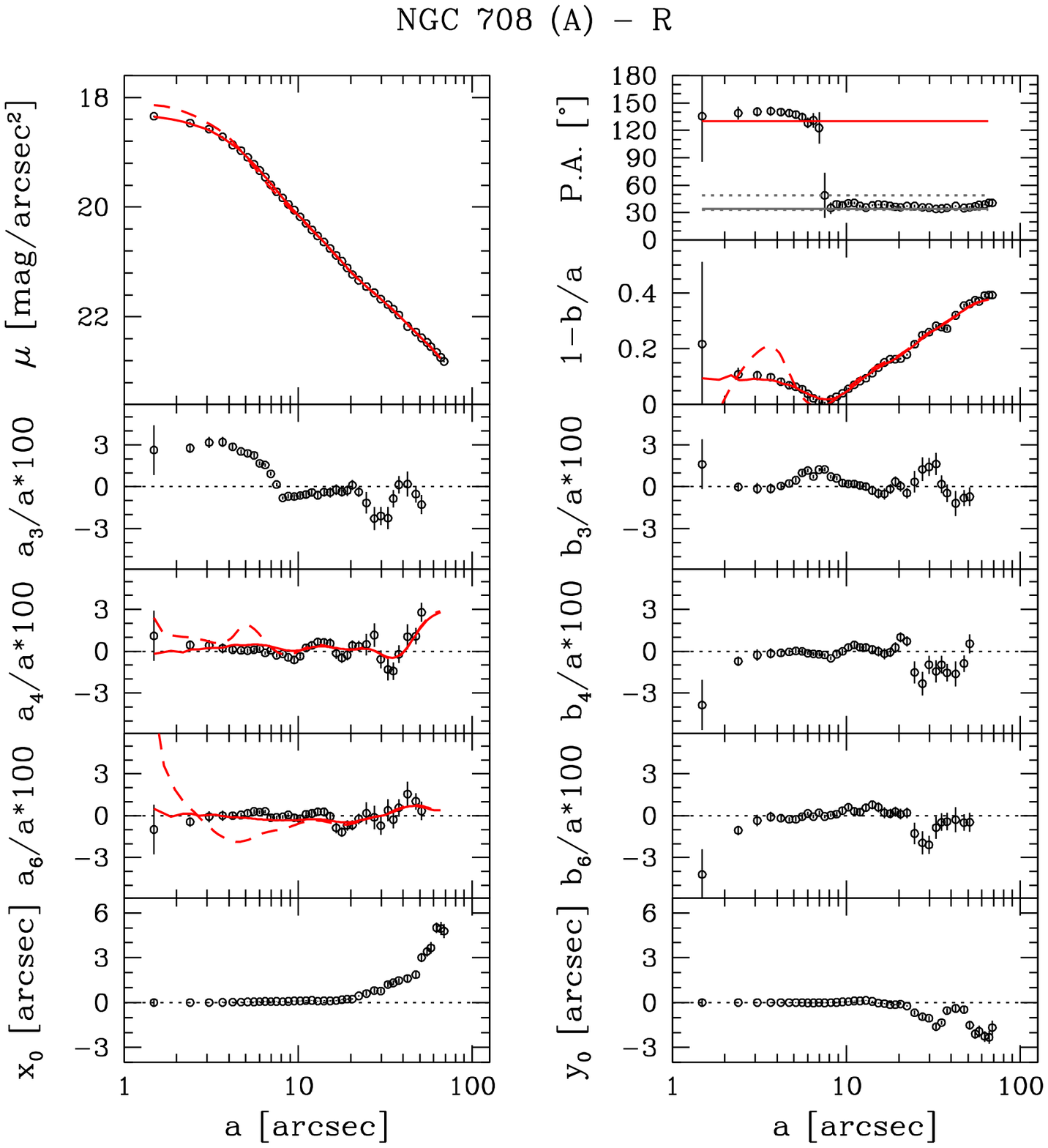} 
\includegraphics[width=0.5\textwidth]{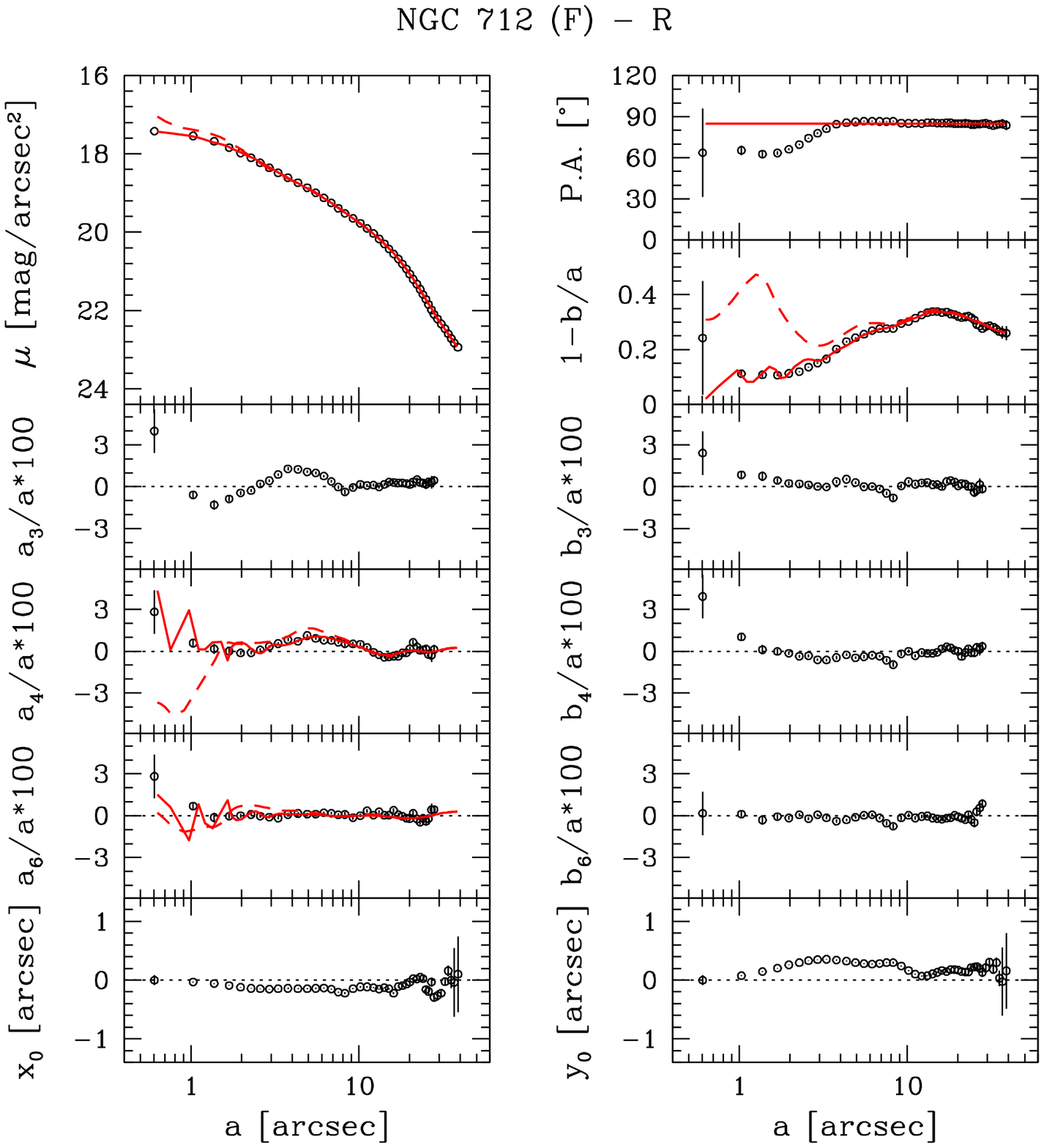} 
\caption{{\em Continued.}}
\end{figure*}  

\addtocounter{figure}{-1} 
\begin{figure*}[t!]  
\epsscale{1.0}  
\includegraphics[width=0.5\textwidth]{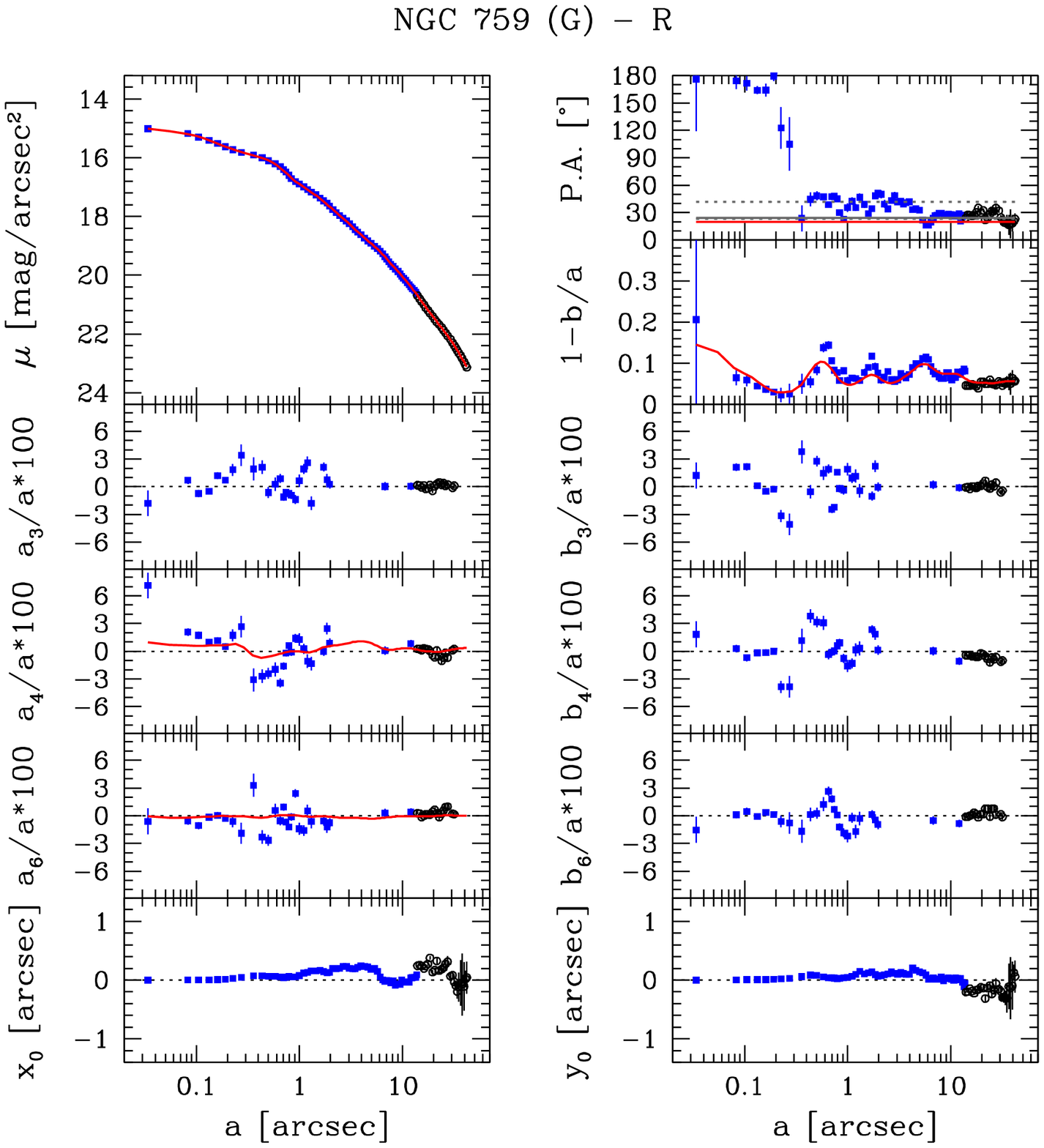} 
\includegraphics[width=0.5\textwidth]{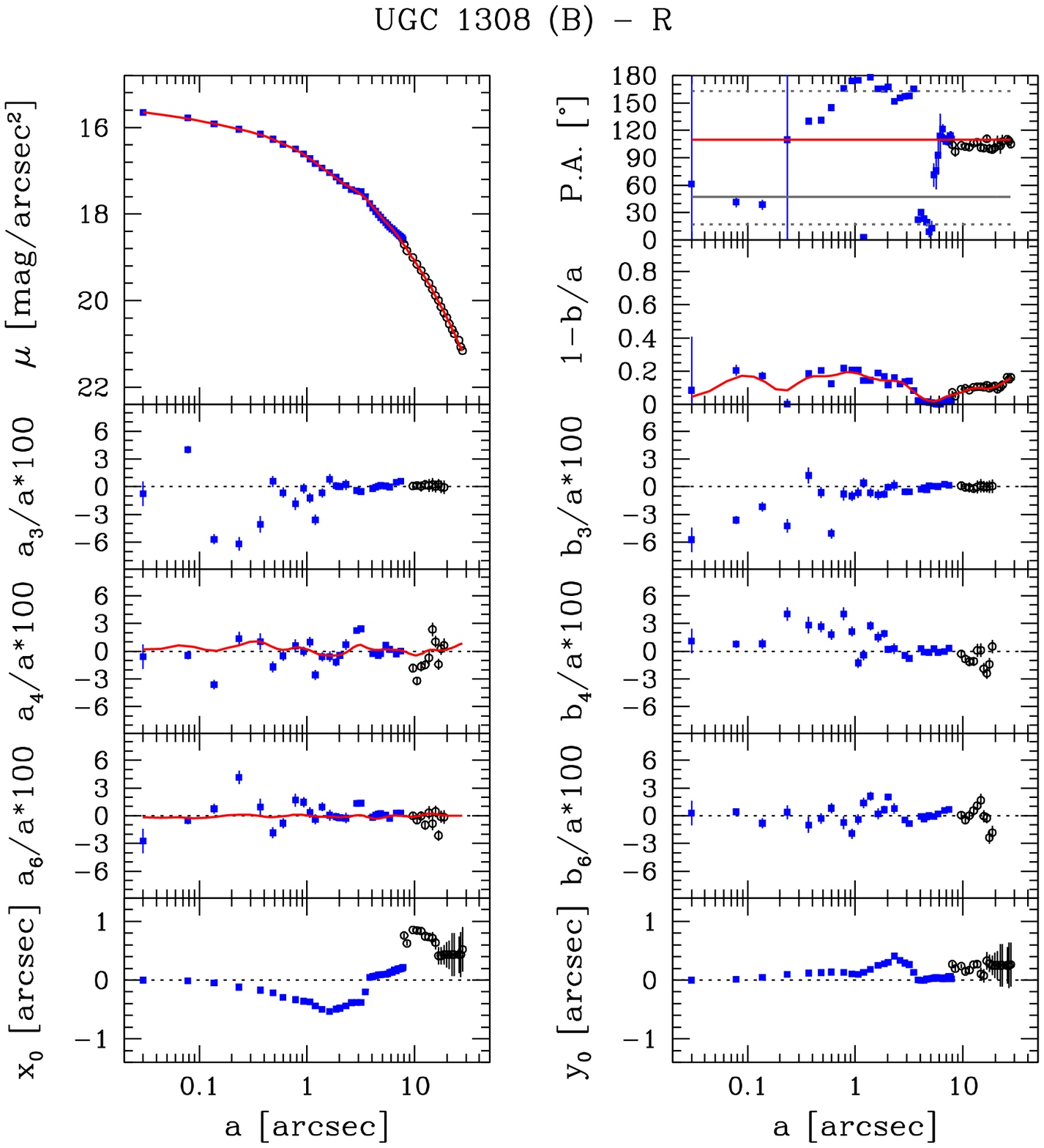} 
\caption{{\em Continued.}}
\end{figure*}  

\section{Observations, data reduction, and analysis: spectroscopy} 
\label{sec:spectroscopy} 
 
\subsection{Observations and data reduction} 
 
Long-slit spectroscopic data of the sample galaxies were obtained with
with the 2.4-m Hiltner telescope of the MDM Observatory at Kitt Peak,
Arizona on 2003 November 14-23 (run 1), 2005 October 25-31 (run 2),
and 2006 November 10-16 (run 3).
 
In runs 1 and 2 the telescope mounted the Moderate Resolution
Spectrograph with a plane reflection grating with 1200 grooves
mm$^{-1}$ blazed at 5000 \AA\ at the first order and a
$1\farcs9\,\times\,9\farcm6$ slit. The ``Echelle'' CCD was adopted as
detector. It is a thinned and backside illuminated Site CCD with
$2048\times2048$ pixels of $24\,\times\,24$ $\mu$m$^2$. The gain and
read-out noise are 2.7 $e^-$ ADU$^{-1}$ and 7.9 $e^-$ (rms),
respectively.
In run 3 the Boller \& Chivens CCD Spectrograph was mounted with a
plane reflection grating with 600 grooves mm$^{-1}$ blazed at 5875
\AA\ in the second order. A $1\farcs7\,\times\,5\farcm2$ slit with a
LG370 order blocking filter was used.  The ``Ohio State University
Loral C'' CCD was the detector. The latter is a thinned and backside
illuminated Loral CCD with $1200\,\times\,800$ pixels of
$15\,\times\,15$ $\mu$m$^2$. The gain and read-out noise are 2.1 $e^-$
ADU$^{-1}$ and 7.0 $e^-$ (rms), respectively.
No pixel binning was adopted. The wavelength range from 4670 to 6541
\AA\ was covered with a reciprocal dispersion of 0.870
\AA\ pixel$^{-1}$ in run 1, from 4675 to 6530 \AA\ with 0.906
\AA\ pixel$^{-1}$ in run 2, and from 6270 to 7175 \AA\ with 0.755
\AA\ pixel$^{-1}$ in run 3. The spatial scale was $0\farcs606$
pixel$^{-1}$ in runs 1 and 2, and $0\farcs41$ pixel$^{-1}$ in run 3.
 
In runs 1 and 2 the spectra were obtained along the major, minor, and
a diagonal axis for all the sample galaxies, except for UGC~1308 which
was observed only along the major and minor axis.
The major axis of IC~171 and NGC~679 and the minor axis of UGC~1308
were observed in both runs to perform a consistency check between the
measurements of stellar kinematics and line-strength indices of the
two runs.
In run 3 the spectra were obtained along a diagonal axis. Only NGC~708
was also observed also along the major and minor axes. For NGC~759 the
spectra were also taken close to the major axis and along the minor
axis.
The integration time of the single galaxy spectra was 3600 s in
  runs 1 and 2 and 1800 s in run 3. Total integration times and slit
position angle of the galaxy spectra as well as the log of the
spectroscopic observations are given in Table~\ref{tab:log}.

\begin{table*}[t!]  
\caption{Log of the spectroscopic observations \label{tab:log}}     
\begin{tabular}{lcrcccc}    
\tableline  
\tableline   
\multicolumn{1}{c}{Object} & \multicolumn{1}{c}{Run} &  
\multicolumn{1}{c}{P.A.} & \multicolumn{1}{c}{Position} &   
\multicolumn{1}{c}{Single Exp. Time} & \multicolumn{1}{c}{Total Exp. Time} &  
\multicolumn{1}{c}{Q}\\   
\multicolumn{1}{c}{} & \multicolumn{1}{c}{} &  
\multicolumn{1}{c}{($^\circ$)} & \multicolumn{1}{c}{} &  
\multicolumn{1}{c}{(s)} & \multicolumn{1}{c}{(h)} &  
\multicolumn{1}{c}{} \\   
\multicolumn{1}{c}{(1)} & \multicolumn{1}{c}{(2)} &   
\multicolumn{1}{c}{(3)} & \multicolumn{1}{c}{(4)} &   
\multicolumn{1}{c}{(5)} & \multicolumn{1}{c}{(6)} &   
\multicolumn{1}{c}{(7)} \\ 
\tableline    
\object{IC   171 (E)} & 3 &   0 & DG & $3\times1800$ & 1.5 & 4 \\ 
                      & 2 &  10 & DG & $3\times3600$ & 3.0 & 2 \\ 
                      & 1 & 110 & MJ & $4\times3600$ & 4.0 & 2 \\ 
                      & 2 & 110 & MJ & $1\times3600$ & 1.0 & 2 \\ 
                      & 2 & 145 & DG & $3\times3600$ & 3.0 & 2 \\ 
\object{NGC  679 (D)} & 1 &   0 & DG & $3\times3600$ & 3.0 & 1 \\ 
                      & 2 &   0 & DG & $1\times3600$ & 1.0 & 1 \\ 
                      & 3 &   0 & DG & $2\times1800$ & 1.0 & 4 \\ 
                      & 2 &  90 & DG & $3\times3600$ & 3.0 & 2 \\ 
                      & 2 & 135 & MJ & $4\times3600$ & 4.0 & 2 \\ 
\object{NGC  687 (C)} & 1 &   0 & DG & $3\times3600$ & 3.0 & 1 \\ 
                      & 3 &   0 & DG & $2\times1800$ & 1.0 & 4 \\ 
                      & 2 &  20 & MN & $3\times3600$ & 3.0 & 1 \\ 
                      & 2 & 110 & MJ & $3\times3600$ & 3.0 & 1 \\ 
\object{NGC  703 (I)} & 2 &   0 & DG & $3\times3600$ & 3.0 & 2 \\ 
                      & 3 &   0 & DG & $2\times1800$ & 1.0 & 4 \\ 
                      & 1 &  45 & MJ & $3\times3600$ & 3.0 & 1 \\
                      & 2 & 135 & MN & $4\times3600$ & 4.0 & 2 \\ 
\object{NGC  708 (A)} & 1 &   0 & DG & $1\times3600$ & 1.0 & 3 \\ 
                      & 3 &   0 & DG & $3\times1800$ & 1.5 & 3 \\ 
                      & 2 &  40 & MN & $3\times3600$ & 3.0 & 3 \\ 
                      & 3 &  40 & MN & $3\times1800$ & 1.5 & 4 \\ 
                      & 2 & 130 & MJ & $3\times3600$ & 3.0 & 2 \\
                      & 3 & 130 & MJ & $3\times1800$ & 1.5 & 4 \\ 
\object{NGC  712 (F)} & 3 &   0 & DG & $3\times1800$ & 1.5 & 4 \\ 
                      & 1 &  95 & MJ & $3\times3600$ & 3.0 & 2 \\ 
                      & 2 & 125 & DG & $3\times3600$ & 3.0 & 2 \\ 
                      & 2 & 170 & MN & $4\times3600$ & 4.0 & 2 \\                       
\object{NGC  759 (G)} & 3 &   0 & DG & $3\times1800$ & 1.5 & 4 \\ 
                      & 1 &  11 & MJ & $3\times3600$ & 3.0 & 1 \\ 
                      & 2 & 100 & MN & $3\times3600$ & 3.0 & 2 \\ 
                      & 3 & 100 & MN & $3\times1800$ & 1.5 & 4 \\ 
                      & 2 & 145 & DG & $3\times3600$ & 3.0 & 2 \\ 
                      & 3 & 145 & DG & $3\times1800$ & 1.5 & 4 \\ 
\object{UGC 1308 (B)} & 3 &   0 & DG & $2\times1800$ & 1.5 & 4 \\ 
                      & 1 &  28 & MN & $3\times3600$ & 3.0 & 2 \\ 
                      & 2 &  28 & MN & $2\times3600$ & 2.0 & 3 \\ 
                      & 2 & 120 & MJ & $3\times3600$ & 3.0 & 2 \\ 
\tableline    
\end{tabular}     
\tablecomments{ Col. 1: Name.  Col. 2: Observing run.  Col. 3: Slit
  position angle measured North through East.  Col. 4: Slit
  position. MJ = major axis (or close to the major axis); MN = minor
  axis (or close to the minor axis); DG = diagonal axis.  Col. 5:
  Number and exposure time of the single exposures.  Col. 6: Total
  exposure time.  Col. 7: Estimated quality of the resulting spectrum.
  1: excellent; 2: very good; 3: good; 4: fair (see
  Fig.~\ref{fig:quality}).}
\end{table*}    

At the beginning of each exposure the galaxy was centered on the slit
using the guiding camera which looks onto the slit.
In runs 1 and 2 several spectra of giant stars with spectral type
ranging from late-G to early-K were obtained for templates in
measuring stellar kinematics and line-strength indices. The template
stars were selected from \citet{Faber1985} and
\citet{Gonzalez1993}. At least one flux standard star per night was
observed to calibrate the flux of the spectra before line-strength
indices were measured. Spectra of the comparison arc lamp were taken
before and/or after object exposures.
  
{ All the spectra were bias subtracted, flatfield corrected,
  cleaned of cosmic rays, corrected for bad columns and wavelength and
  flux calibrated using standard IRAF routines.}
Each spectrum was rebinned using the wavelength solution obtained from
the corresponding arc-lamp spectrum. The difference between the
measured and predicted wavelengths in the comparison arc lamp spectra
have a rms of $0.05$ \AA , corresponding to $3\;\kms$ at 5170
\AA\ (i.e., the wavelength of the \MgI\ absorption triplet) in runs 1
and 2 and $2\;\kms$ at \Ha\ in run 3.
Systematic errors of the absolute wavelength calibration
($\leq10\;\kms$) were estimated from the
brightest night-sky emission lines in the observed spectral range
\citep{Osterbrock1996}.
The instrumental resolution in each run was derived as the mean of the
Gaussian FWHMs measured for a number of unblended arc-lamp lines which
were distributed over the whole spectral range of a
wavelength-calibrated spectrum. The mean FWHM of the arc-lamp lines
was 2.3 \AA\ in run 1 and 2.4 \AA\ in run 2 and the corresponding
instrumental resolution derived at 5170 \AA\ is $\sigma_{\rm
  inst}\,\simeq\,60\;\kms$. In run 3 the mean FWHM was 3.2
\AA\ corresponding to $\sigma_{\rm inst}\,\simeq\,60\;\kms$ at \Ha .
All the galaxy and stellar spectra were corrected for CCD
misalignment. The sky contribution was determined by interpolating
along the outermost $10''-30''$ at the two edges of the slit, where
the galaxy or stellar light was negligible, and then subtracted. A sky
subtraction better than $1\%$ was achieved.
Each spectrum was flux calibrated using the sensitivity function
obtained from the flux standard star spectrum of the corresponding
night.
In each run the spectra obtained for the same galaxy along the same
axis were coadded using the center of the stellar continuum as
reference, thus improving the signal-to-noise ratio ($S/N$) of the
final two-dimensional spectrum.  The spectra of the template stars
were deredshifted to laboratory wavelengths.

\subsection{Stellar kinematics} 
 
The stellar kinematics of the galaxies was measured from the galaxy
absorption features present in the wavelength range and centered on
the \MgI\ line triplet ($\lambda\lambda\,5164,5173,5184$ \AA) using
the Penalized Pixel-Fitting \citep[pPXF,][]{Cappellari2004} and Gas
and Absorption Line Fitting \citep[GANDALF,][]{Sarzi2006}
IDL\footnote{The Interactive Data Language is distributed by ITT
  Visual Information Solutions.} codes adapted for dealing with MDM
spectra.
 
\begin{figure}[t!]  
\centering  
\includegraphics[width=0.6\textwidth,angle=90]{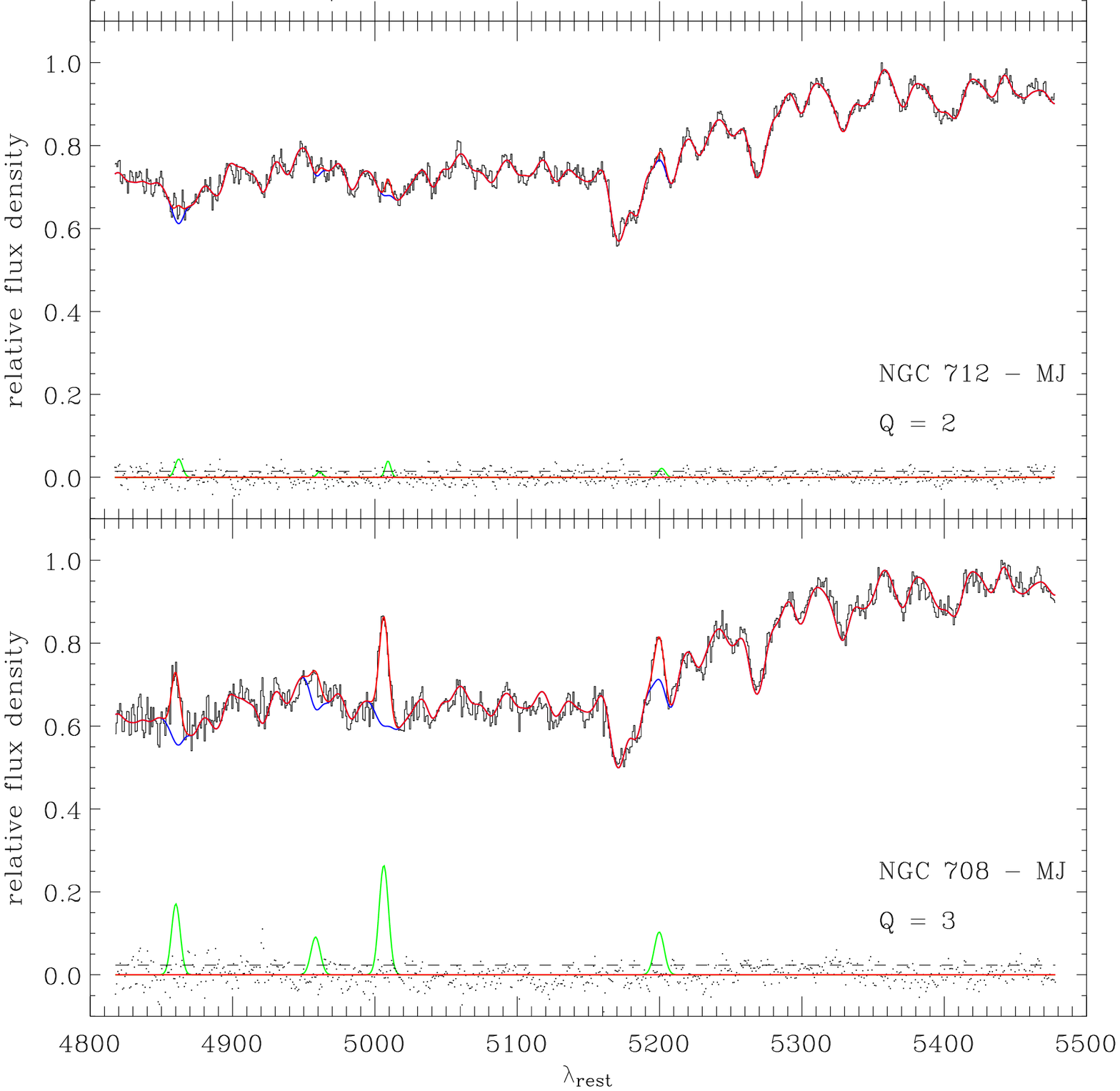}  
\caption{Example of central spectra covering the range of quality 
  classes of runs 1 and 2. Relative fluxes have false zero points for 
  viewing convenience. In each panel the best-fitting model (red line) 
  is the sum of the spectra of the ionized-gas (green line) and 
  stellar component (blue line).  The latter is obtained convolving 
  the synthetic templates with the best-fitting LOSVD and multiplying 
  them by the best-fitting Legendre polynomials. The residuals (dots) 
  are obtained by subtracting the model from the spectrum. The dashed 
  line corresponds to the rms of the residuals.} 
\label{fig:quality}  
\end{figure}

The spectra of runs 1 and 2 were rebinned along the dispersion
direction to a logarithmic scale, and along the spatial direction to
obtain a $S/N\geq40$ per resolution element. It decreases to
$S/N\approx20$ per resolution element at the outermost radii.  The
quality of the final spectrum depends on the resulting $S/N$.
Examples of central spectra obtained in runs 1 and 2 covering the
quality classes listed in Table~\ref{tab:log} are shown in
Fig.~\ref{fig:quality}. The quality parameter is 1 for $S/N \geq 100$
per resolution element, 2 for $50 \leq S/N <100$, and 3 for $30 \leq
S/N <50$.

At each radius a linear combination of template stellar spectra from
the empirical library by \citet[][i.e., the MILES
library]{Sanchez-Blazquez2006} was convolved with the line-of-sight
velocity distribution (LOSVD) and fitted to the observed galaxy
spectrum by $\chi^2$ minimization in pixel space. The LOSVD was
assumed to be a Gaussian plus third- and fourth-order Gauss-Hermite
polynomials ${\cal H}_3$ and ${\cal H}_4$, which describe the
asymmetric and symmetric deviations of the LOSVD from a pure Gaussian
profile \citep{vanderMarel1993, Gerhard1993}.
This allowed us to derive profiles of the line-of-sight velocity
($v$), velocity dispersion ($\sigma$), and third- ($H_3$) and
fourth-order ($H_4$) Gauss-Hermite moments of the stars.

\begin{figure*}[hp!]  
\includegraphics[width=0.5\textwidth]{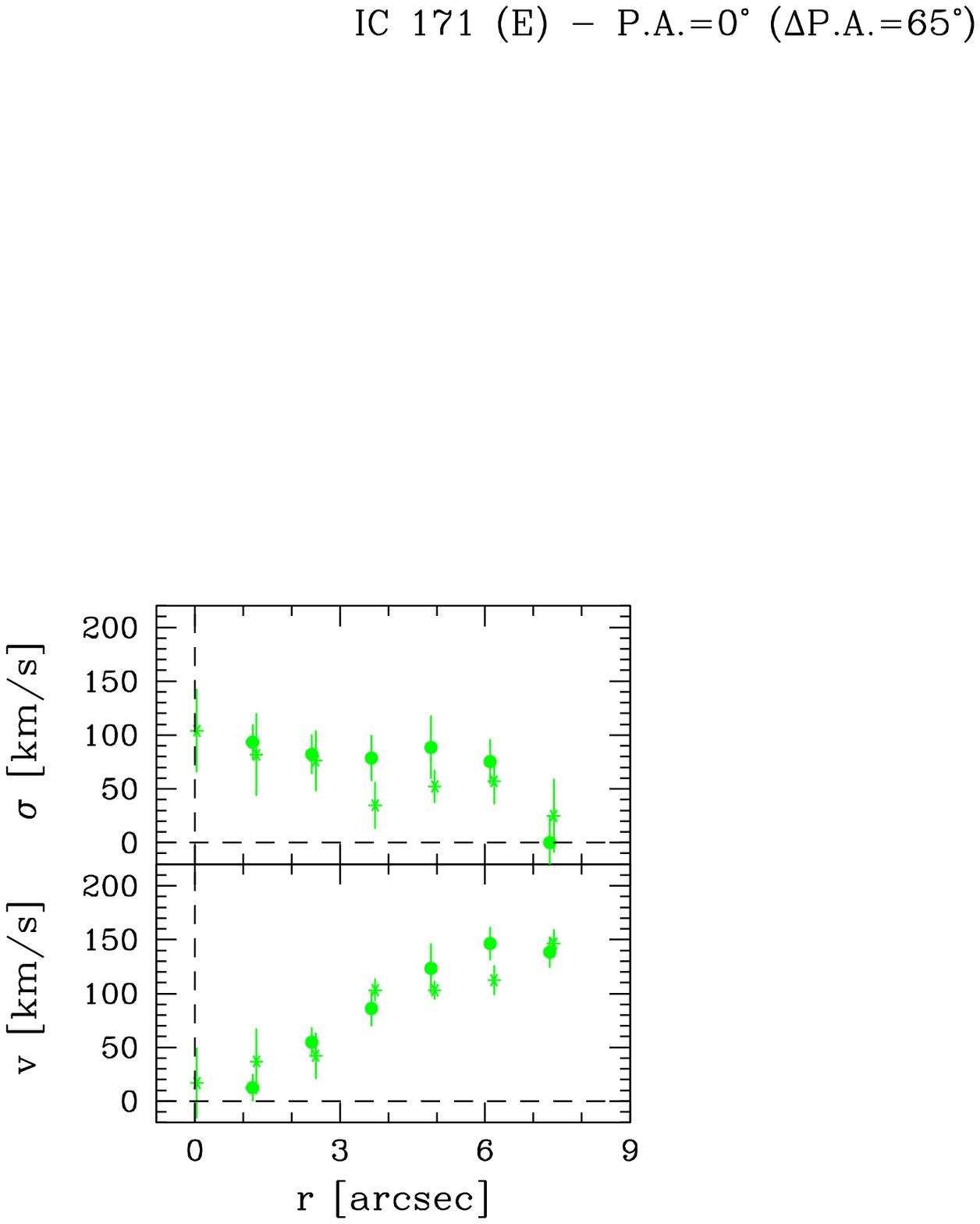}
\includegraphics[width=0.5\textwidth]{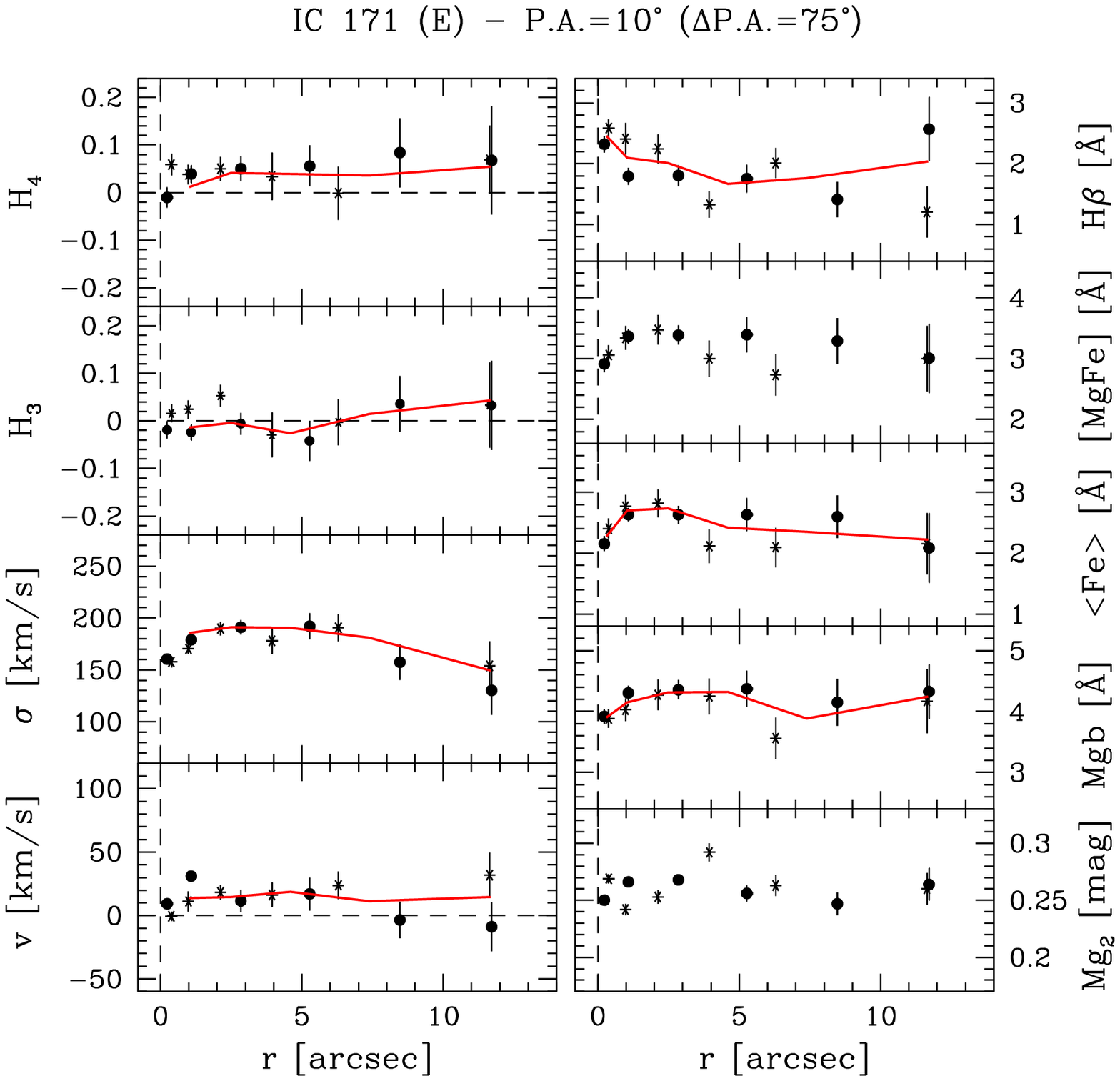} 
\includegraphics[width=0.5\textwidth]{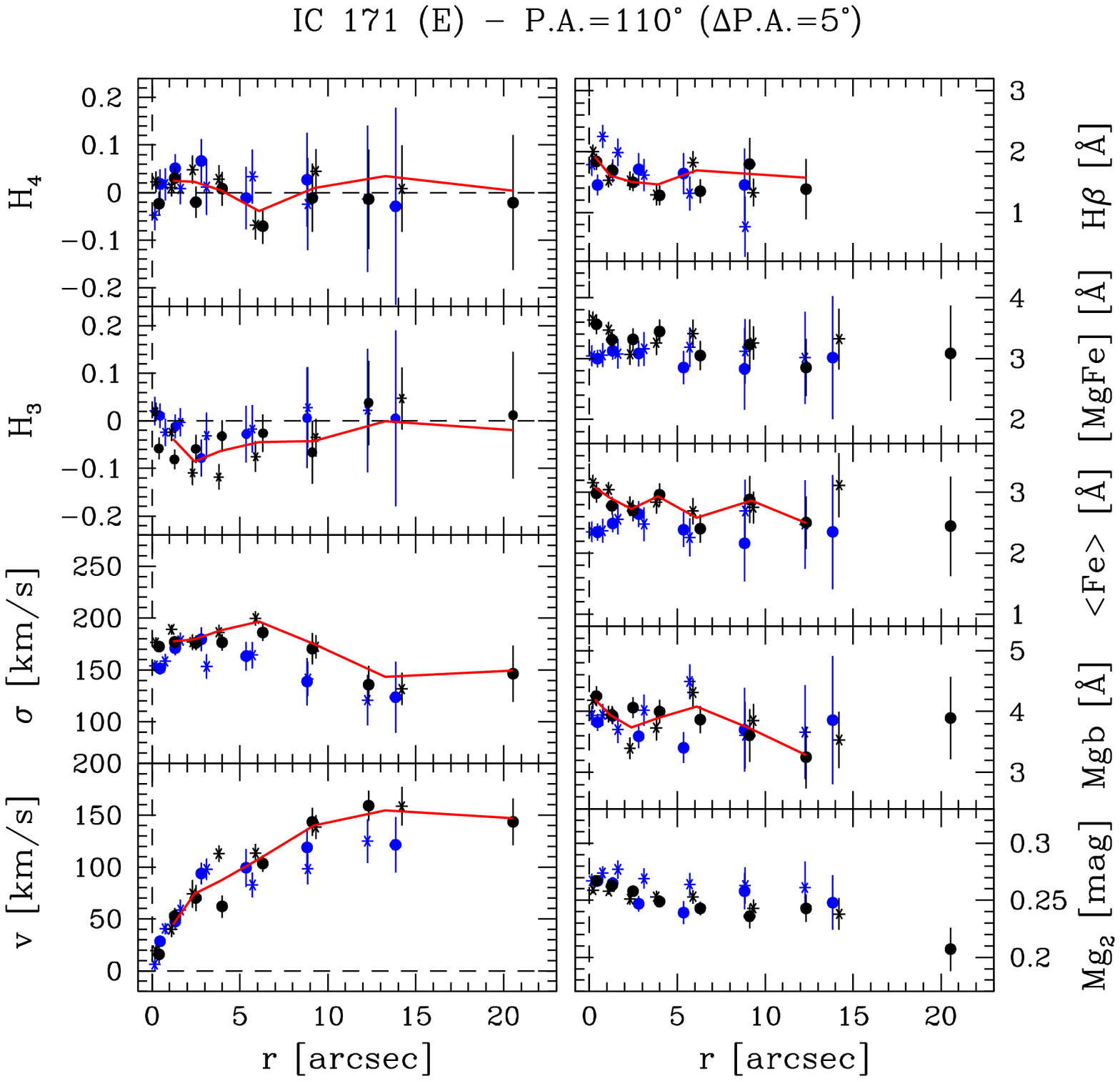} 
\includegraphics[width=0.5\textwidth]{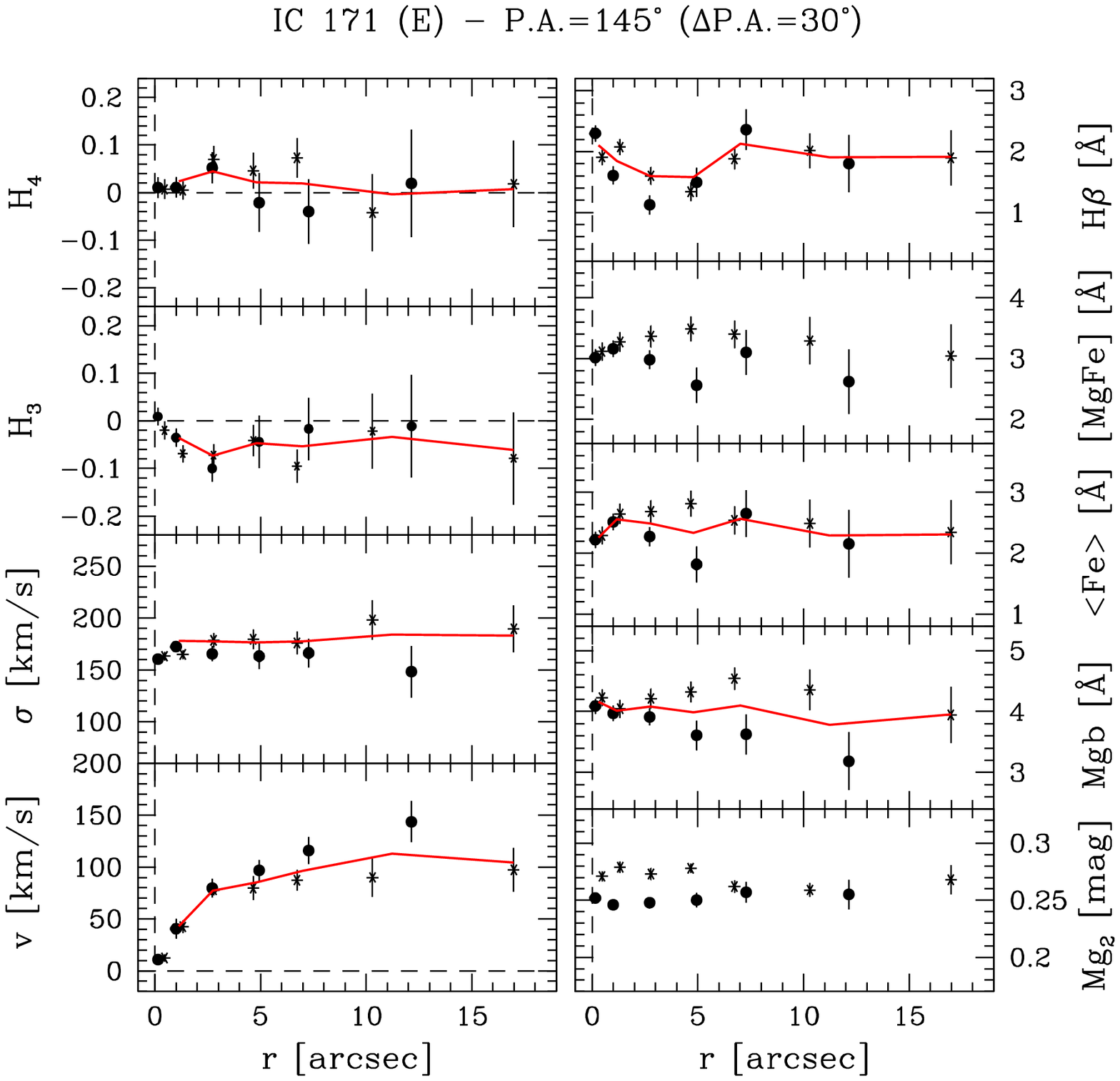}
\caption{Kinematic parameters of the stars (run 1: black symbols; run
  2: blue symbols) and ionized gas (run 3: green symbols) and the
  line-strength indices measured along the observed axes of the sample
  galaxies. For each axis the curves are folded around the
  nucleus. Circles and asterisks (or squares) refer to data measured
  along the receding and approaching side, respectively. The radial
  profiles of the line-of-sight velocity ($v$) after the subtraction
  of systemic velocity, velocity dispersion ($\sigma$), third, and
  fourth order coefficient of the Gauss-Hermite decomposition of the
  LOSVD ($H_3$ and $H_4$) are shown in the left panels (from top to
  bottom). The red solid lines correspond to the stellar kinematic
  parameters of the best fitting dynamical model. The difference
  between the position angle of the observed axis and that adopted in
  the dynamical model for the galaxy line of nodes
  (Table~\ref{tab:dynres}) is given in the top label ($\Delta {\rm
    P.A.}$).
  The radial profiles of the line-strength indices \Hb, [MgFe], \Fe ,
  \Mgb , and \Mgd\ are plotted in the right panels (from top to
  bottom). The red solid lines correspond to the line-strength indices
  derived from SSP models.}
\label{fig:kinematics}    
\end{figure*}  
  
\addtocounter{figure}{-1} 
\begin{figure*}[ht!]   
\includegraphics[width=0.5\textwidth]{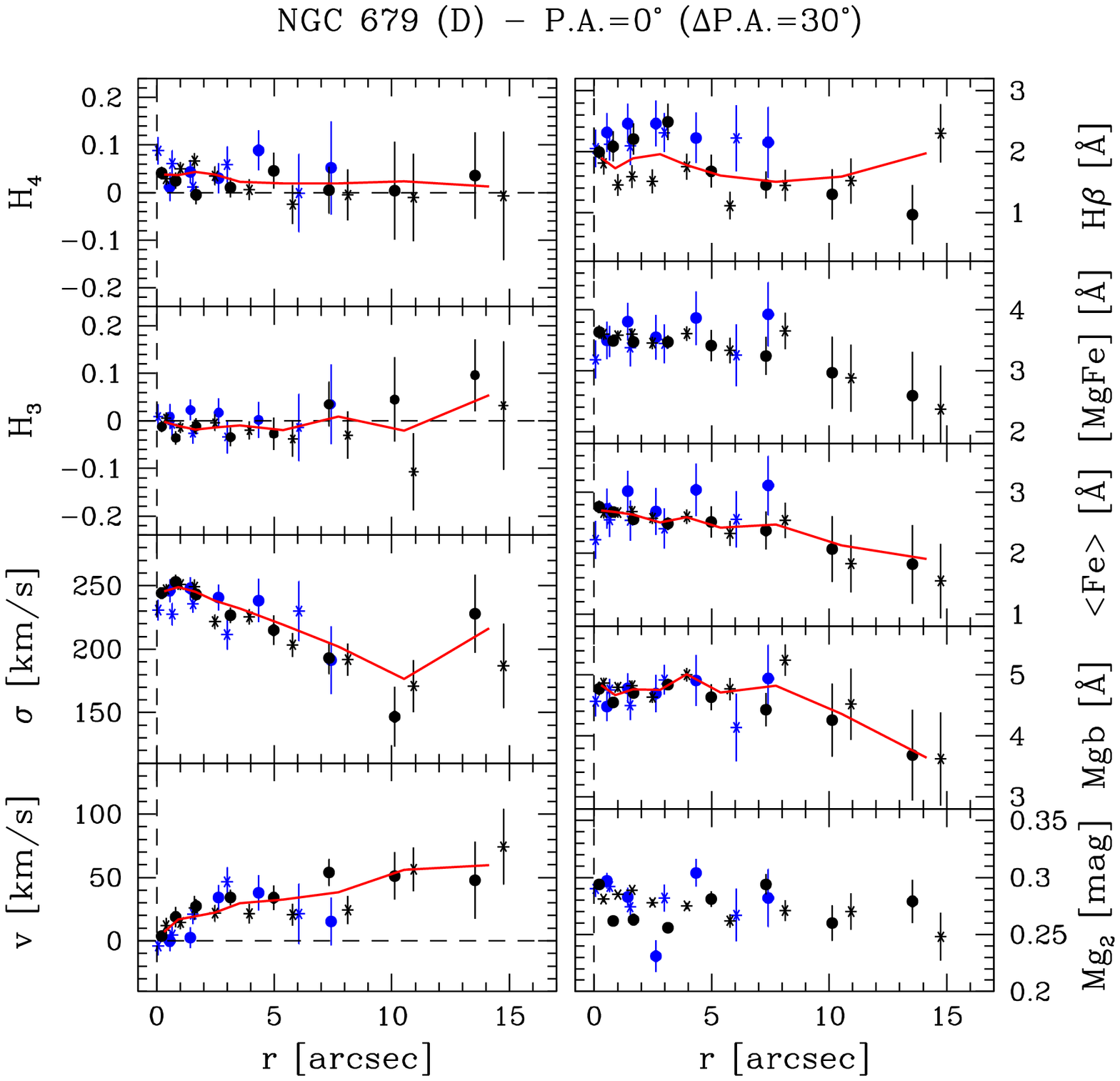} 
\includegraphics[width=0.5\textwidth]{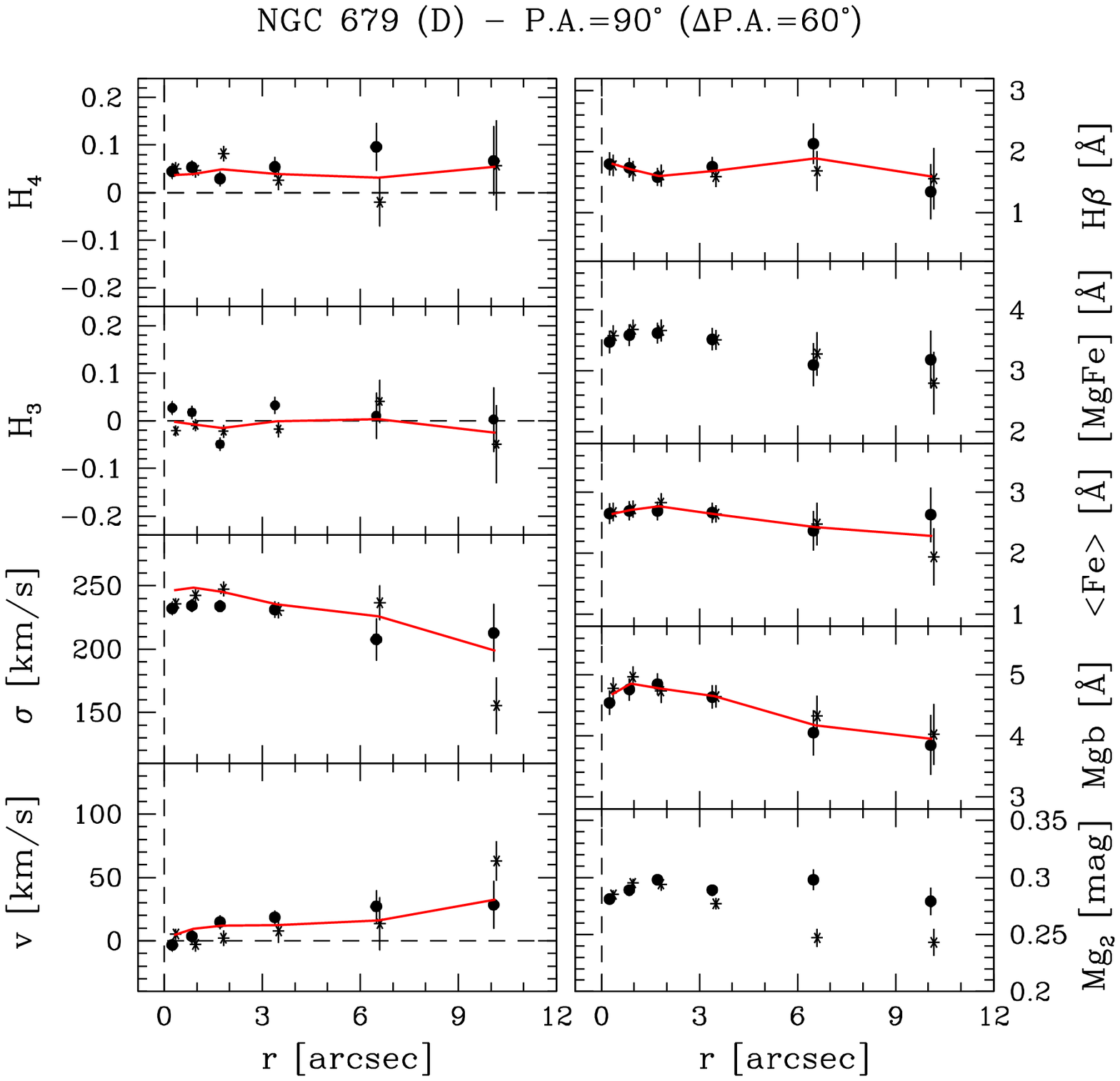} 
\includegraphics[width=0.5\textwidth]{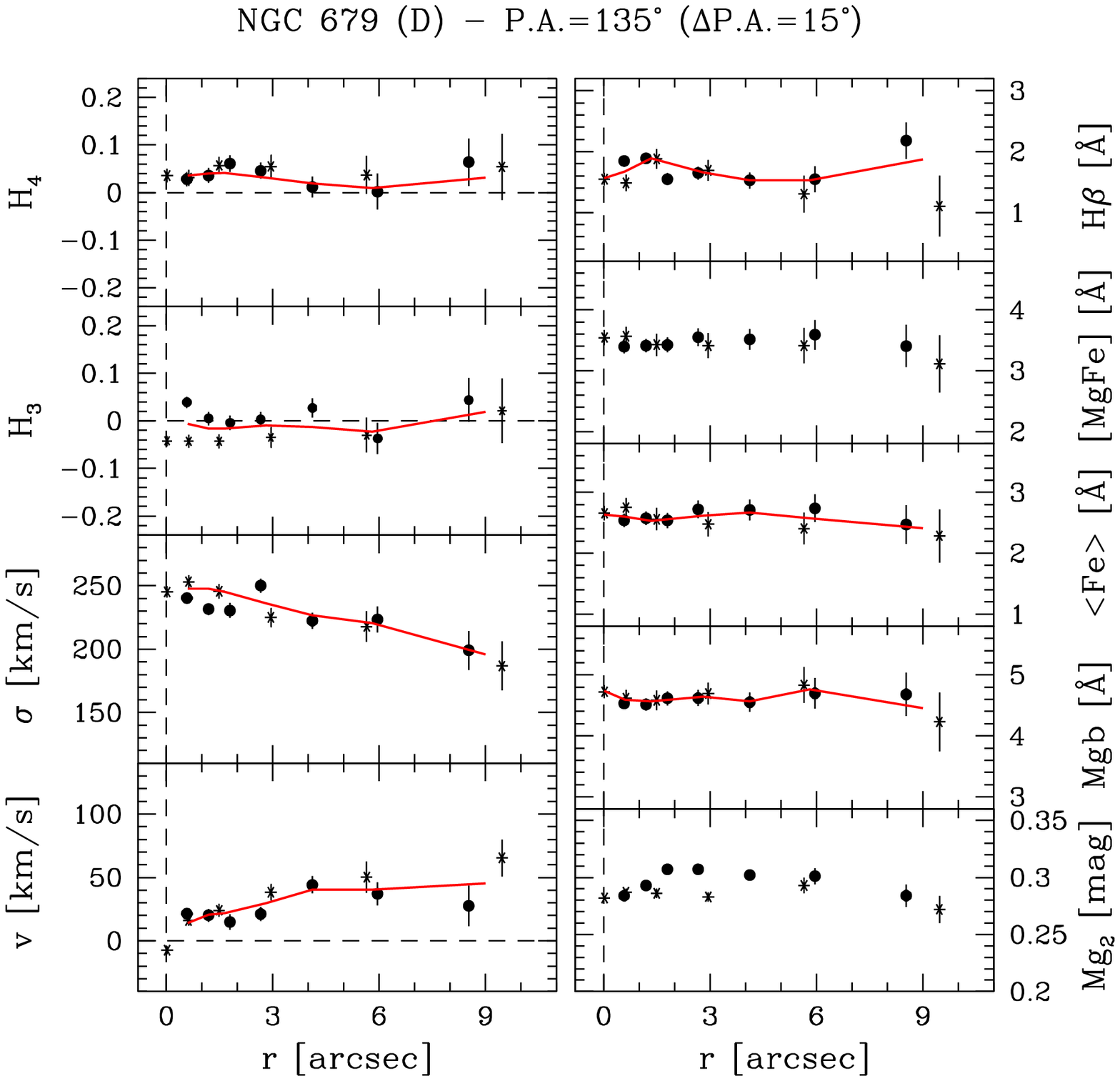} 
\includegraphics[width=0.5\textwidth]{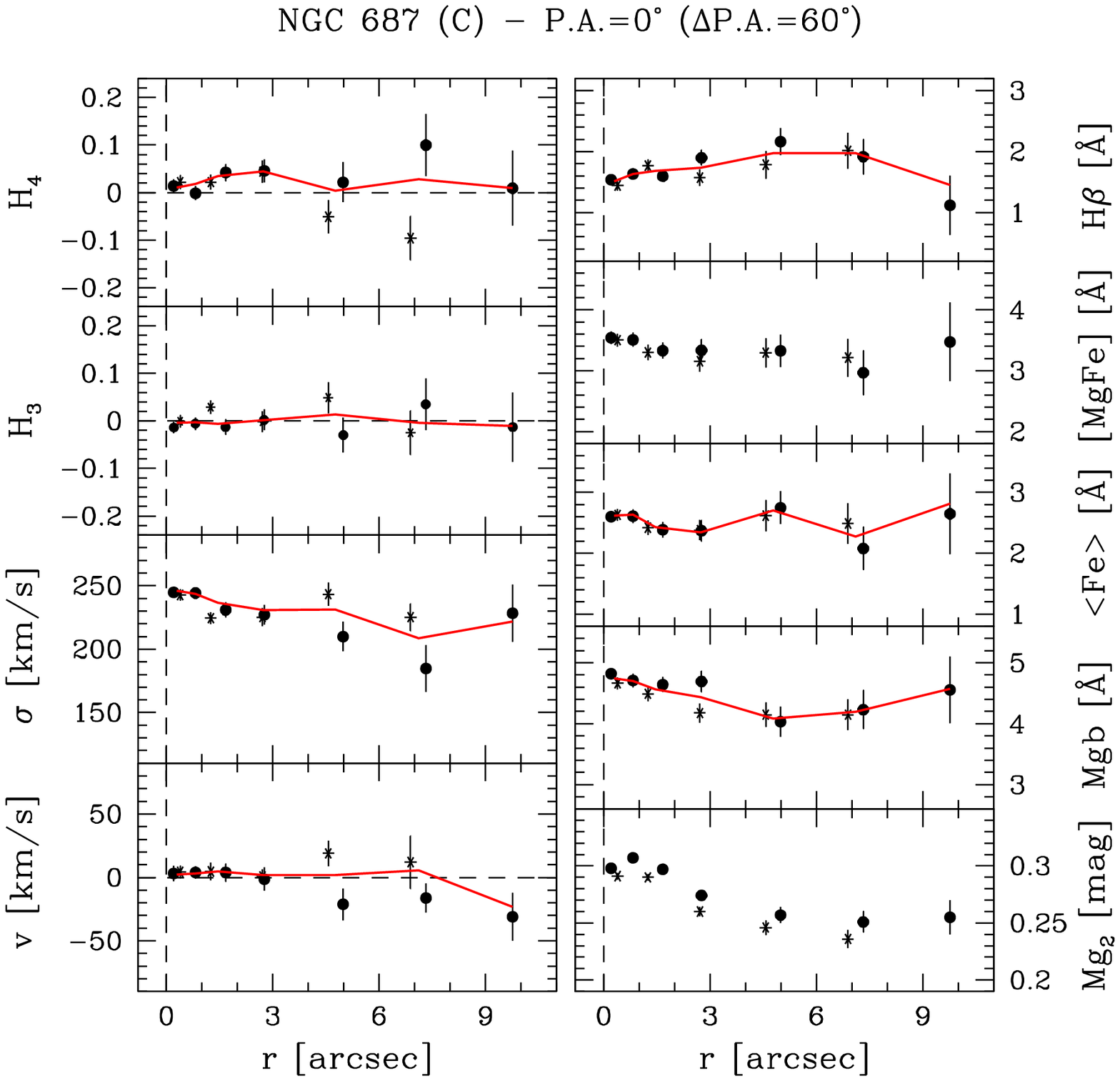} 
\caption{{\em Continued.}} 
\end{figure*} 
 
\addtocounter{figure}{-1} 
\begin{figure*}[ht!]  
\includegraphics[width=0.5\textwidth]{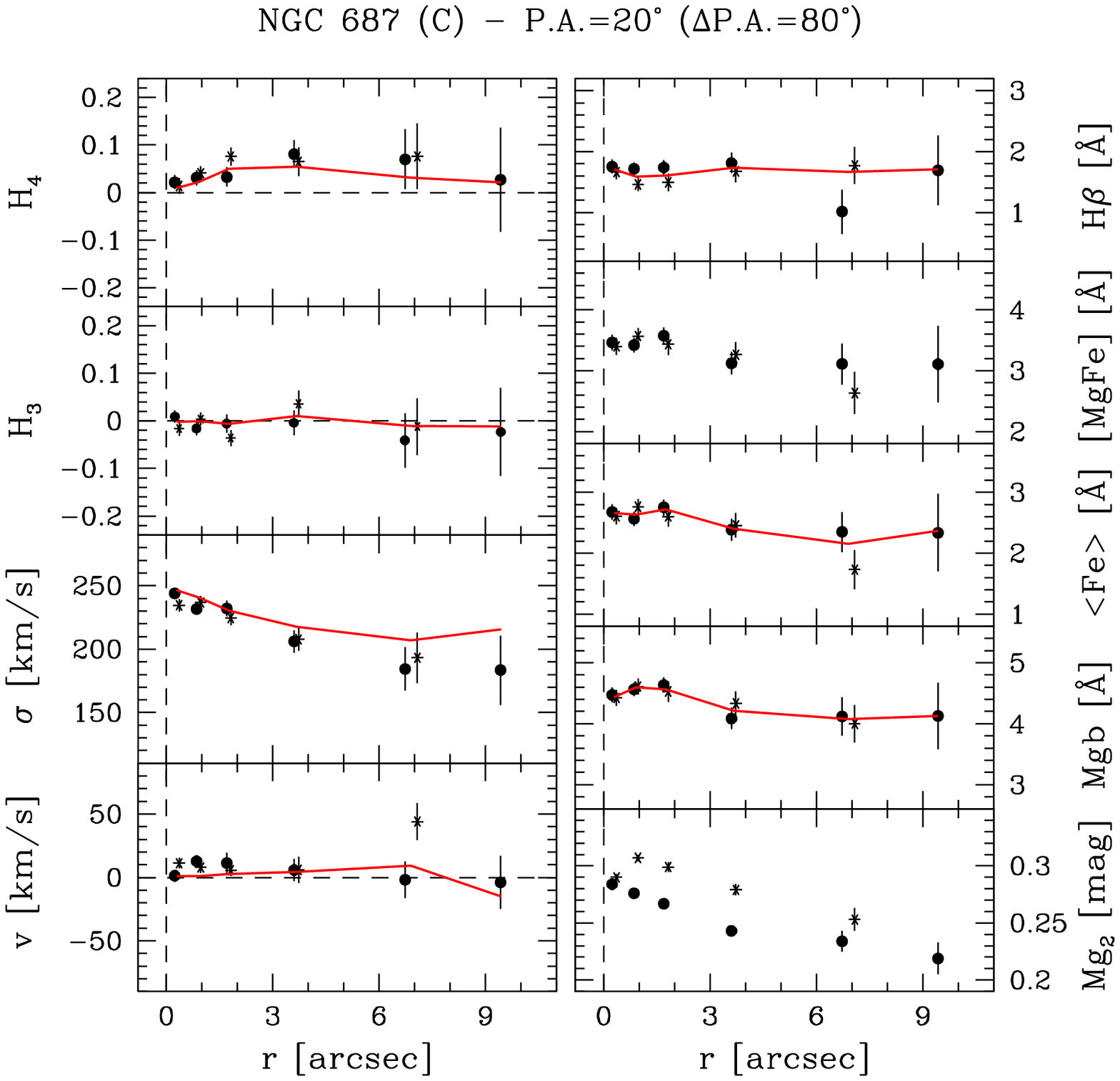} 
\includegraphics[width=0.5\textwidth]{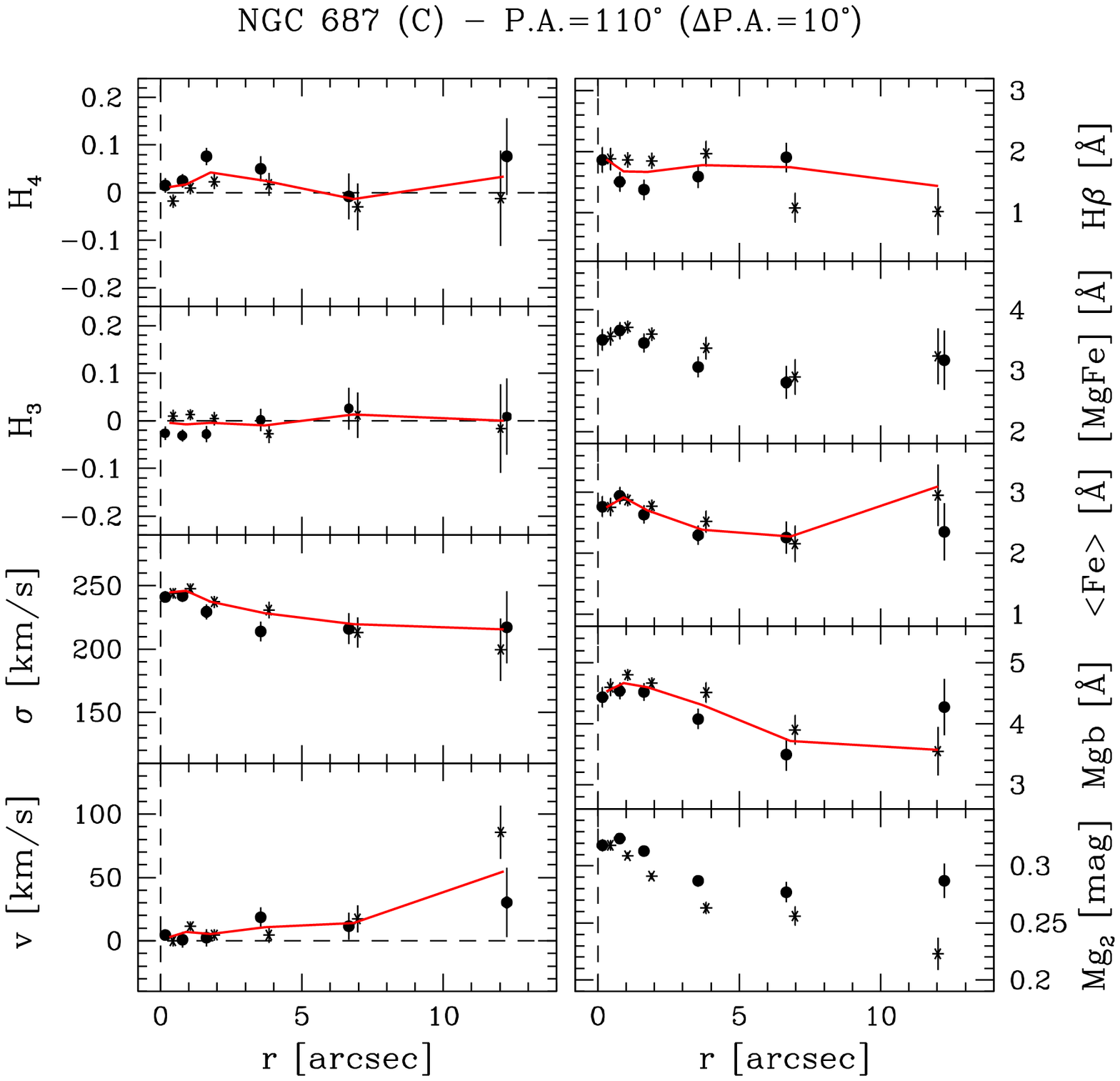} 
\includegraphics[width=0.5\textwidth]{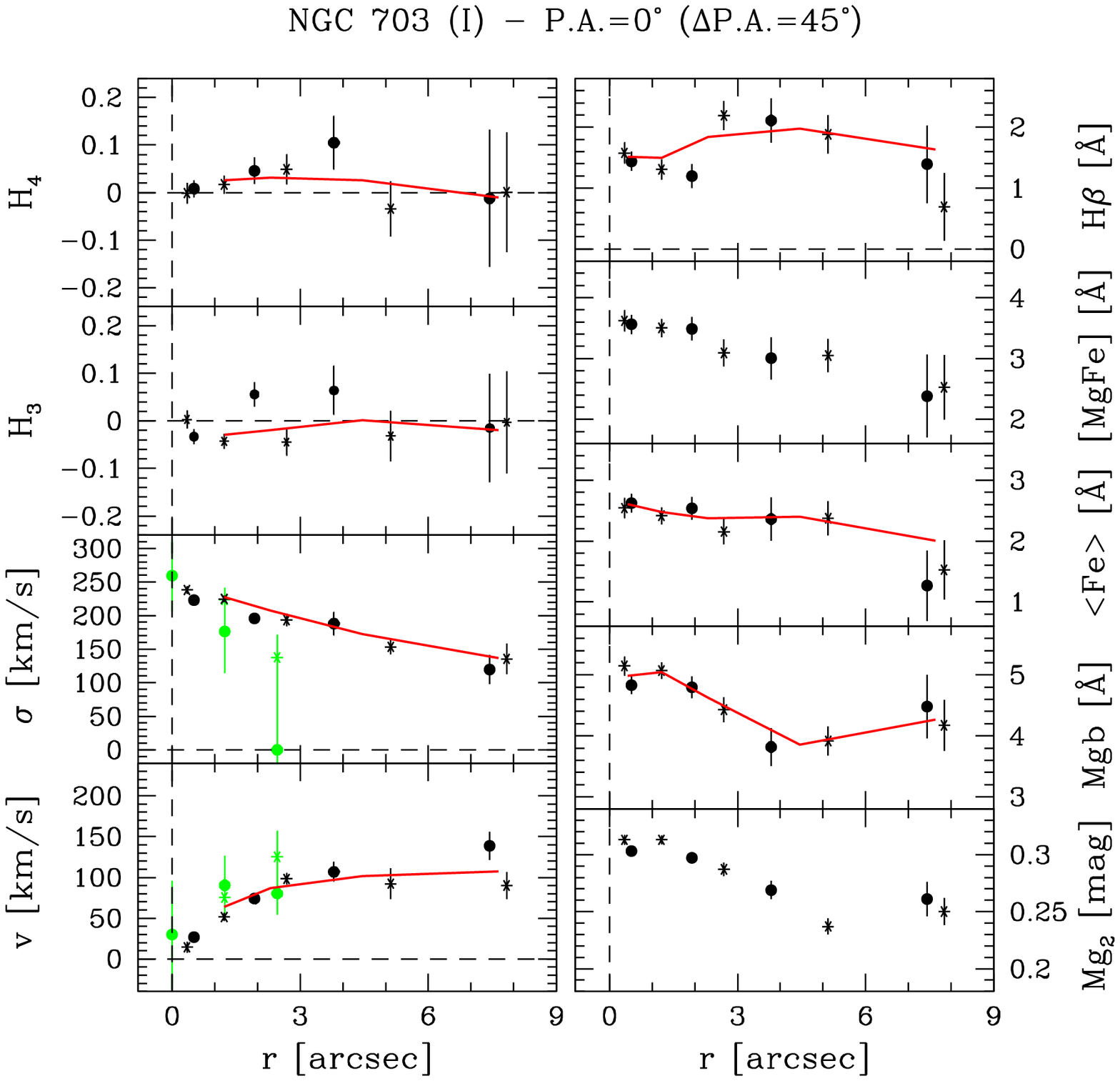} 
\includegraphics[width=0.5\textwidth]{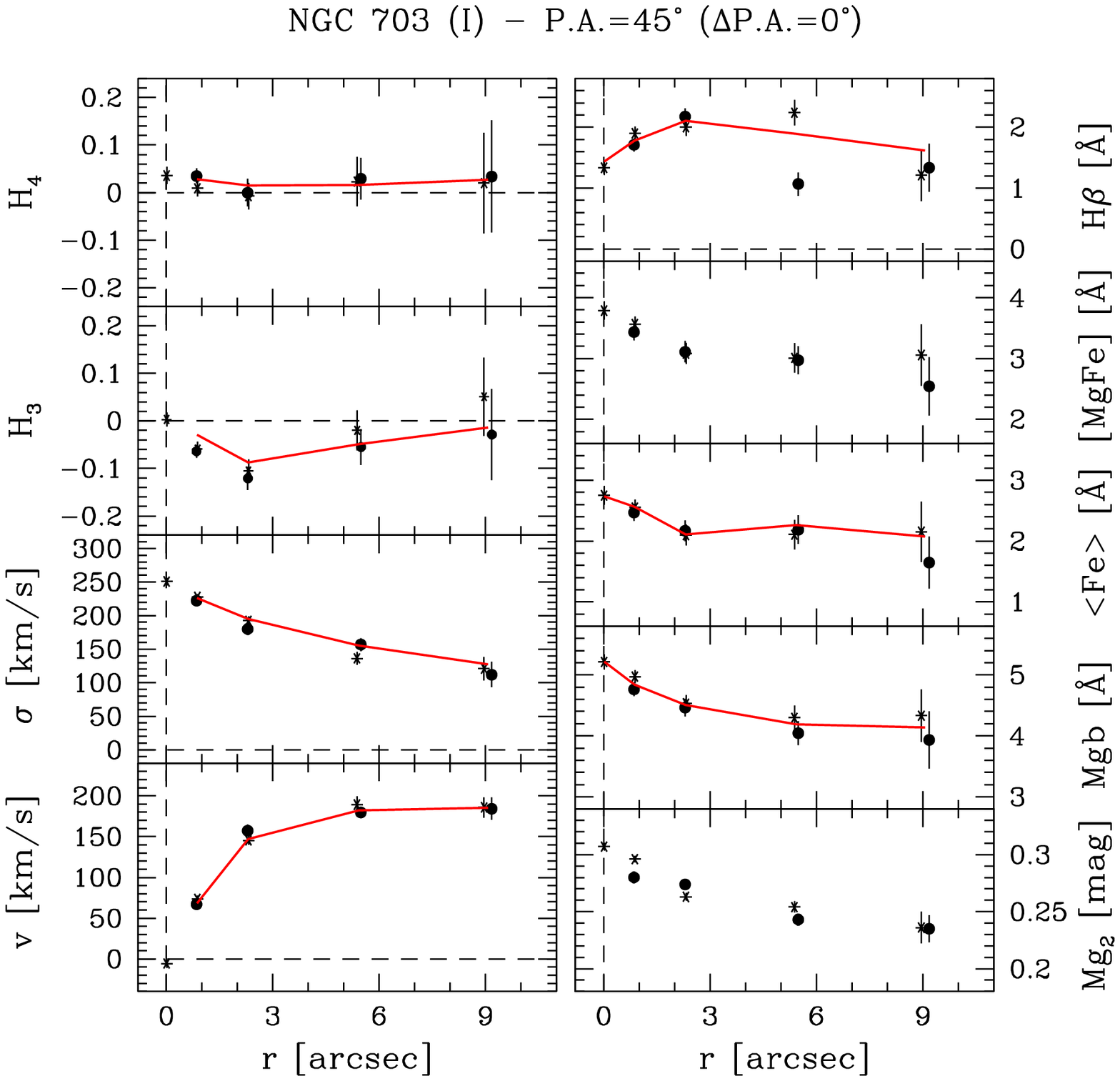} 
\caption{{\em Continued.}} 
\end{figure*} 
 
\addtocounter{figure}{-1} 
\begin{figure*}[ht!]   
\includegraphics[width=0.5\textwidth]{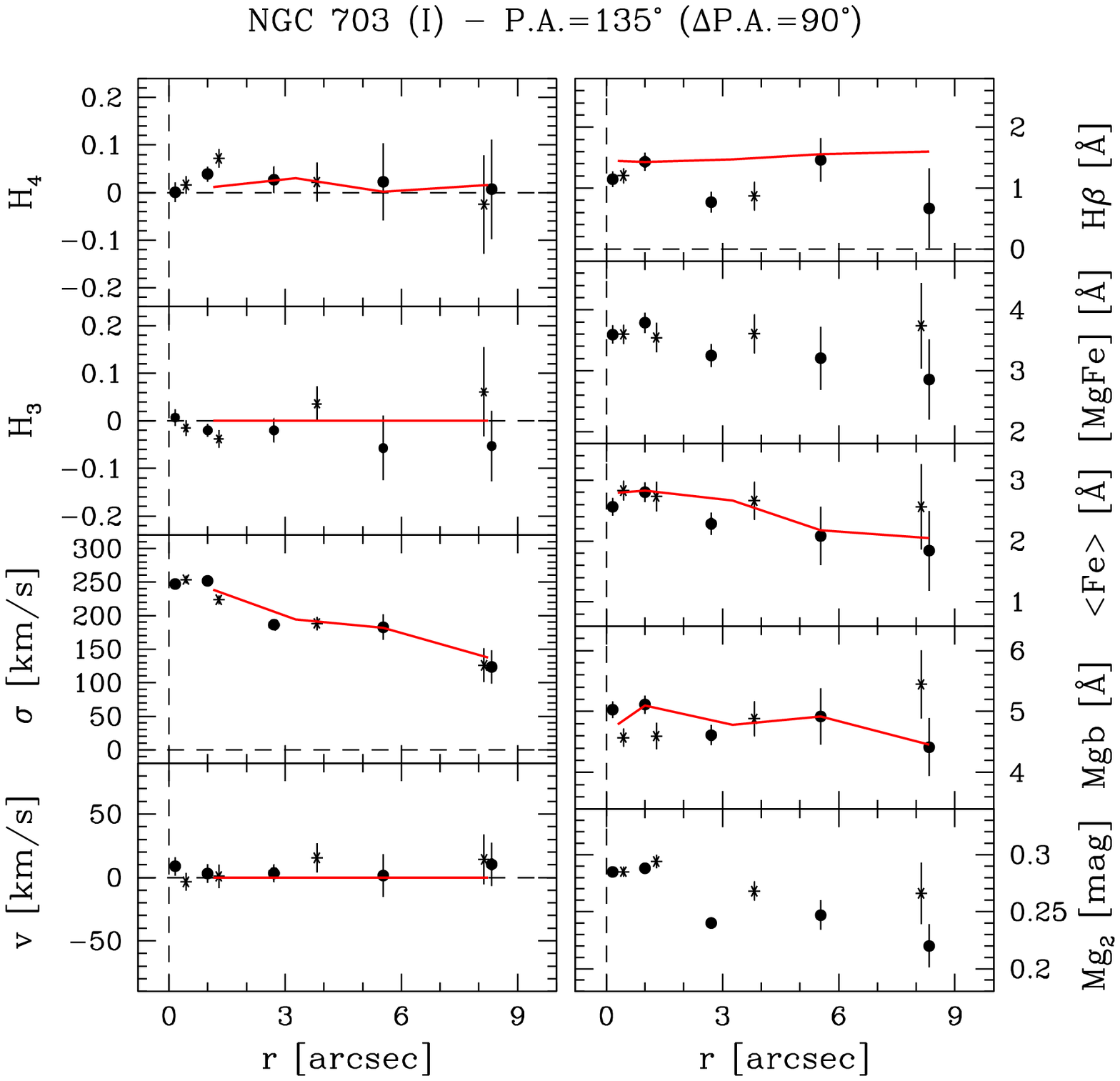} 
\includegraphics[width=0.5\textwidth]{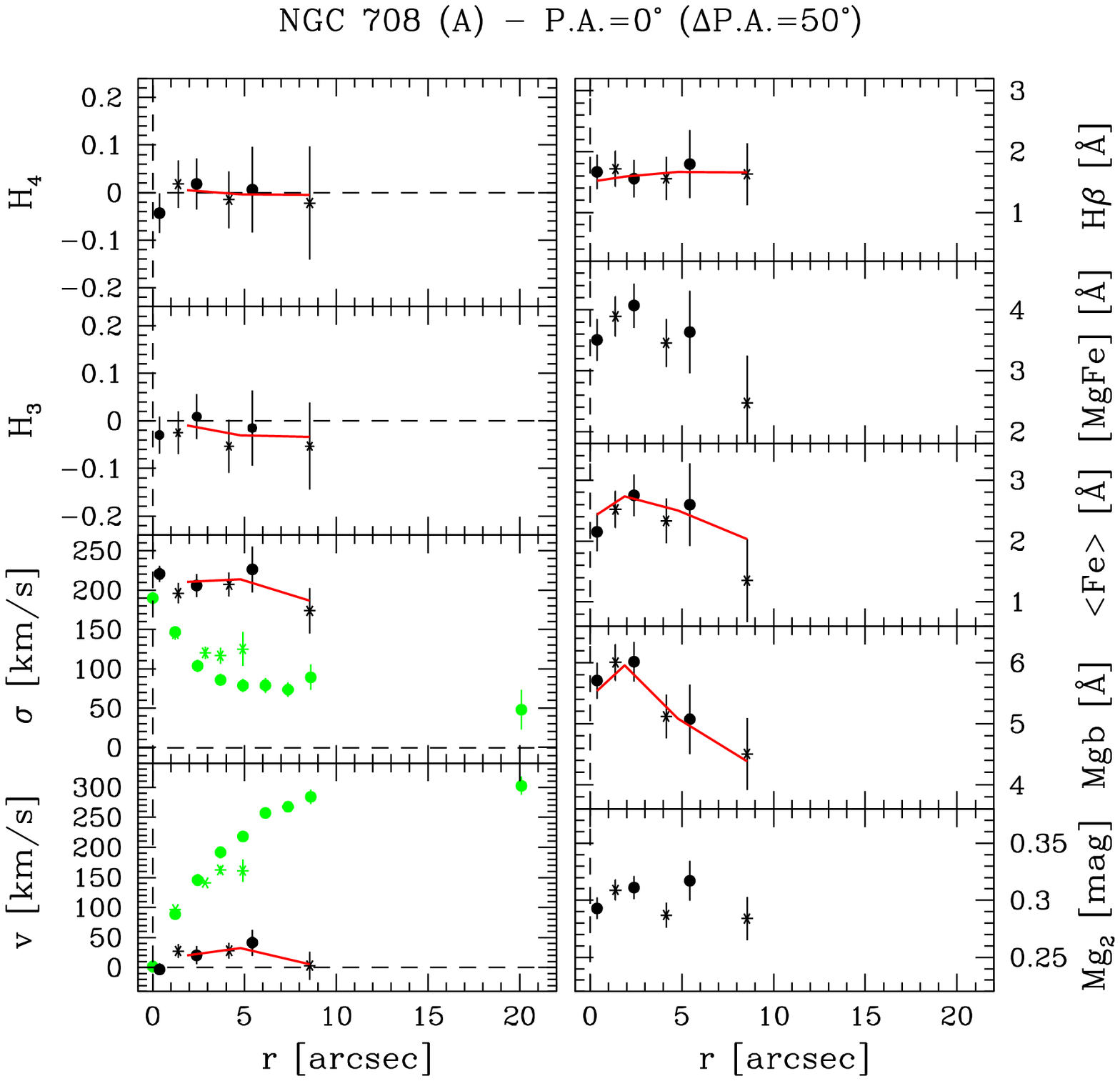} 
\includegraphics[width=0.5\textwidth]{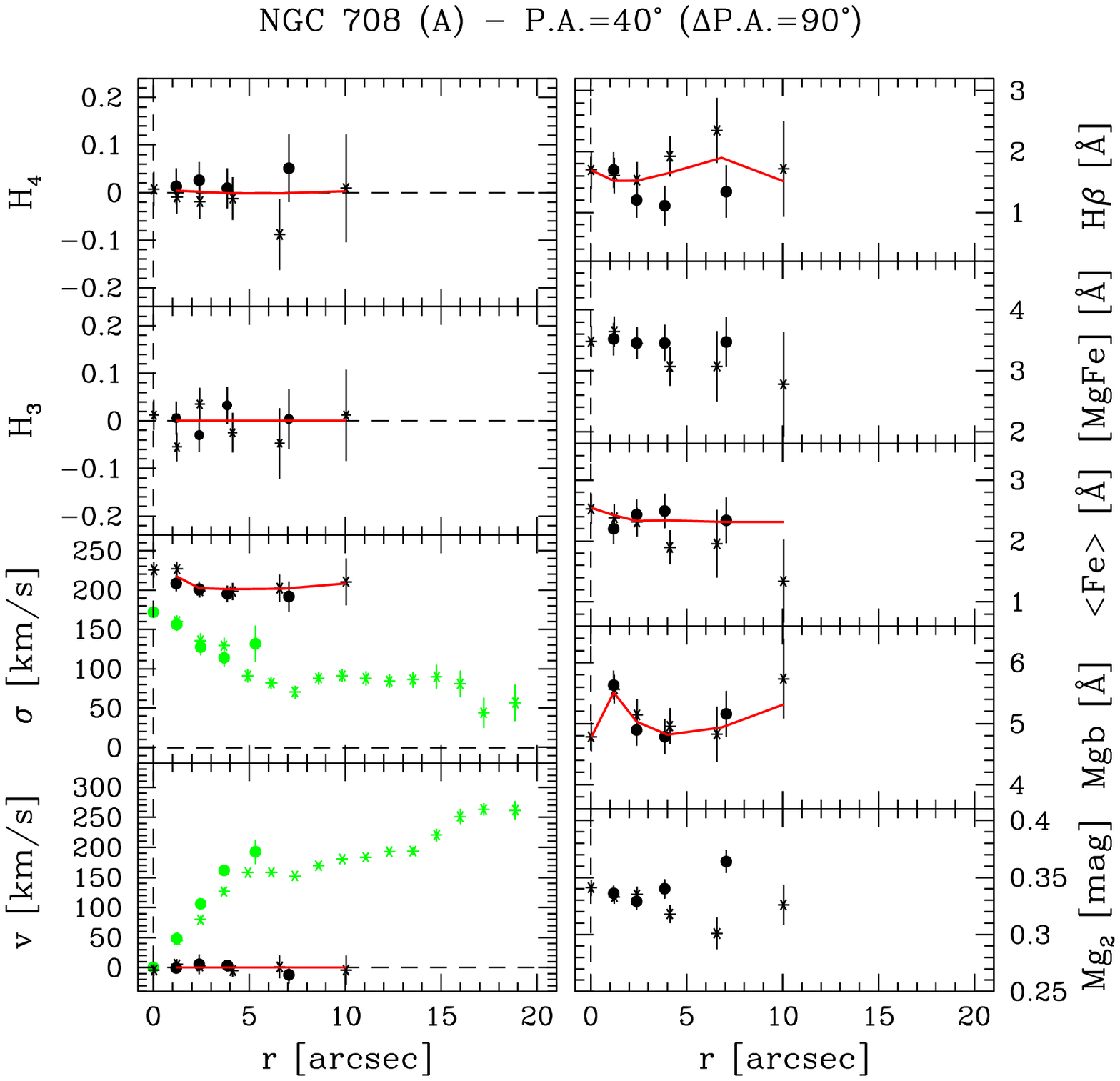} 
\includegraphics[width=0.5\textwidth]{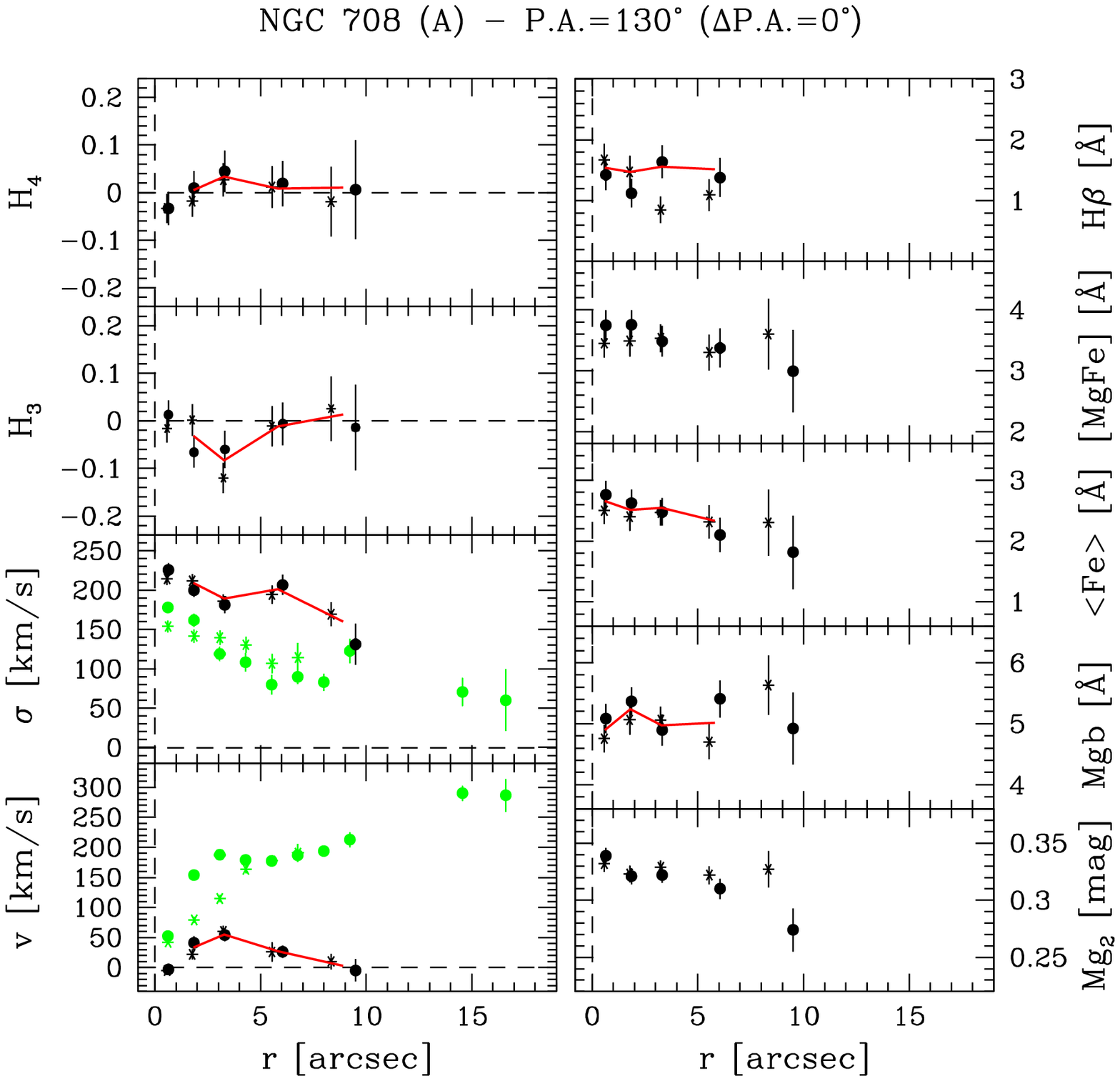} 
\caption{{\em Continued.}} 
\end{figure*} 
 
\addtocounter{figure}{-1} 
\begin{figure*}[ht!]  
\includegraphics[width=0.5\textwidth]{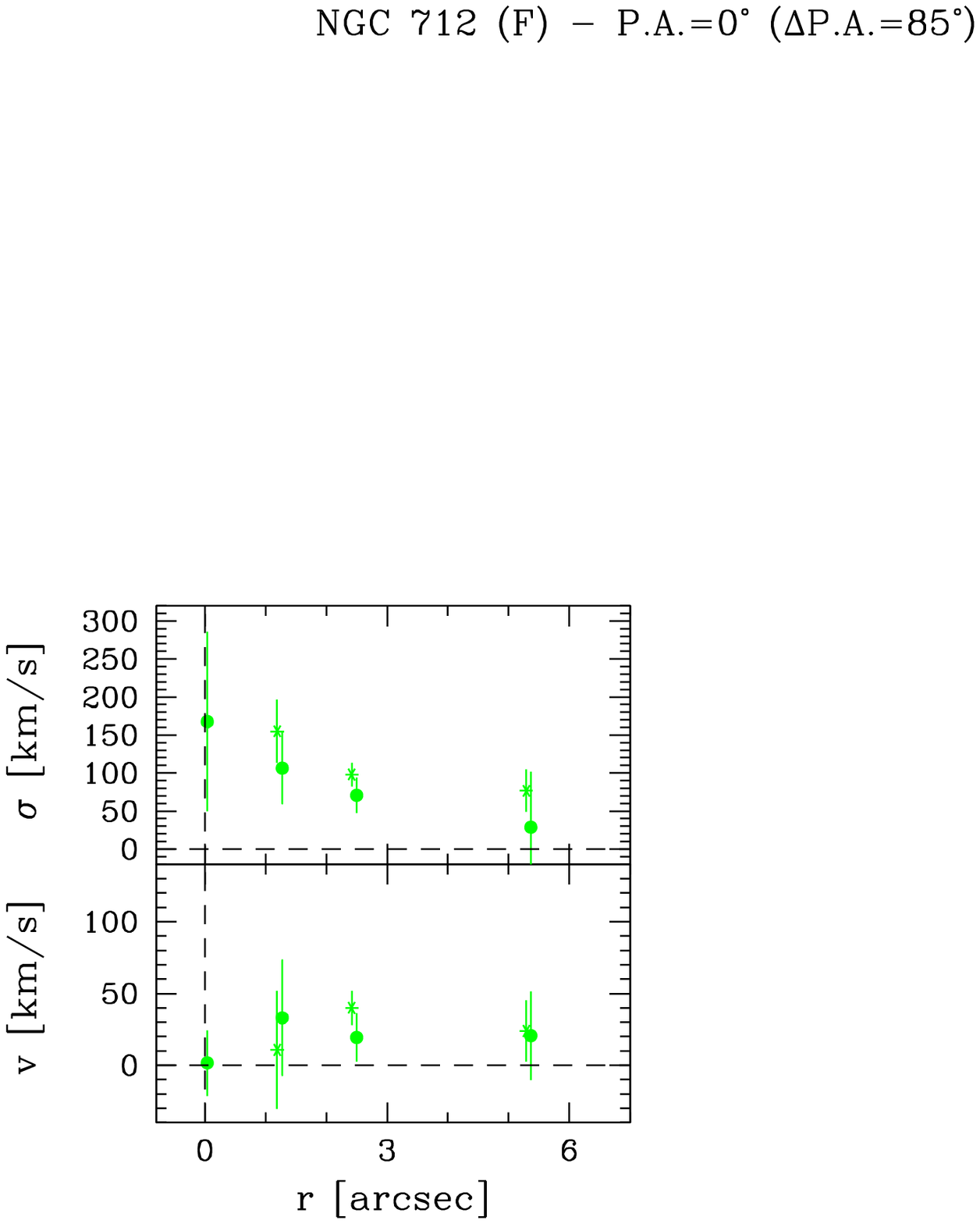} 
\includegraphics[width=0.5\textwidth]{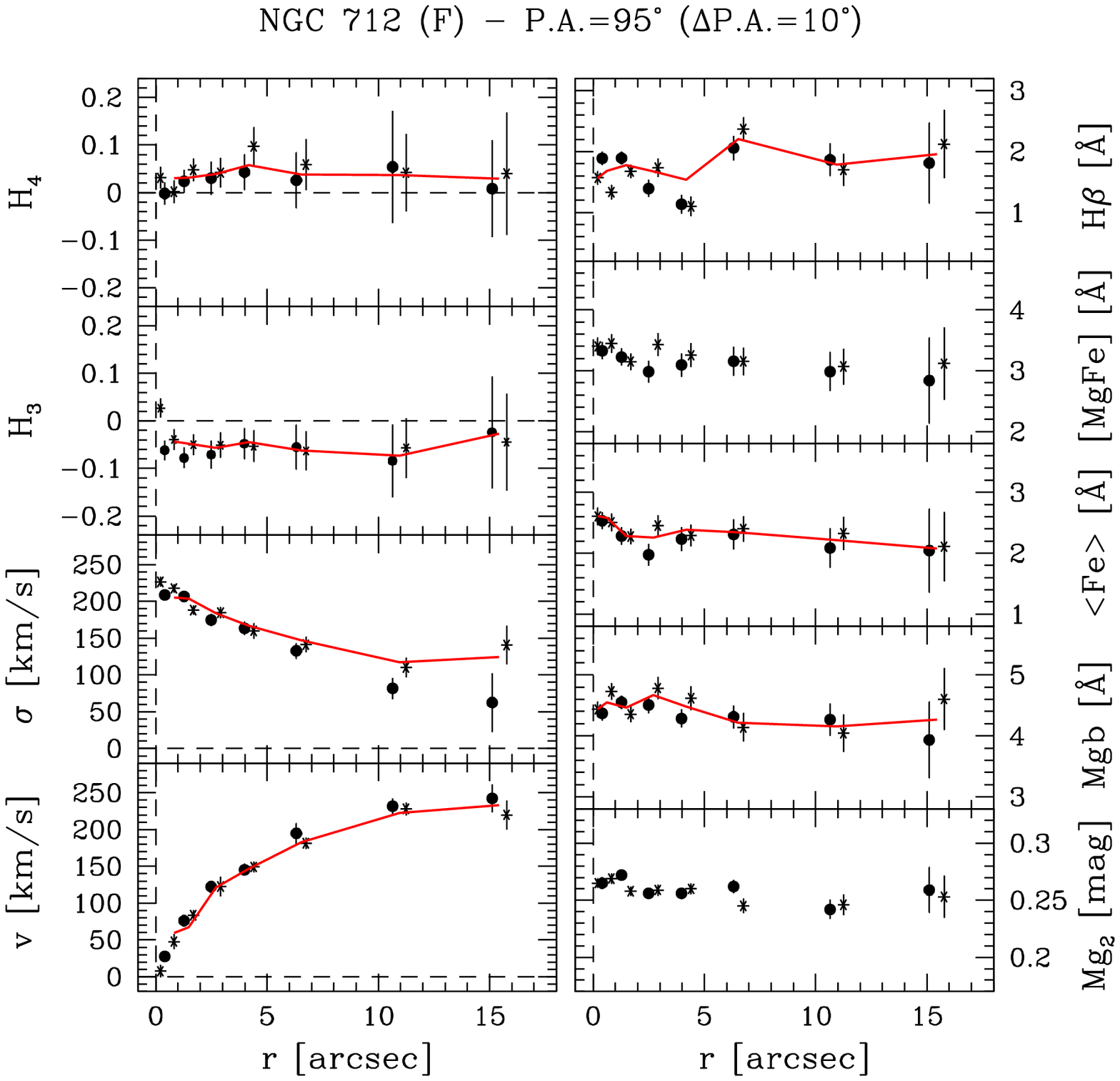}
\includegraphics[width=0.5\textwidth]{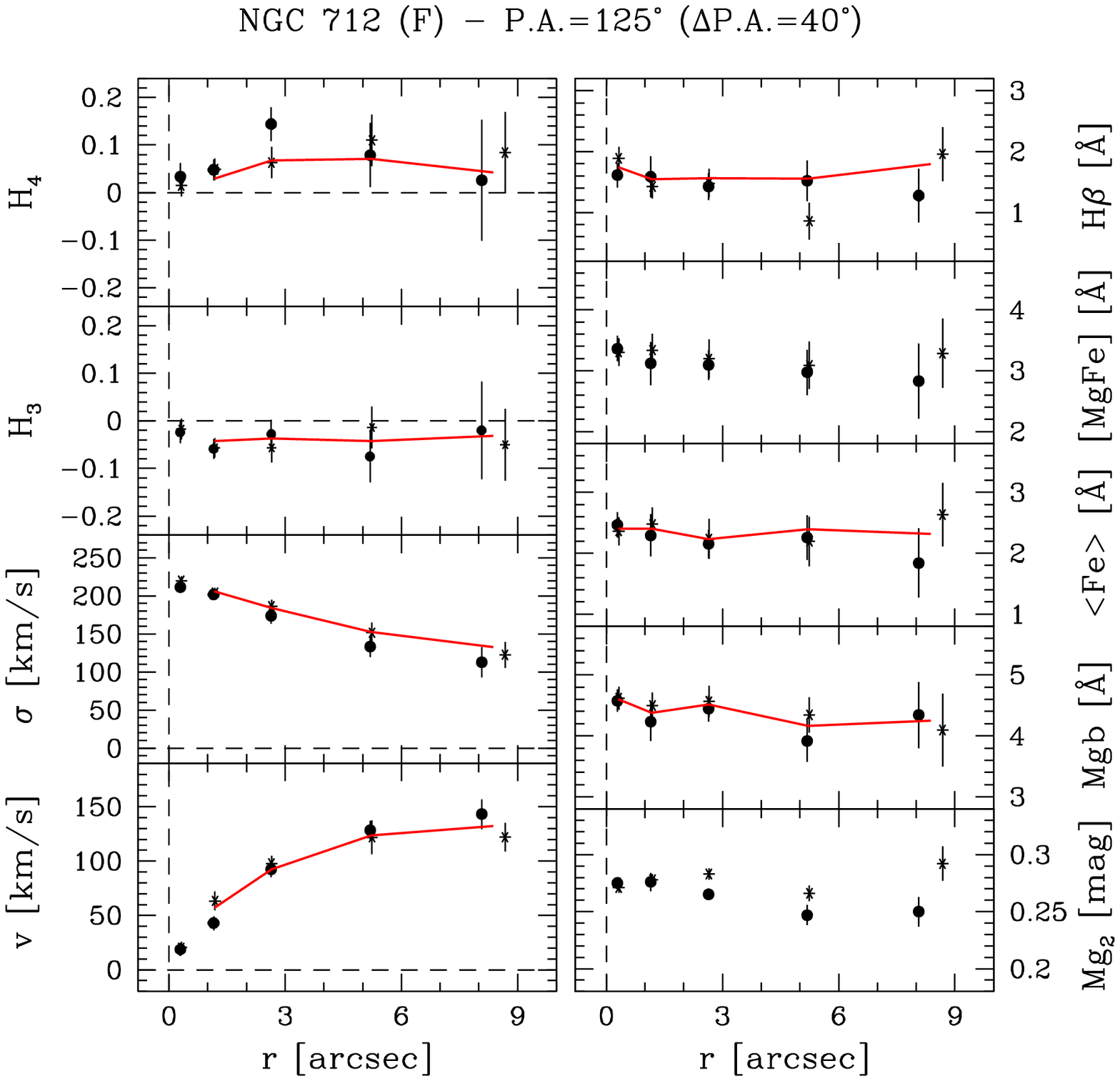} 
\includegraphics[width=0.5\textwidth]{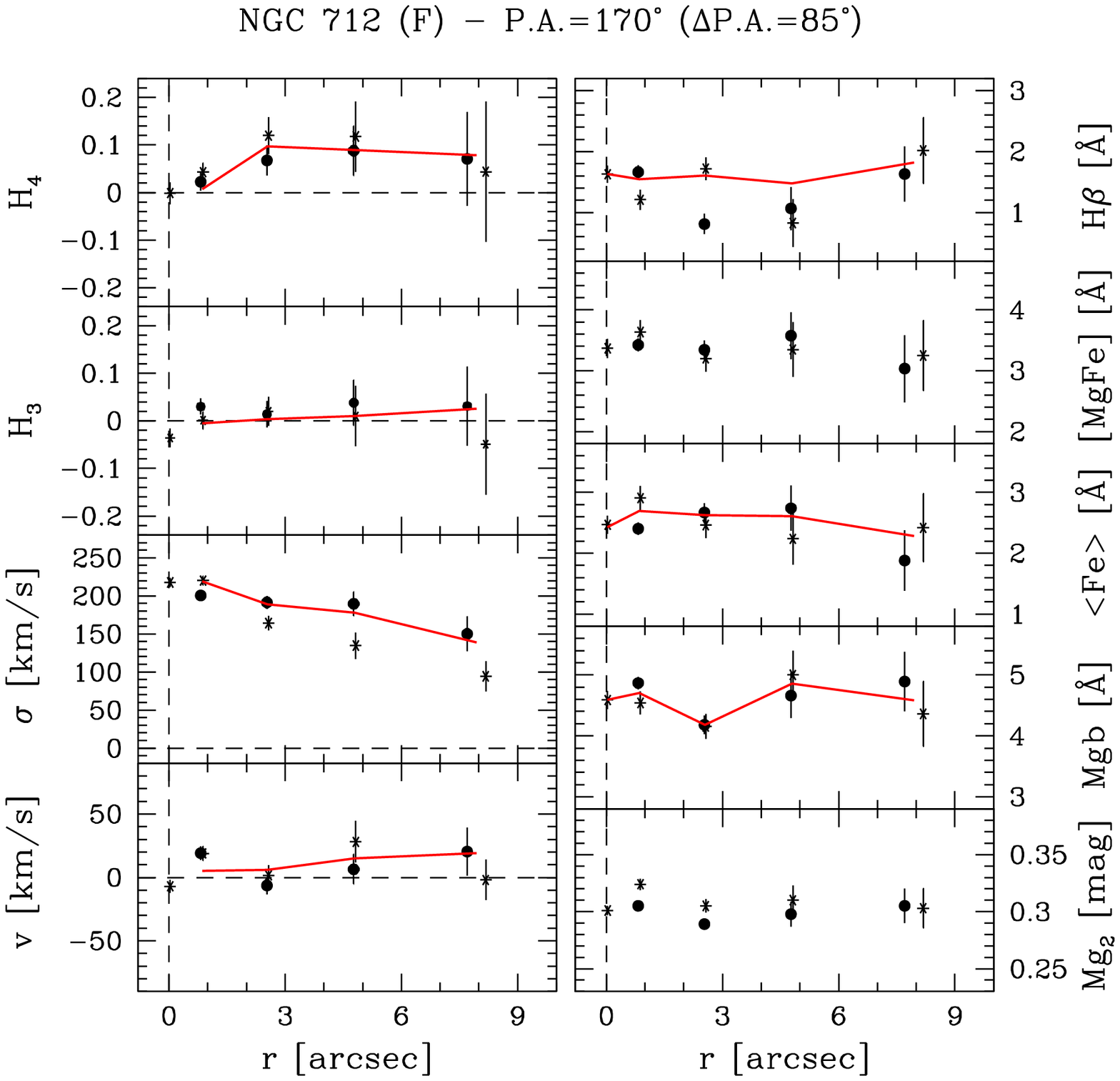} 
\caption{{\em Continued.}} 
\end{figure*} 
 
\addtocounter{figure}{-1} 
\begin{figure*}[ht!]
\includegraphics[width=0.5\textwidth]{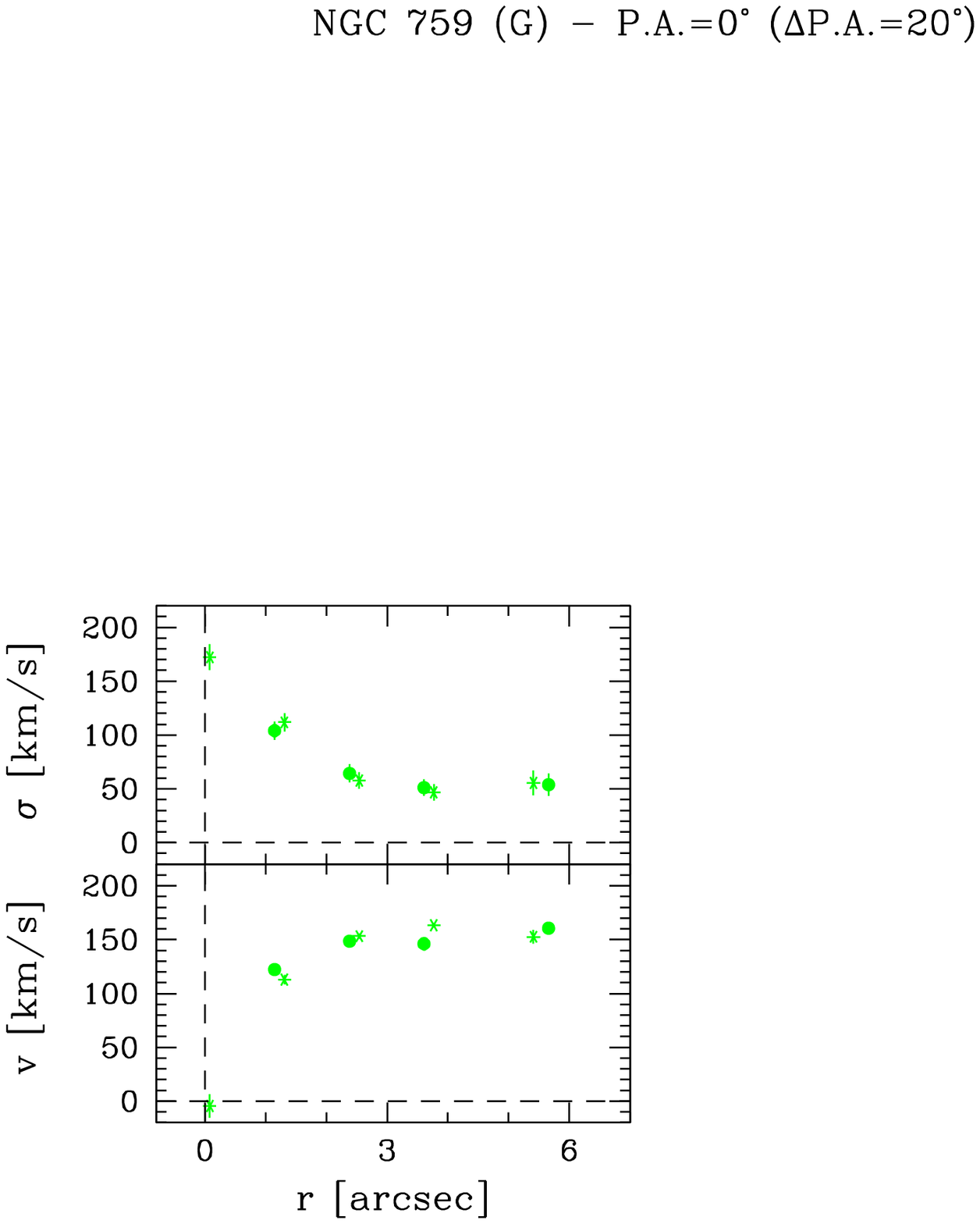} 
\includegraphics[width=0.5\textwidth]{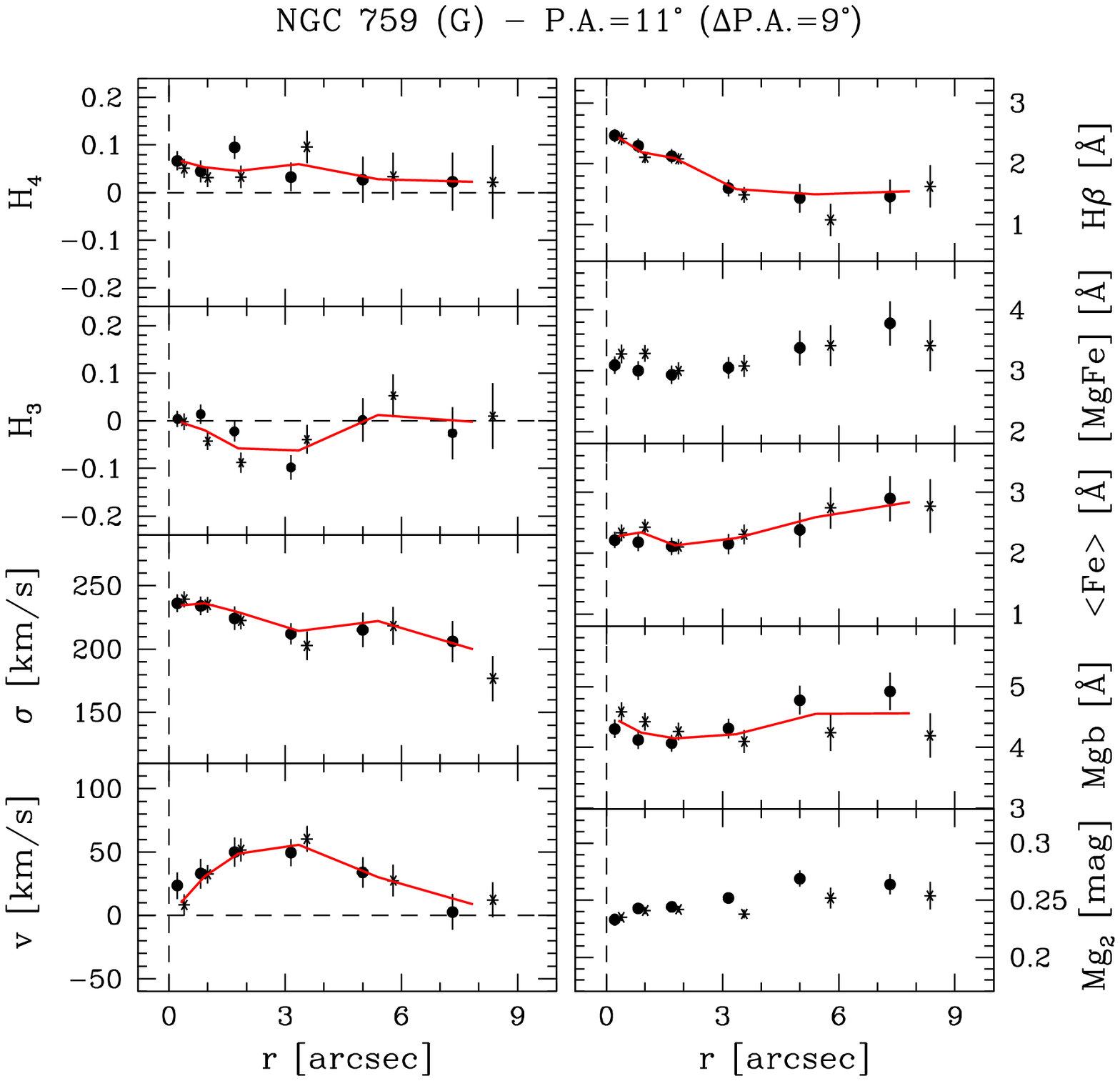} 
\includegraphics[width=0.5\textwidth]{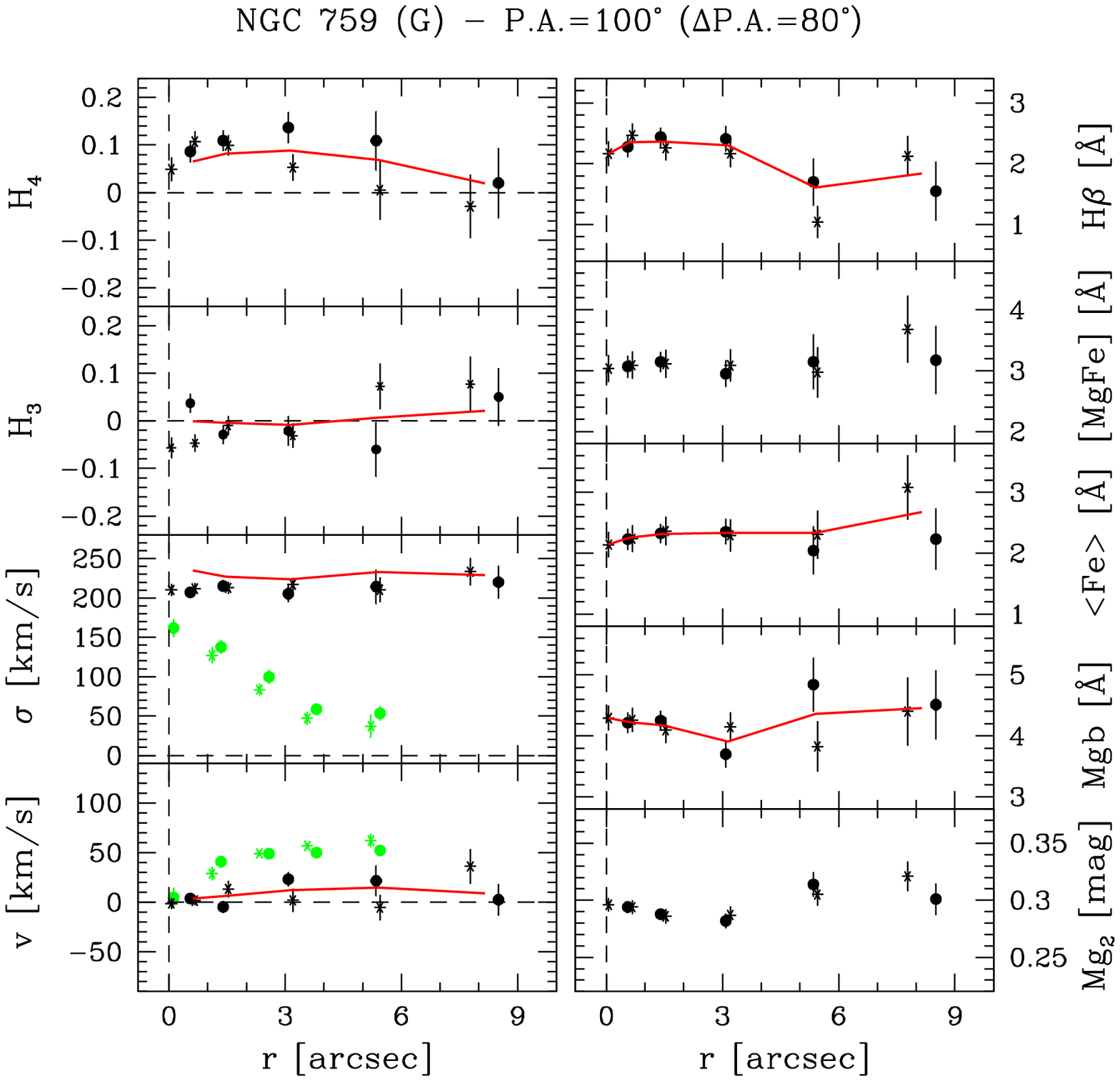} 
\includegraphics[width=0.5\textwidth]{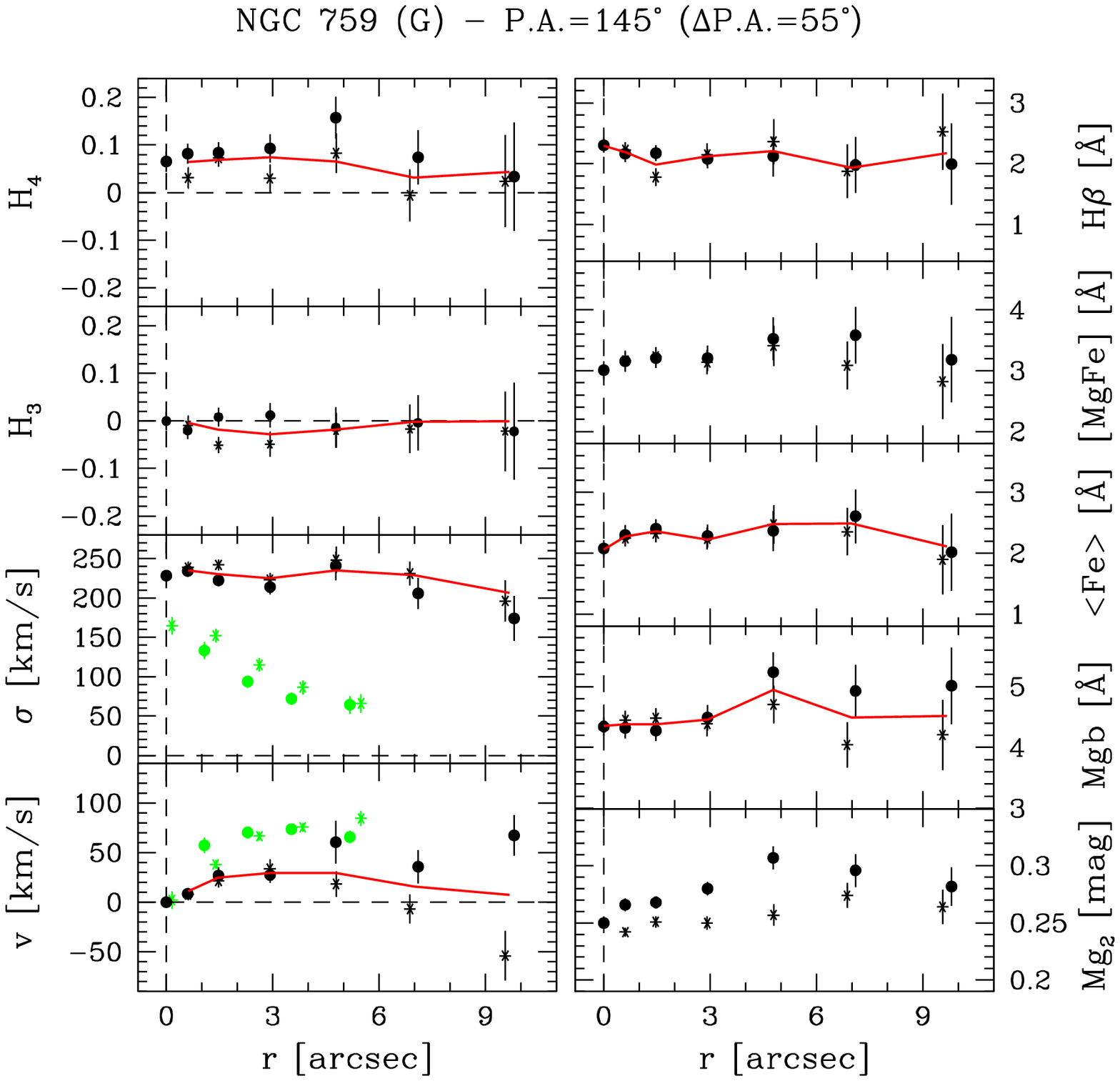}  
\caption{{\em Continued.}} 
\end{figure*} 

\addtocounter{figure}{-1} 
\begin{figure*}[ht!] 
\includegraphics[width=0.5\textwidth]{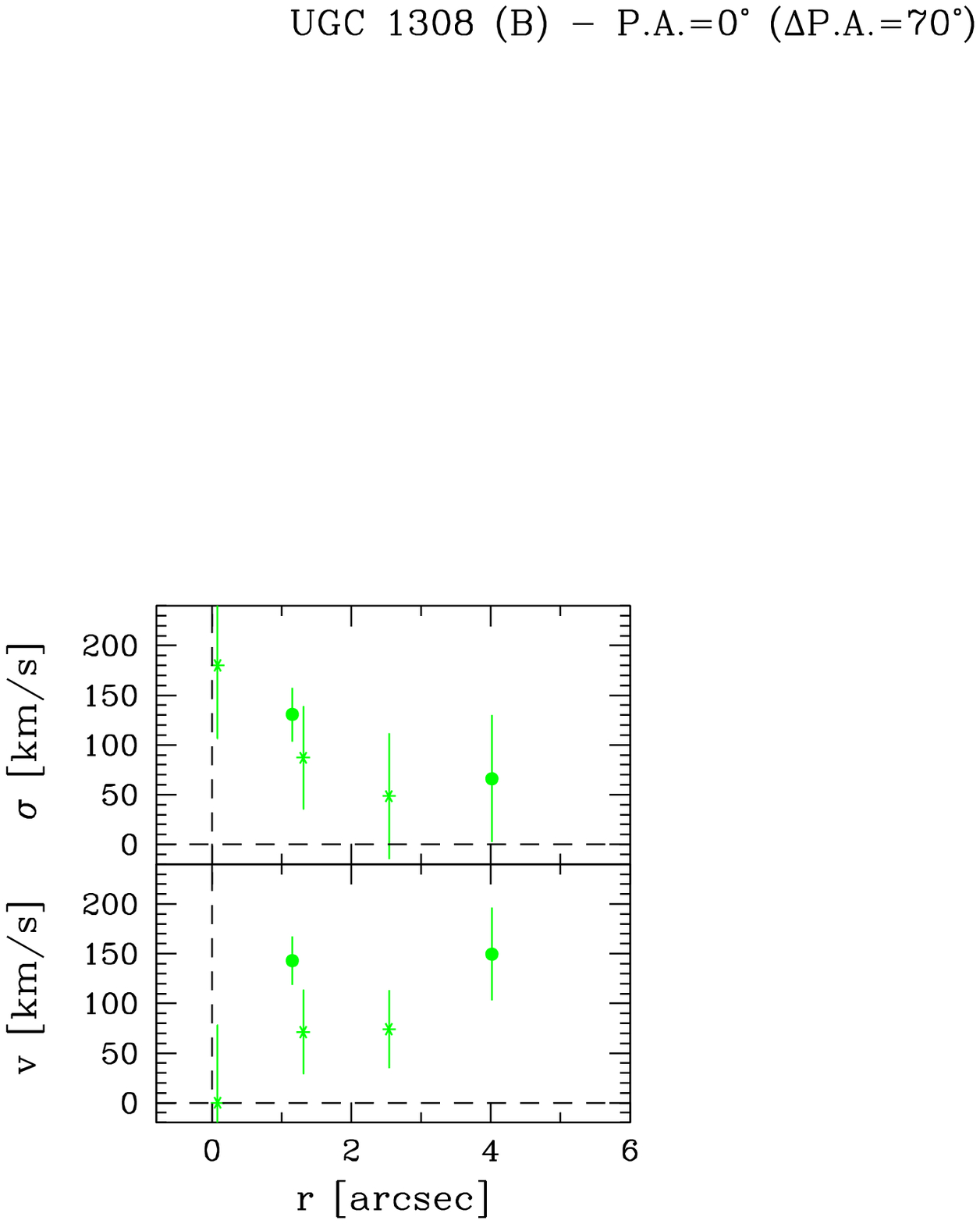}
\includegraphics[width=0.5\textwidth]{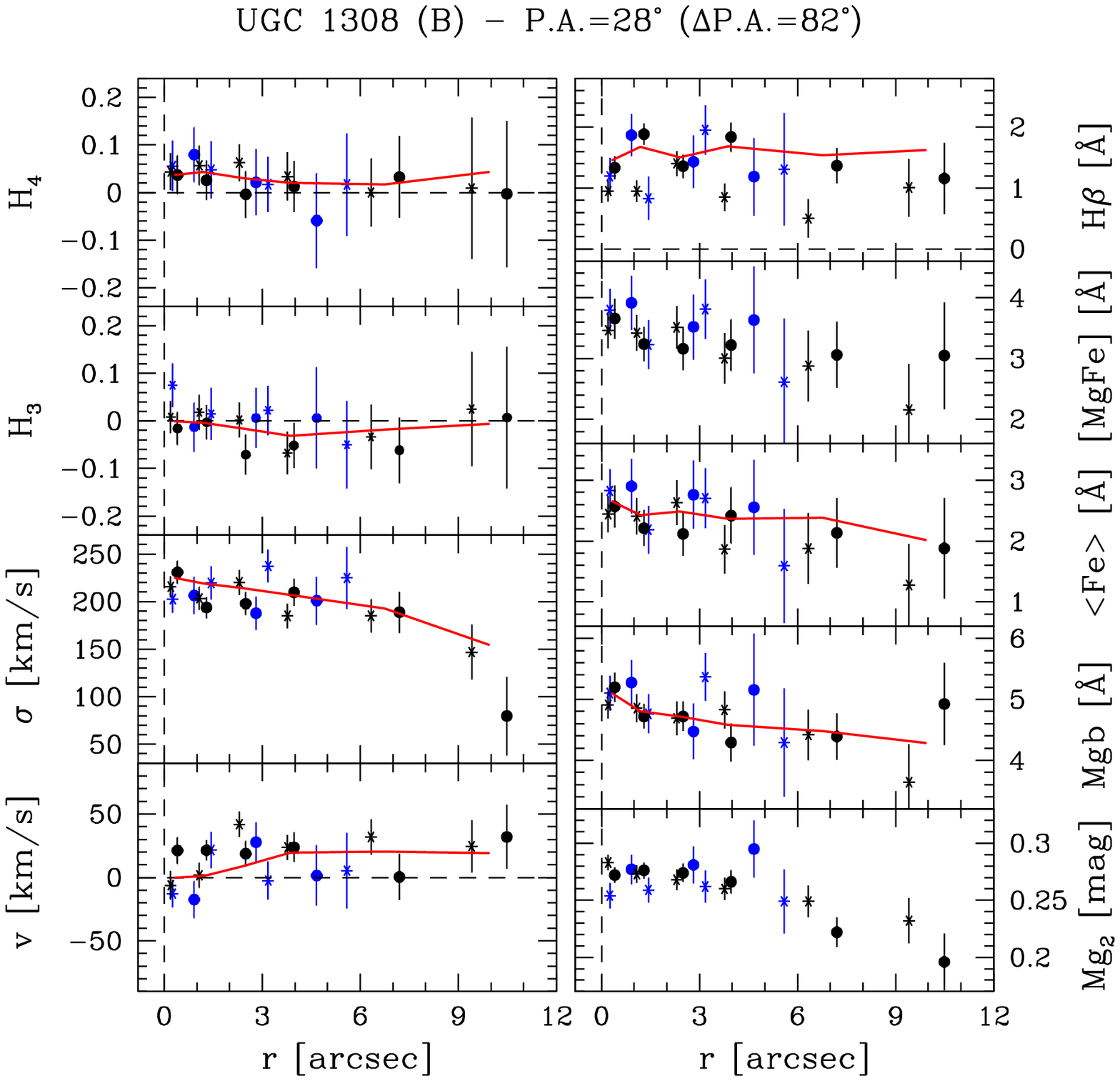} 
\includegraphics[width=0.5\textwidth]{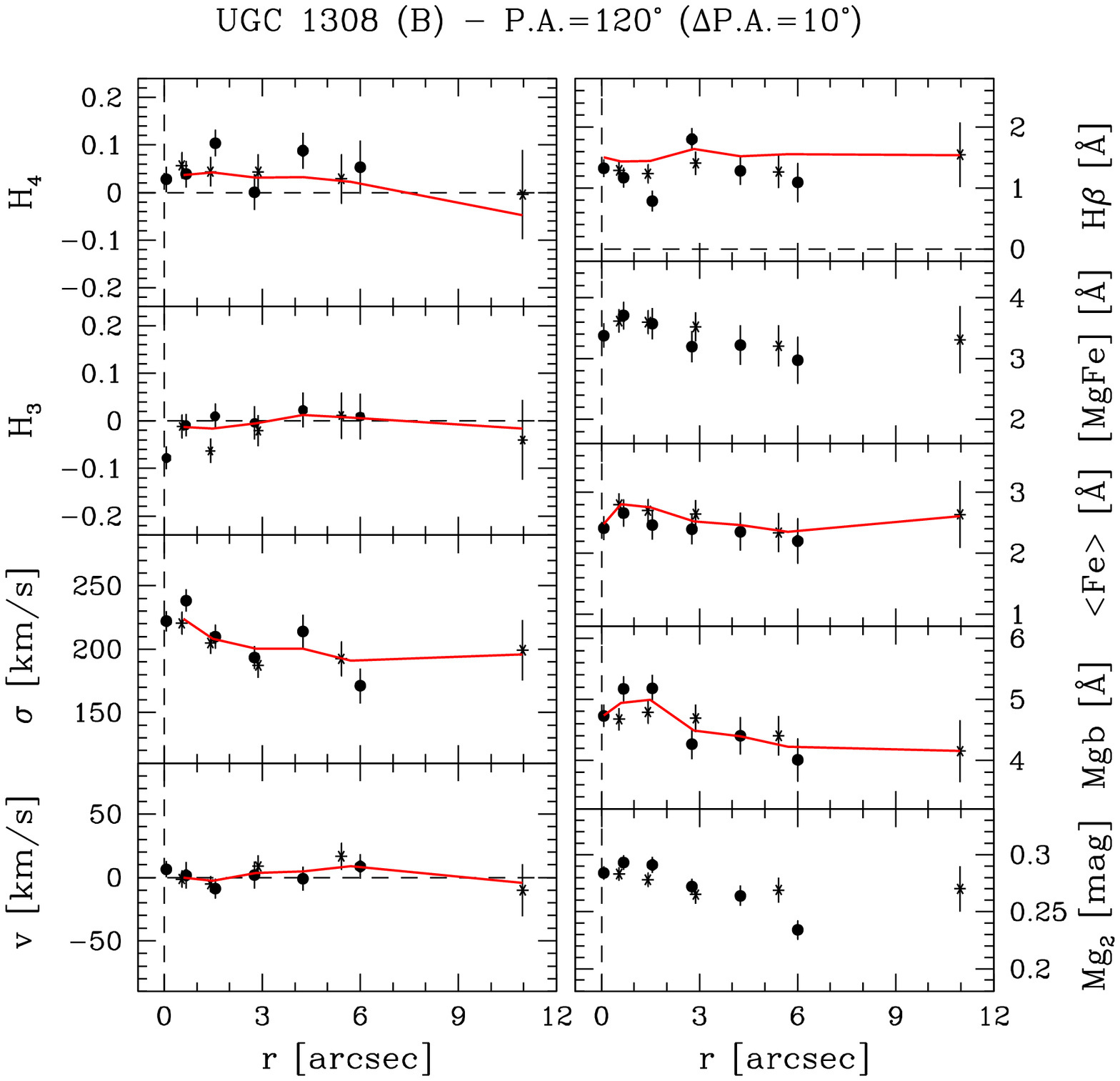}  
\caption{{\em Continued.}} 
\end{figure*} 

The galaxy spectra were convolved with a Gaussian function to match
the MILES spectral resolution \citep[FWHM=2.5 \AA,][]{Beifiori2010}.
Bad pixels coming from imperfect subtraction of cosmic rays and sky
emission lines were properly masked and excluded from the fitting
procedure. Ionized-gas emission lines were simultaneously fitted and a
fourth-order additive Legendre polynomial was added to correct for the
different shape of the continuum in the galaxy and template spectra.
 
The uncertainties on the kinematic parameters were estimated by Monte
Carlo simulations. The simulated spectra were obtained by convolving
the template spectra with a LOSVD parametrized as a Gauss-Hermite
series and adding photon, readout, and sky noise. The simulated
spectra were measured as if they were real.  Extensive testing was
performed to provide an estimate of the biases of the pPXF method with
the adopted instrumental setup and spectral sampling.  No bias was
found in the ranges of $S/N$ and $\sigma$ which characterize the
spectra of the sample galaxies. The values of $H_3$ and $H_4$ measured
for the simulated spectra differ from the intrinsic ones only within
the measured errors.
 
The measured stellar kinematics is reported in
Table~\ref{tab:kinematics} where velocities are relative to the galaxy
centers. The folded kinematic profiles are plotted in
Fig.~\ref{fig:kinematics}.
The multiple observations of IC~171, NGC~679, and UGC~1308 agree 
within the errors. 
  
\begin{figure}[t!]  
\includegraphics[bb=40 190 520 590, width=0.45\textwidth]{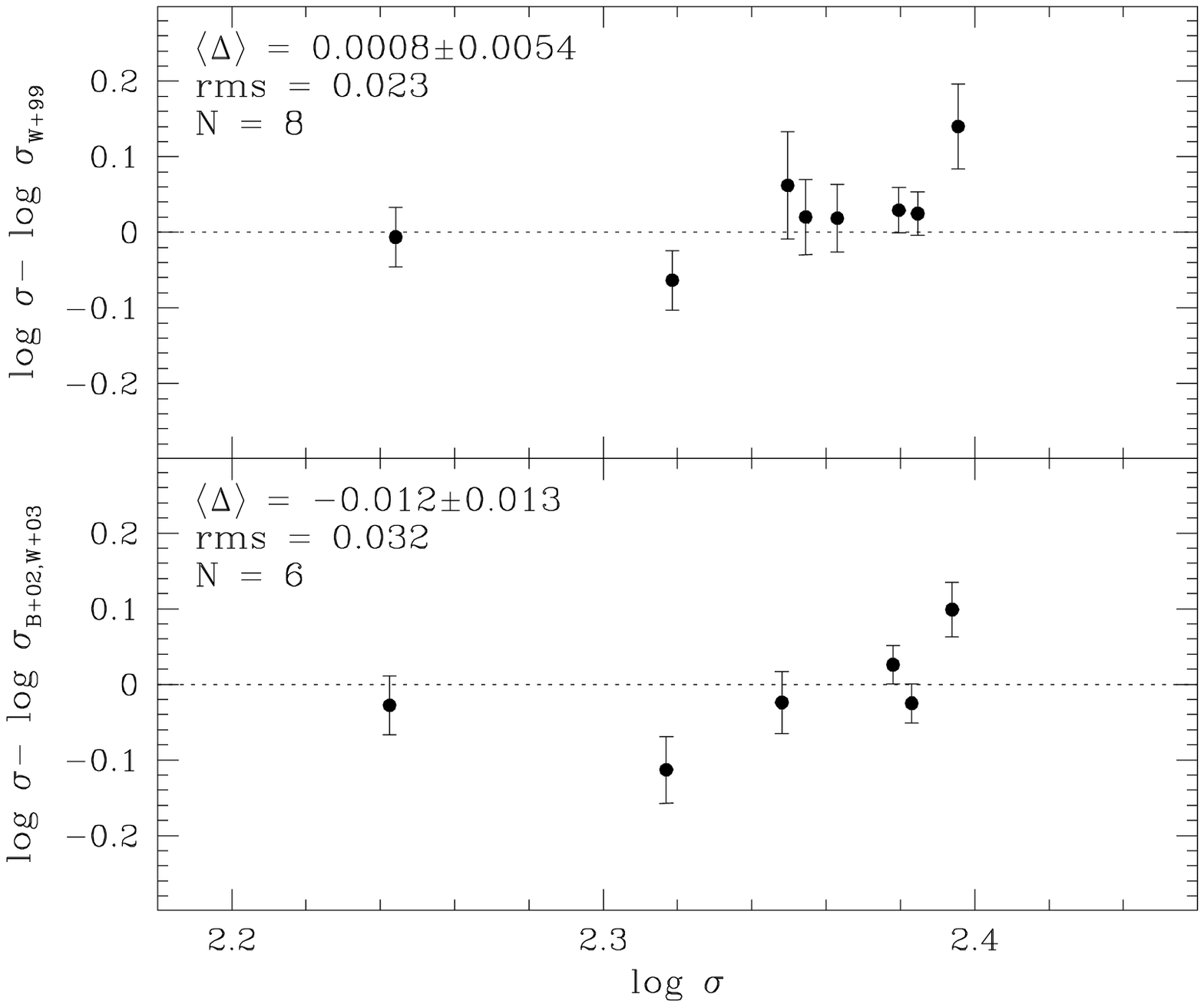}  
\caption{Central values of the stellar velocity dispersion
  $\sigma$ measured along the major axis of the sample galaxies are
  compared to those available in \citet[W$+$99]{Wegner1999},
  \citet[B$+$02]{Bernardi2002}, and \citet[W$+$03]{Wegner2003}.}
\label{fig:comparison}  
\end{figure}

Fig.~\ref{fig:comparison} shows the comparison between the
measurements of $\sigma$ obtained along the major axis of the sample
galaxies and values available in literature.
The mean difference in velocity dispersion with respect to the
measurements by \citet{Wegner1999} is $\langle \Delta
\log{\sigma}\rangle = 0.0008\pm0.0054$ and has a scatter
$\sigma_{\Delta \log{\sigma}} = 0.023$.  The central velocity
dispersion of all the sample galaxies, except for NGC~712 and
UGC~1308, were measured by \citet{Bernardi2002} and
\citet{Wegner2003}, too.  The mean difference is $\langle \Delta
\log{\sigma}\rangle = -0.012\pm0.013$ with $\sigma_{\Delta
  \log{\sigma}} = 0.032$.  { Our values thus agree with published
  data sets within their measured errors.}
 
Fig.~\ref{fig:central} shows the central values of $\sigma$ obtained
along the minor and diagonal axes of the sample galaxies { as a
  variance-weighted mean of the values available within an aperture of
  $3''$}. They are plotted as a function of the corresponding values
obtained along the major axis. Most of the data are consistent within
$3\sigma$ errors. No systematic effects are observed for the remaining
ones.

\subsection{Ionized-gas kinematics} 
  
The ionized-gas kinematics was measured from the emission lines
present in the spectra of run 3, namely \NII\ $\lambda\lambda$6548,
6583 \AA, \Ha , \SII\ $\lambda\lambda$6716, 6731 \AA. Each observed
emission line was fitted by a Gaussian, while describing the stellar
continuum with a low-order polynomial. The Gaussians were assumed to
share the same line-of-sight velocity ($v_{\rm gas}$) and velocity
dispersion ($\sigma_{\rm gas}$). A flux ratio of 1:2.96 was assumed
for the \NII\ doublet, as dictated by atomic physics
\citep[e.g.,][]{Osterbrock1989}.
 
The best-fitting Gaussian parameters were derived using a nonlinear 
least-squares minimization based on the robust Levenberg-Marquardt 
method by \citet{More1980}. The actual computation has been done using 
the MPFIT algorithm \citep{Markwardt2009} under the IDL environment. 
We averaged adjacent spectral rows to increase the $S/N$ of the
relevant emission lines. All the spectra of run 3 have $S/N<30$ per
resolution element in the continuum and were assigned to quality class
4 (Table~\ref{tab:log}).  We checked that the error in the kinematic
parameter determination derived by Monte Carlo simulations did not
differ significantly from the formal errors given as output by the
least-squares fitting routine. We therefore decided to assume the
latter as error bars on the gas kinematics.

The measured ionized-gas kinematics is reported in Table~\ref{tab:gas}
where velocities are given relative to galaxy centers and velocity
dispersions are corrected for instrumental FWHM. The folded kinematic
profiles are plotted in Fig.~\ref{fig:kinematics}.

\subsection{Line-strength indices} 
 
The Mg, Fe, and \Hb\ line-strength indices were measured following
\citet{Faber1985} and \citet{Worthey1994} from flux calibrated spectra
obtained in runs 1 and 2.
The average Iron index \Fe $\rm = (Fe_{5270}+Fe_{5335})/2$
\citep{Gorgas1990} and the combined Magnesium-Iron index with $\rm
[MgFe] = \sqrt{Mg{\it b} \langle Fe \rangle}$ \citep{Gonzalez1993}
were computed too.  Spectra were rebinned in the dispersion direction
as well as in the radial direction as before. The difference between
our spectral resolution and the Lick/IDS system one ($\rm FWHM=8.6$
\AA ; \citealt{Worthey1997}) was taken into account by degrading our
spectra to match the Lick/IDS resolution before measuring the
line-strength indices.
The original Lick/IDS spectra are not flux calibrated contrary to
ours. Such a difference in the continuum shape is expected to
introduce small systematic offsets of the measured values of the
indices. To establish these offsets and to calibrate our measurements
to the Lick/IDS system, the values of the line-strength indices
measured for the templates were compared to those obtained by
\citet{Worthey1994}. Fig.~\ref{fig:lick} shows the differences between
our and Lick/IDS line-strength indices for the stars observed in runs
1 and 2. The offsets were evaluated as the mean of the differences
between our and Lick/IDS line-strength values, and neglected if they
were smaller than the mean error of the differences. The latter was
the case of \Hb , \Mgd, and \Mgb\ for which no offset was adopted. On
the contrary, it was applied an offset to the measured values of
Fe$_{5270}$ ($\Delta {\rm Fe_{5270}}=-0.29$ \AA\ for run 1, $-0.28$
\AA\ for run 2) and Fe$_{5335}$ ($\Delta {\rm Fe_{5335}} =-0.48$
\AA\ for run 1, $-0.44$ \AA\ for run 2) to bring our data to the
Lick/IDS system.
No focus correction was applied because atmospheric seeing was the
dominant effect during observations (see \citealt{Mehlert1998} for
details).
Errors on indices were derived from photon statistics and CCD read-out 
noise, and calibrated by means of Monte Carlo simulations. 
  
\begin{figure*}[t!] 
\includegraphics[width=0.75\textwidth,angle=270]{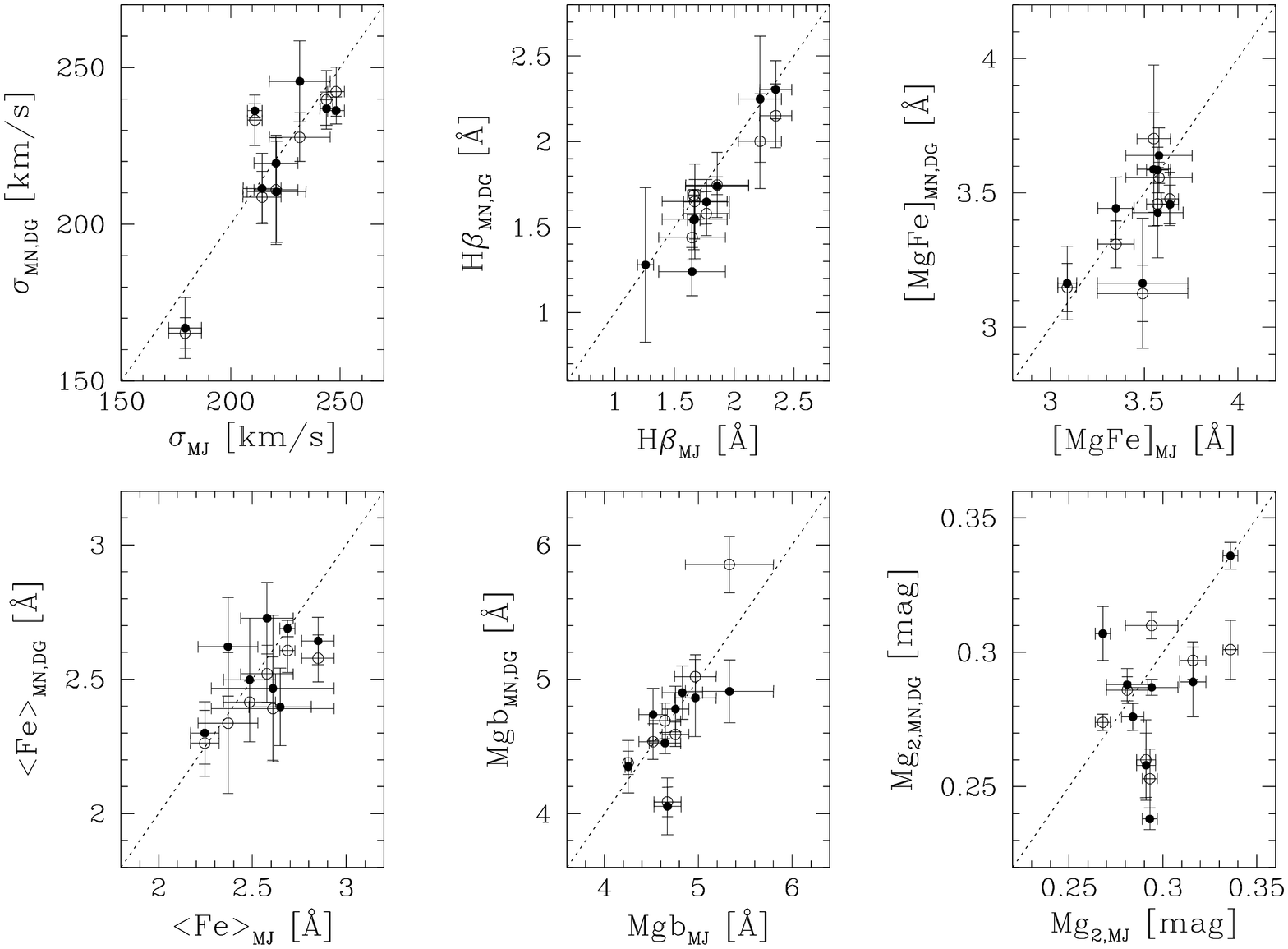}  
\caption{Central values of the stellar velocity dispersion
  $\sigma$ and line-strength indices \Hb , [MgFe], \Fe , \Mgb , and
  \Mgd\ measured within an aperture of $3''$ along the minor (filled
  circles) and diagonal axis (open circles) are compared to those
  measured along the major axis.}
\label{fig:central}  
\end{figure*}

\begin{figure*}[t!]  
\includegraphics[bb=20 145 590 540, width=\textwidth]{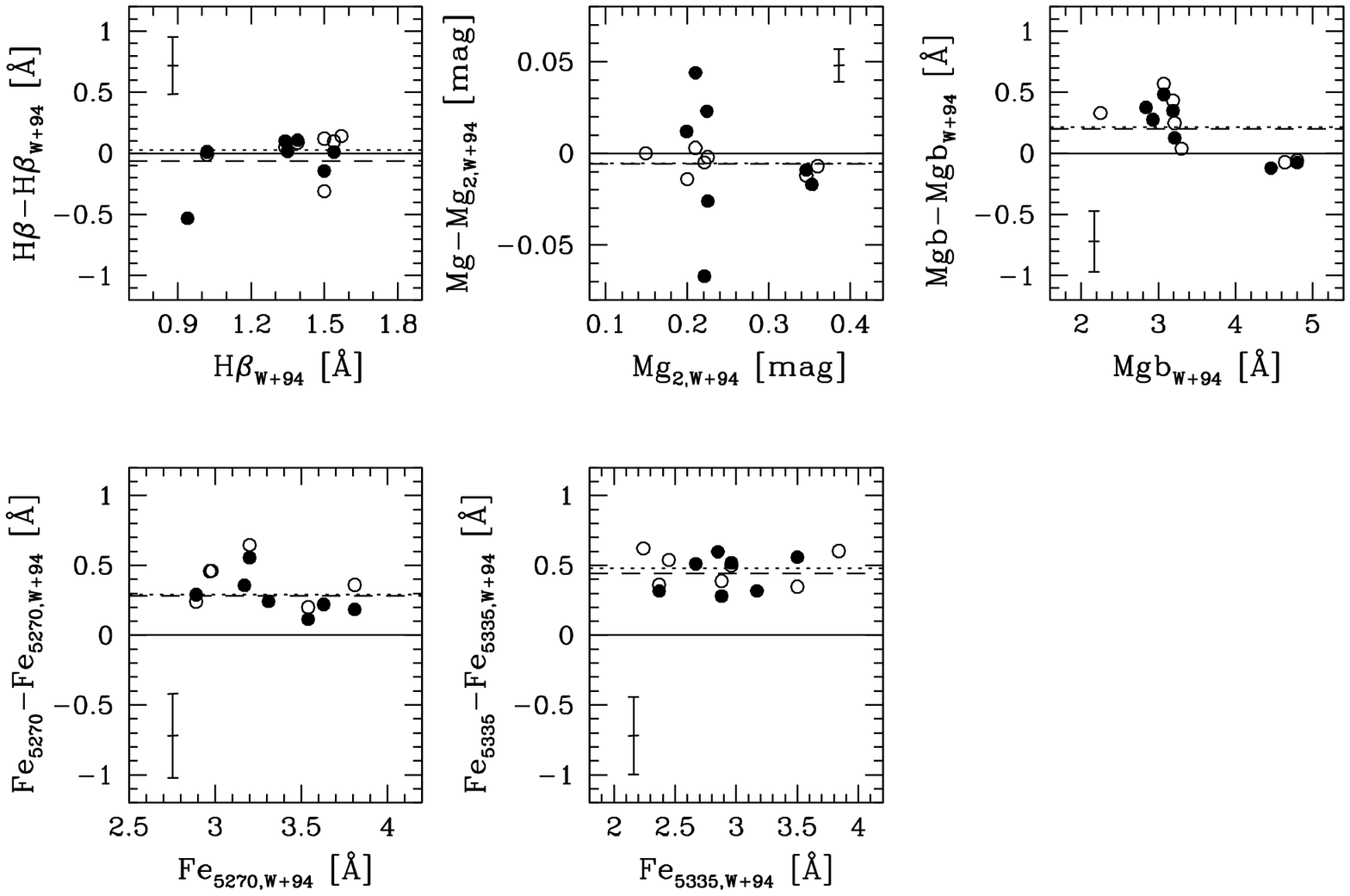}  
\caption{Comparison of the line-strength indices \Hb , \Mgd , \Mgb ,
  Fe$_{\rm 5270}$, and Fe$_{\rm 5335}$ measured for the stars in
  common with \citet[W$+$94]{Worthey1994}.  In each panel the error
  bar in the upper left corner indicates the mean error of the
  difference between the our values and those by \citet{Worthey1994},
  the continuous line shows the line of correspondence, the dotted and
  dashed lines correspond to the average difference of data in run 1
  (open circles) and in run 2 (filled circles), respectively.}
\label{fig:lick}    
\end{figure*}

The measured values of \Hb , [MgFe], \Fe, \Mgb , and \Mgd\ are listed
in Table~\ref{tab:indices} and plotted in Fig.~\ref{fig:kinematics}.
Fig.~\ref{fig:central} shows the central values of \Hb , [MgFe], \Fe ,
\Mgb , and \Mgd\ obtained along the minor and diagonal axes of the
sample galaxies as { a variance-weighted mean of the values
  available within an aperture of $3''$} and those obtained along the
corresponding major axes. Most of the data are consistent within
$3\sigma$ errors. No systematic effects are observed for the remaining
ones.

\subsection{Properties of the stellar populations}
\label{sec:SSP}

{ The traditional and effective method of studying stellar
  population properties uses diagrams of different pairs of
  line-strength indices. Here,} the modeling of the line-strength
indices measured in the previous section followed the method described
in \citet{Saglia2010a} and \citet{Pu2010}.  { Among the measured
  line-strength indices, \Hb, \Fe , and \Mgb\ were considered since
  \Hb\ is sensitive to early-type stars and thus useful as an age
  indicator, \Fe\ gives the metal abundance, and \Mgb\ covers the
  $\alpha-$element abundance.}  The SSP models of \citet{Maraston1998,
  Maraston2005} and the models of Lick/IDS line-strength indices with
$\alpha-$element overabundance of \citet{ThomasD2003} were
spline-interpolated on a fine grid in age $t$ (up to 15 Gyr { on
  steps of 0.1 Gyr}), metallicity [$Z$/H] (from $-2.25$ to $0.67$ dex
{ on steps of 0.02 dex}) and overabundances [$\alpha$/Fe] (from
$-0.3$ to $0.5$ dex { on steps of 0.05 dex}). At each radius $r$ of
the galaxies the best-fit values of $\tau$, [$Z$/H], and [$\alpha$/Fe]
were determined by minimizing the $\chi^2$ function
\begin{equation}
\label{eq:chi2}
\chi^2(r)= \Delta^2 {\rm H}\beta(r) + \Delta^2 {\rm Mg}b(r)
   + \Delta^2 \langle {\rm Fe} \rangle (r),
\end{equation}
where
\begin{equation}
\label{eq:delta2}
\Delta^2 {\rm I} (r) = \left( \frac{{\rm I}(r) - {\rm I}_{\rm
    SSP}(\tau,[Z/{\rm H}],[\alpha/{\rm Fe}])}{\sigma_{\rm I}(r)} \right)^2.
\end{equation}
Here $\sigma_{\rm I}(r)$ is the error on the index ${\rm I}(r)$
measured at a distance $r$ and ${\rm I}_{\rm SSP}(\tau,[Z/{\rm
    H}],[\alpha/{\rm Fe}])$ is the value predicted by the SSP models
for the given set of age, metallicity, and overabundance. As in
\citet{Saglia2010a} and \citet{Pu2010}, { the best-fit values of
  $\tau$, [$Z$/H], and [$\alpha$/Fe] are derived to the same accuracy
  through the $\chi^2$ minimization by giving the same weight to \Hb,
  \Fe , and \Mgb . We} did not extrapolate to values outside the model
grid. Therefore, the minimum $\chi^2$ solutions are sometimes found at
the edges of the parameter space. Once the set of best-fit values,
i.e.  the set $(\tau_{\rm min},[Z/{\rm H}]_{\rm min},[\alpha/{\rm
    Fe}]_{\rm min})$ that minimizes $\chi^2(r)$ was determined, we
computed the values of the mass-to-light ratios of the corresponding
SSP models for the Kroupa IMF ($\kroupa$).  The resulting values of
$\kroupa$, stellar-population ages, metallicities and overabundances
(inside $\reff$) are given in Table~\ref{tab:sspres}.  Errors on all
quantities were derived by considering the minimal and maximal
parameter variations compatible with $\Delta \chi^2 =
\chi^2-\chi^2_{\rm min}=1$. { Note, however, that errors on age and
  metallicity are correlated to some extent \citep{Worthey1994b}.}

\section{Dynamical modeling} 
\label{sec:dynamics} 
 
As discussed in Sec.~\ref{sec:introduction}, the dynamical modeling
follows our previous work as in \citet{Thomas2007a, Thomas2007b,
  Thomas2009a} and is based on the orbit superposition technique
\citep{Schwarzschild1979} as implemented in \citet{Thomas2004,
  Thomas2005}.  We first determined the stellar luminosity density
from the surface brightness profile by deprojection into a
three-dimensional, axisymmetric distribution with specified
inclination, { using the simulated annealing scheme implemented by
  \citet{Magorrian1999}. This algorithm reconstructs the luminosity
  density non-parametrically on a set of $\approx 250$ sampling points
  in a plane perpendicular to the equatorial one and crossing the
  galactic center.}
 
The adopted inclinations of the galaxies are given in
Table~\ref{tab:dynres}. The nearly round dust features in NGC~679,
NGC~759 and UGC~1308 suggest that these galaxies are almost face-on
and we deprojected them at the lowest inclination compatible with a
system that is intrinsically not flatter than an E7 galaxy.  In all
other galaxies, except NGC~687, the projected ellipticity reaches
$\epsilon \approx 0.4$ at some radius and we assumed that these
galaxies are edge-on ($i=90^\circ$). The inclination of NGC~687 is
unclear, but the galaxy is slightly flattened without showing any
significant rotation at almost all radii and we assumed that this
galaxy is edge-on, too.
For the galaxies where no HST images were available, we deprojected
the ground-based surface brightness taking into account seeing
convolution as in \citet{Rusli2010}. The solid lines in the
Fig.~\ref{fig:photometry} show the photometric profiles obtained by
projecting the deprojected stellar luminosity density.
   
{ As a second step, we modeled the total mass distribution of the
  galaxies as the sum of two components
\begin{equation}
\rho = \rholum + \rhodm
\label{eq:rho}
\end{equation}
where $\rholum$ is the density of matter (either luminous or dark)
that is distributed like the stars and $\rhodm$ is the density of dark
matter distributed in a halo.  The mass-to-light ratio of the matter
distributed like the stars is
\begin{equation}
\label{refseq2.rho}
\mllum = \rholum/\nu 
\end{equation}
where $\nu$ is the deprojected stellar luminosity density. Note that
$\rholum$ measures all the matter following the light (it may include
dark matter as well as stars). In contrast, $\rhodm$ is the halo
density (some of the dark matter may not follow this profile) and its
distribution is assumed to be either a cored logarithmic profile
\citep{Binney1987}
\begin{equation} 
\label{eq:log} 
\rho_{\rm LOG}(r) = \frac{V_{\rm c}^2}{4 \pi G} 
  \frac{3 r_{\rm c}^2+r^2}{(r_{\rm c}^2+r^2)^2} 
\end{equation} 
or a NFW profile \citep{Navarro1996}
\begin{equation}
\rho_\mathrm{NFW}(r) \propto \frac{1}{r(r+r_{\rm s})^2},
\end{equation}
implemented as in \citet{Thomas2007a}. In either case the halo is
assumed spherical. Cored logarithmic halos have an asymptotically
constant circular velocity $V_{\rm c}$ and a flat density core inside
$r < r_{\rm c}$.}

Then, the gravitational potential $\Phi$ was computed by integrating
Poisson's equation. Thousands of orbits were calculated in this fixed
potential.  The orbits were superposed to fit the observed LOSVDs,
following the luminosity density constraint. { The maximum entropy
  technique of \citet{Richstone1988} was used to fit the kinematic
  data.  The maximum entropy technique is a method for modeling
  galaxies from limited observations. While many distribution
  functions are consistent with the collisionless Boltzmann equation
  and the observational data, statistical mechanics and information
  theory suggest that the most natural ones have the largest entropy.
  The function
\begin{equation} 
\label{eq:entropy} 
\hat{S} = S - \alpha \chi^2, 
\end{equation} 
is maximized,
where $S$ is an approximation to the Boltzmann entropy and $\chi^2$ is
the sum of the squared residuals to the kinematic data. The smoothing
parameter $\alpha$ controls the influence of the entropy $S$ on the
orbital weights. \citet{Thomas2005} used Monte-Carlo simulations to
derive the optimal $\alpha$ for a typical long-slit setup.  }

\begin{table*}[t!] 
\caption{Stellar-population parameters averaged inside $\reff$
    \label{tab:sspres}} 
\begin{tabular}{lccccc} 
\tableline  
\tableline  
\multicolumn{1}{c}{Object} & 
\multicolumn{1}{c}{$\kroupa$} &
\multicolumn{1}{c}{$\tau$} & 
\multicolumn{1}{c}{[Z/H]} &
\multicolumn{1}{c}{[$\alpha$/Fe]} &
\multicolumn{1}{c}{$\zstars$} \\  
\multicolumn{1}{c}{} & 
\multicolumn{1}{c}{(\msun\ \lsun$^{-1}$)} &
\multicolumn{1}{c}{(Gyr)} &
\multicolumn{1}{c}{} & 
\multicolumn{1}{c}{} & 
\multicolumn{1}{c}{} \\   
\multicolumn{1}{c}{(1)} & \multicolumn{1}{c}{(2)} &   
\multicolumn{1}{c}{(3)} & \multicolumn{1}{c}{(4)} &   
\multicolumn{1}{c}{(5)} & \multicolumn{1}{c}{(6)} \\ 
\tableline    
\object{IC   171 (E)} & $3.22\pm 0.70$ & $9.0\pm 2.4$ & $0.12 \pm 0.10$ & $0.08 \pm 0.05$ & $1.3^{+1.3}_{-0.6}$ \\ 
 
\object{NGC  679 (D)} & $3.65\pm 0.29$ & $9.7\pm 1.0$ & $0.21 \pm 0.05$ & $0.31 \pm 0.02$ & $1.5^{+0.5}_{-0.3}$ \\ 
 
\object{NGC  687 (C)} & $3.72\pm 0.44$ & $9.9\pm 1.6$ & $0.17 \pm 0.07$ & $0.22 \pm 0.03$ & $1.6^{+1.1}_{-0.5}$ \\  
  
\object{NGC  703 (I)} & $2.15\pm 0.32$ & $5.2\pm 1.2$ & $0.23 \pm 0.05$ & $0.45 \pm 0.02$ & $0.5^{+0.2}_{-0.1}$ \\  
  
\object{NGC  708 (A)} & $4.17\pm 1.05$ & $11.0\pm 3.1$ & $0.19 \pm 0.14$ & $0.39 \pm 0.05$ & $2.2^{+\infty}_{-1.2}$ \\  
  
\object{NGC  712 (F)} & $2.42\pm 0.37$ & $6.2\pm 1.4$ & $0.20 \pm 0.08$ & $0.39 \pm 0.03$ & $0.7^{+0.2}_{-0.2}$ \\  
 
\object{NGC  759 (G)} & $2.75\pm 0.31$ & $7.1\pm 0.9$ & $0.16 \pm 0.07$ & $0.35 \pm 0.04$ & $0.8^{+0.2}_{-0.1}$ \\  
 
\object{UGC 1308 (B)} & $3.07\pm 1.06$ & $8.8\pm 4.4$ & $0.07 \pm 0.16$ & $0.30 \pm 0.10$ & $1.2^{+5.0}_{-0.8}$ \\  
\tableline    
\end{tabular}       
\tablecomments{ Col. 1: Name.  Col. 2: SSP mass-to-light ratio
$\kroupa$ in the Kron-Cousins $R$ band.  Col. 3: SSP age
$\tau$. Col. 4: SSP metallicity [Z/H].  Col. 5: SSP $\alpha$-elements
overabundance [$\alpha$/Fe].  Col. 6: Star-formation redshift
$\zstars$.}
\end{table*}    

The best-fit dark halo densities, dark-halo mass fractions, and
mass-to-light ratios $\mldyn$ of the mass that is tied to the light
are given in Table~\ref{tab:dynres}. The corresponding best fitting
lines to the kinematic data are shown in Fig.~\ref{fig:kinematics}.

\begin{table*}[t!] 
\caption{Mass-to-light ratio, parameters of the dark matter
  halo, stellar velocity dispersion, and geometric parameters adopted 
  for the dynamical model.
    \label{tab:dynres}} 
\begin{tabular}{lccccccccc} 
\tableline  
\tableline  
\multicolumn{1}{c}{Object} & 
\multicolumn{1}{c}{Model} &  
\multicolumn{1}{c}{$\mldyn$} &
\multicolumn{1}{c}{$\dmfrac$} &  
\multicolumn{1}{c}{$\log_{10} \langle \rhodm \rangle$} &  
\multicolumn{1}{c}{$\zdm$} &
\multicolumn{1}{c}{$\sigeff$} & 
\multicolumn{1}{c}{$\istars$} &
\multicolumn{1}{c}{$\pastars$} \\   
\multicolumn{1}{c}{} & 
\multicolumn{1}{c}{} &  
\multicolumn{1}{c}{(\msun\ \lsun$^{-1}$)} & 
\multicolumn{1}{c}{} & 
\multicolumn{1}{c}{(\msun\ pc$^{-3}$)} &  
\multicolumn{1}{c}{} &
\multicolumn{1}{c}{($\kms$)} &
\multicolumn{1}{c}{($^\circ$)} & 
\multicolumn{1}{c}{($^\circ$)} \\  
\multicolumn{1}{c}{(1)} & \multicolumn{1}{c}{(2)} &   
\multicolumn{1}{c}{(3)} & \multicolumn{1}{c}{(4)} &   
\multicolumn{1}{c}{(5)} & \multicolumn{1}{c}{(6)} & 
\multicolumn{1}{c}{(7)} & \multicolumn{1}{c}{(8)} & 
\multicolumn{1}{c}{(9)} \\ 
\tableline    
\object{IC   171 (E)} & NFW & $3.5^{+0.5}_{-0.5}$  & $0.41^{+0.18}_{-0.12}$ 
 & $-2.17^{+0.33}_{-0.14}$ 
 & $0.7$ & $211.4\pm 1.5$ & $90$ & $115$\\ 
 
\object{NGC  679 (D)} & NFW & $4.0^{+0.5}_{-0.5}$  & $0.27^{+0.13}_{-0.11}$ 
 & $-1.00^{+0.25}_{-0.04}$ 
 & $1.3$ & $252.7\pm 0.9$ & $30$ & $150$\\ 
 
\object{NGC  687 (C)} & LOG & $3.5^{+0.5}_{-1.0}$  & $0.36^{+0.21}_{-0.07}$ 
 & $-1.03^{+0.08}_{-0.21}$ 
 & \nodata & $237.4\pm 1.3$ & $90$ & $120$\\  
  
\object{NGC  703 (I)} & LOG & $9.5^{+0.5}_{-0.5}$  & $0.001^{+0.062}_{-0.001}$ 
 & $-3.12^{+1.56}_{-\infty}$ 
 & $3.7$ & $233.3\pm 1.2$ & $90$ & $ 45$\\  
  
\object{NGC  708 (A)} & SC & $7.5^{+1.0}_{-2.5}$  & $<0.41$ 
 & $<-2.11$ 
 & $0.8$ & $189.2\pm 1.6$ & $90$ & $130$\\  
  
\object{NGC  712 (F)} & SC & $8.5^{+0.5}_{-0.5}$  & $<0.06$ 
 & $<-1.84$ 
 & $4.0$ & $225.2\pm 0.9$ & $90$ & $ 85$\\  
 
\object{NGC  759 (G)} & LOG & $4.5^{+0.5}_{-0.5}$  & $0.48^{+0.04}_{-0.18}$ 
 & $-1.22^{+0.17}_{-0.25}$ 
 & $2.8$ & $231.0\pm 1.1$ & $40$ & $ 20$\\  
 
\object{UGC 1308 (B)} & SC & $2.6^{+0.1}_{-0.1}$  & $<0.20$ 
 & $<-1.95$ 
 & \nodata & $202.6\pm 1.8$ & $50$ & $110$\\  
\tableline    
\end{tabular}     
%
\tablecomments{ Col. 1: Name.  Col. 2: Type of best-fit mass model
  (LOG: logarithmic halo, NFW: NFW halo, SC: self-consistent, i.e. no
  dark matter halo).  Col. 3: Dynamical { mass-to-light ratio
    $\mldyn$ (including all the mass that follows the light)} in the
  Kron-Cousins $R$ band.  Col. 4: Spherically averaged { halo-mass}
  fraction within $\reff$. Col. 5: Average { halo} density within
  $2 \, \reff$.  Col. 6: Dark halo assembly redshift $\zdm$.  Col. 7:
  Velocity dispersion $\sigeff$ inside $\reff$.  Col. 8: Assumed
  galaxy inclination $\istars$.  Col. 9: Position angle $\pastars$ of
  the line of nodes measured North through East adopted for the
  dynamical model.}
\end{table*}    

\section{Comparison with strong gravitational lenses
and kinematics of the ionized gas}
\label{sec:comparison}

{ The statistical significance and accuracy of the methods employed
  here have been previously tested. \citet{Thomas2005} checked the
  axisymmetric orbit superposition code with mock observations of Coma
  galaxies and used Monte Carlo simulations to show that internal
  velocity moments could be reconstructed to $\approx\,15\%$.  In
  \citet{Thomas2007a}, the mass-to-light ratio, halo-mass fraction,
  and shape of the circular velocity curve from the dynamical
  modeling of 17 Coma early-type galaxies are shown to be robust
  against the choice of the regularization parameter $\alpha$.

A more severe modeling uncertainty is the unknown flattening of the
galaxies along the line-of-sight. Even if all galaxies were perfectly
axisymmetric their flattening along the line-of-sight is not directly
observable and can only be inferred from the apparent flattening if
the inclination is given. For early-type galaxies this is in general
not the case and, moreover, even with the help of stellar kinematics
the inclination remains only weakly constrained
\citep[e.g.,][]{Krajnovic2005, vandenBosch2009}. However, dynamical
mass-to-light ratios turn out to be robust against the inclination
mismatch \citep[e.g.,][]{Krajnovic2005, Thomas2005,
  vandenBosch2009}. Only if the flattening along the line-of-sight is
significantly underestimated (as in the case of an almost face-on disk
modeled as a spheroid), the dynamical masses can be biased by up to a
factor of two (\citealt{Thomas2007b}). In addition, the mass recovery
in the axisymmetric approximation becomes dependent on viewing angle
and shape for significantly triaxial galaxies. End-on views of
prolate/triaxial models give $20-30\%$ overestimated masses, while for
highly flattened face-on systems masses are underestimated by up to
$50\%$ \citep{Thomas2007b}.

Both, published masses of gravitational strong lenses that are
structurally similar to our galaxies as well as our limited
emission-line data for the galaxies in Abell 262 provide valuable
checks on our foregoing analysis.}

\subsection{Comparison with strong gravitational lensing} 
\label{sec:lens} 

\begin{figure}[t!]  
\centering \includegraphics[width=0.45\textwidth]{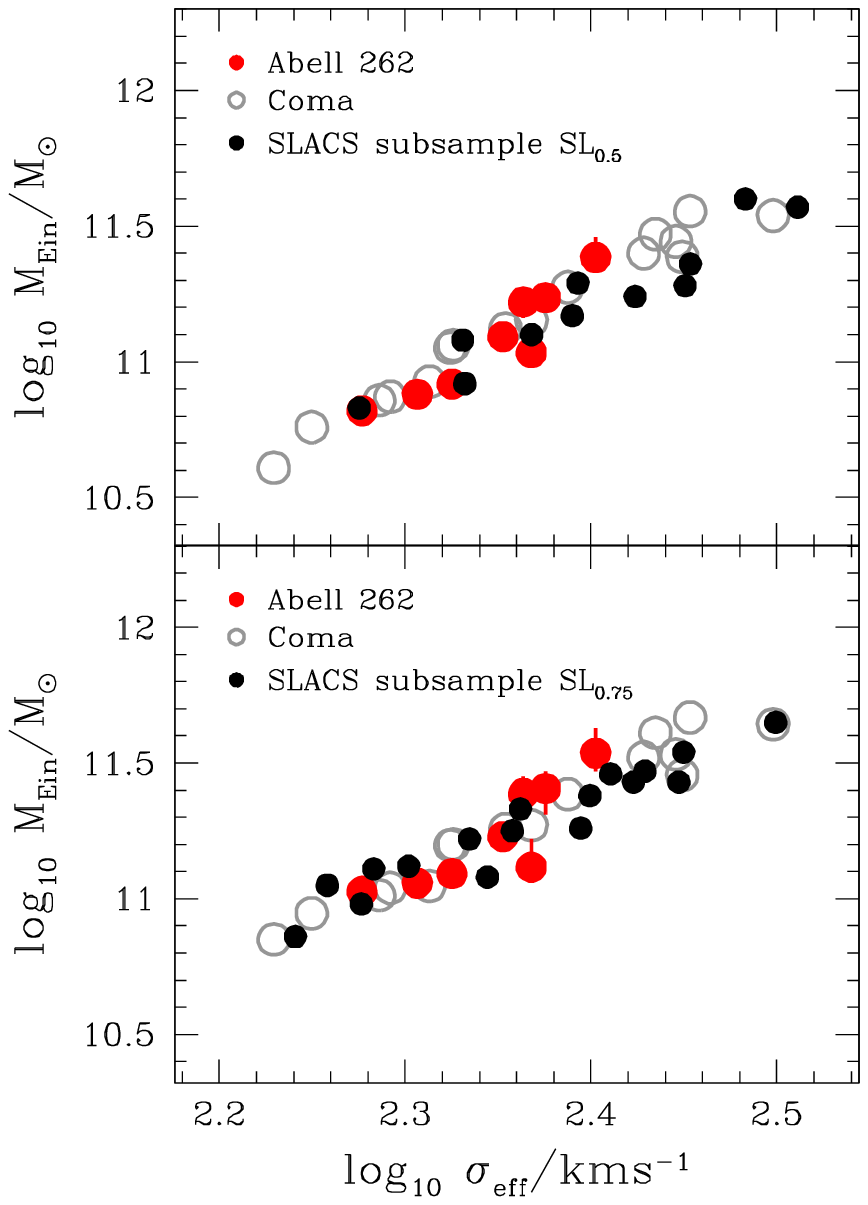}
\caption{Projected total mass, $M_{\rm Ein}$, within a fiducial
  Einstein radius, $r_{\rm Ein}$, as a function of the effective
  velocity dispersion, $\sigeff$, for the galaxies in Abell~262 (red
  filled circles; { the error bars are smaller than the symbols in
    most cases}), Coma Cluster \citep[open circles,][]{Thomas2011},
  and SLACS survey (\citealt{Auger2009}; black filled circles).  The
  dynamical masses from two-component models with dark halos ($\rho =
  \mldyn \times \nu + \rhodm$) are shown. The lensing galaxies are
  divided into the SL$_{0.5}$ (with $\langle r_{\rm Ein}/\reff \rangle
  \approx 0.5$, top panel) and SL$_{0.75}$ (with $\langle r_{\rm
    Ein}/\reff \rangle \approx 0.75$, bottom panel) subsamples,
  respectively.}
\label{fig:projmass_tot}  
\end{figure}

{ Fig.~\ref{fig:projmass_tot} compares the projected dynamical
  masses of the Abell~262 galaxies with strong gravitational lenses
  from the SLACS survey \citep{Auger2009}. The comparison is done
  exactly as in \citet{Thomas2011} and we refer the reader to this
  paper for further details. We find no significant differences
  between the Abell~262 galaxies, Coma galaxies, and SLACS lenses,
  respectively. Note that the scatter in the dynamical masses is not
  larger than in the lensing masses.  This implies that possible
  biases in the dynamical modeling related to spatially limited
  kinematic observations and/or symmetry assumptions are unlikely (as
  they would increase the scatter in dynamical masses of galaxies
  observed at random viewing angles). That the Coma and Abell~262
  galaxies must have nearly axisymmetric shapes is consistent with the
  lack of isophotal twists or minor-axis rotation in the photometric
  and kinematic observations, respectively (except for IC~171 and
  NGC~708; cf. Sec.~\ref{sec:gasresults}).}

\subsection{Comparison between circular velocity and 
ionized-gas rotation velocity}
\label{sec:gasresults}

Fig.~\ref{fig:rotation} compares the circular velocity
$v_\mathrm{circ}$ obtained from stellar dynamics and rotation velocity
$v_\mathrm{rot}$ of the ionized gas.  The comparison is restricted to
radii $r_\mathrm{gas}$ where the pressure support in the gas component
is small ($\sigma_\mathrm{gas} \le 60 \; \kms$) and therefore the gas
velocity is supposed to trace $v_\mathrm{circ}$ \citep{Bertola1995,
  Pignatelli2001, Dalcanton2010}.  The line-of-sight gas velocity
$v_\mathrm{gas}$ is derived from the observed one after subtracting
the systemic velocity and folding around the galaxy center. It was
deprojected into $v_\mathrm{rot}$ by assuming that the gaseous
component rotates in circular orbits in an infinitesimally thin
disk. The orientation of the gaseous disk with respect to the line of
sight is given by its inclination ($\igas$) and the position angle of
its line of nodes ($\pagas$). Therefore, it is
\begin{equation}
v_\mathrm{rot}\,=\,\frac{v_\mathrm{gas}}{\sin{\igas}\,\cos{\vartheta}},
\end{equation}
where 
\begin{equation}
\vartheta\,=\,\arctan{\frac{\tan{\Delta \pa_\mathrm{gas}}}{\cos{\igas}}}
\end{equation}
and 
\begin{equation}
\Delta \pa_\mathrm{gas}\,=\,\pagas - \pa_\mathrm{slit}.
\end{equation}
$\pa_\mathrm{slit}$ is the position angle of the slit where
$v_\mathrm{gas}$ is measured. Likewise, the deprojected radius is
\begin{equation}
r\,=\,r_\mathrm{gas}\,\cos{\vartheta}\,\cos{\Delta \pa_\mathrm{gas}}.
\end{equation}

In an axisymmetric spheroidal system, the position angle $\pastars$ of
the stellar isophotes is constant with radius and a rotating gaseous
disk is confined to the equatorial plane. Hence, the unique stellar
$\pastars$ coincides with the $\pagas$ of the gaseous disk. The same
holds for the inclinations. For the ionized-gas component of the
Abell~262 galaxies we only have observations along a few slits and can
neither verify the assumption of an infinitesimally thin disk nor,
given the case, derive its orientation (i.e., $\pagas$ and $\igas$)
directly. Therefore, we performed two different deprojections of the
gas kinematics for each galaxy.

Firstly, we assumed that the galaxy is an axisymmetric spheroid and
deprojected the gas kinematics with $\igas=\istars$ and
$\pagas=\pastars$ from the dynamical model
(Tab.~\ref{tab:dynres}). The corresponding values $v_\mathrm{rot}$ are
open dots in Fig.~\ref{fig:rotation} (when missing, they fall outside
the plot range). We used $\igas=85^\circ$ when the dynamical model has
$\istars=90^\circ$ since an infinitesimally thin and exactly edge-on
gas disk would not be observable at any $\pa$ different from the
galaxy's major axis.

Secondly, since the isophotal $\pa$ of the Abell~262 galaxies show
some scatter, we varied $\igas$ and $\pagas$ and minimized the
$\chi^2$ difference between $v_\mathrm{rot}$ and
$v_\mathrm{circ}$. The $\pagas$ was changed between the minimum and
maximum values measured for the $\pa$ of the galaxy isophotes.  The
best-fitting values of $v_\mathrm{rot}$ are shown by the filled
circles in Fig.~\ref{fig:rotation}. For each galaxy, the figure also
quotes $\igas$ and $\pagas$ for both sets of deprojection. IC~171 was
treated separately, as discussed below.

As Fig.~\ref{fig:rotation} shows, using the same $\pa$ and $i$ for
both the gaseous disks and dynamical models did not yield good
agreement between $v_\mathrm{rot}$ and $v_\mathrm{circ}$. However,
varying $\igas$ and $\pagas$ allowed matching the gas kinematics with
predictions of the dynamical models quite well.
The corresponding inclinations $\igas$ differ from the $\istars$
values of the dynamical models by less than $10^\circ$. The
inclination of an elliptical galaxy can, at best, be derived
dynamically with an uncertainty of $\pm 20^\circ$
\citep[e.g.,][]{Krajnovic2005, Thomas2005, vandenBosch2009}. Thus, a
$\Delta i \le 10^\circ$ is well within the uncertainties of the
dynamical models.

The case for the position angles is more complicated. To ease the
comparison with the orientation of the stars in the galaxies, the
$\pa$ panels in Fig.~\ref{fig:photometry} show the position angles
adopted in the two deprojection sets.  The red solid line corresponds
to $\pagas = \pastars$ and gives the open circles for $v_\mathrm{rot}$
plotted in Fig.~\ref{fig:rotation}. The grey solid line marks the gas
orientation that minimizes the $\chi^2$ between gas rotation and
dynamical $v_\mathrm{circ}$ and leads to the filled circles for
$v_\mathrm{rot}$ given in Fig.~\ref{fig:rotation}. The dotted lines in
Fig.~\ref{fig:photometry} delimit the $68\%$ confidence region of
$\pagas$.

{ The ionized-gas kinematics for our Abell 262 galaxies has limited
  spatial coverage and radial extension because only a few slits were
  observed. Nevertheless, this gives a valuable consistency
  check. Within the above limitations and when $\sigma_\mathrm{gas}
  \le 60 \; \kms$, the gas rotation velocity is consistent with the
  circular velocity obtained from our axisymmetric dynamical models
  and supports to the total mass distributions we derived. This holds
  also for IC~171 and NGC~708, which are not exactly axisymmetric
  systems.

\subsection{Notes on individual galaxies}

In the following we discuss the Abell~262 galaxies individually,
except for NGC~679 and NGC~687 which show no detectable emission line.

\paragraph{IC~171.}
The observed isophotal twist (Fig.~\ref{fig:photometry}) suggests that
IC~171 is not exactly axisymmetric, similarly to NGC~708. Moreover,
$v_\mathrm{rot}$ do not match the dynamical $v_\mathrm{circ}$ for any
$\pa$ within the observed range. If IC~171 is triaxial, the gas might
also be rotating around the long axis of the galaxy. To test this
geometric configuration, we tried a variety of different $\pagas$
tilted by about $90^\circ$ with respect to the photometric major axis,
fixing the inclination at $\igas\,=\,85^\circ$.  A
$\pagas\,=\,13^\circ$ brings $v_\mathrm{rot}$ in agreement with
$v_\mathrm{circ}$. If the gas rotates around the long axis of the
galaxy, then the line of nodes of the gaseous disk is at
$\pa\,=\,103^\circ$. This value is consistent with the orientation of
the galaxy isophotes at the corresponding deprojected radii
($r\,\gtrsim\,3$ kpc, Fig.~\ref{fig:photometry}).

\paragraph{NGC~703.} 
The isophotes show little variation in position angle over all the
observed radii ($30^\circ\,\lesssim\,\pa\,\lesssim\,50^\circ$,
Fig.~\ref{fig:photometry}). We found $\pastars\,\approx\,\pagas$ and
$\istars\,\approx\,\igas$ within the scatter of the data after
minimizing the $\chi^2$-difference between dynamical circular velocity
and gas rotation velocity (Fig.~\ref{fig:rotation}). The photometric
and kinematic data are consistent with our assumptions about both the
stellar and gaseous components.

\paragraph{NGC~708.}
The inner and outer isophotes ($r\,\la 2.5\,\kpc$) have different
orientations (Fig.~\ref{fig:photometry}).  The dynamical model was
adjusted towards the galaxy center, where the stellar kinematics was
measured (Fig.~\ref{fig:kinematics}) and the largest stellar rotation
is at $\pa=130^\circ$. It vanishes at $\pa=40^\circ$, indicating that
the position angle of the galaxy's line of nodes is close to
$\pastars=130^\circ$ adopted for the dynamical model. The gas rotation
differs as it is large along all the observed slits
(Fig.~\ref{fig:kinematics}) and the derived $\pagas$ is close to the
position angle measured for the outer isophotes and $v_\mathrm{rot}$
best matches $v_\mathrm{circ}$ for a nearly edge-on gas disk with
$\pagas=34^\circ$. For $\igas=80^\circ$ the deprojected radii
($r\,\approx\,7$ kpc; Fig.~\ref{fig:rotation}) fall into the region
where the stellar system has a photometric $\pa\,\approx\,40^\circ$
(Fig.~\ref{fig:photometry}). The twist of the isophotes indicates that
the system cannot be exactly axisymmetric. However, as $\pagas$ is
consistent with the stellar orientation at the corresponding radii we
infer that NGC~708 is probably only mildly triaxial and that its
dynamical mass is reliable.

\paragraph{NGC~712.}
The emission-line spectrum was taken close to the galaxy minor-axis,
making deprojection of $v_\mathrm{gas}$ highly uncertain.  The
measured rotation velocities are low (Fig.~\ref{fig:kinematics}) and,
therefore, consistent with a gaseous disk with the same
$\igas\,=\,\istars$ and $\pagas\,=\,\pastars$ as adopted in the
dynamical model of the galaxy.

\paragraph{NGC~759.} 
The position angle of the isophotes of NGC~759 ranges between about
$10^\circ$ and $50^\circ$ beyond radii of 0.1 kpc
(Fig.~\ref{fig:photometry}). The gas kinematics was measured along
three different axes and has a regular and symmetric rotation curve
and velocity dispersion profile (Fig.~\ref{fig:kinematics}). We found
$\pastars\,\approx\,\pagas$ and $\istars\,\approx\,\igas$
(Fig.~\ref{fig:rotation}) confirming the orientation and intrinsic
shape obtained from the stellar dynamics.

\paragraph{UGC~1308.}
The orientation of the isophotes scatters significantly within
$2\;\kpc$, but it is constant at larger radii. For the dynamical model
we adopted $\pastars=110^\circ$, which is representative of the
orientation of the major axis of the isophotes at $r\,\gtrsim\,2$ kpc
(Fig.~\ref{fig:photometry}). A much smaller $\pagas=47^\circ$ is
needed to match $v_\mathrm{rot}$ measured along a diagonal axis with
$v_\mathrm{circ}$ from stellar dynamics. However, the uncertainties in
$\pagas$ are large and the predictions of the dynamical model are
formally consistent with the gas rotation.  Also the measured
photometric $\pa$ has large uncertainties because the galaxy is pretty
round ($\epsilon\lesssim0.2$; Fig.~\ref{fig:photometry}).  }

\section{Results} 
\label{sec:results} 

In this Section we discuss the results from the dynamical and
stellar-population models.

\subsection{Evidence for halo mass not associated to the light} 
\label{subsec:darkmass} 
 
\begin{figure*}[t!] 
\centering 
\includegraphics[bb=18 144 550 700, width=\textwidth]{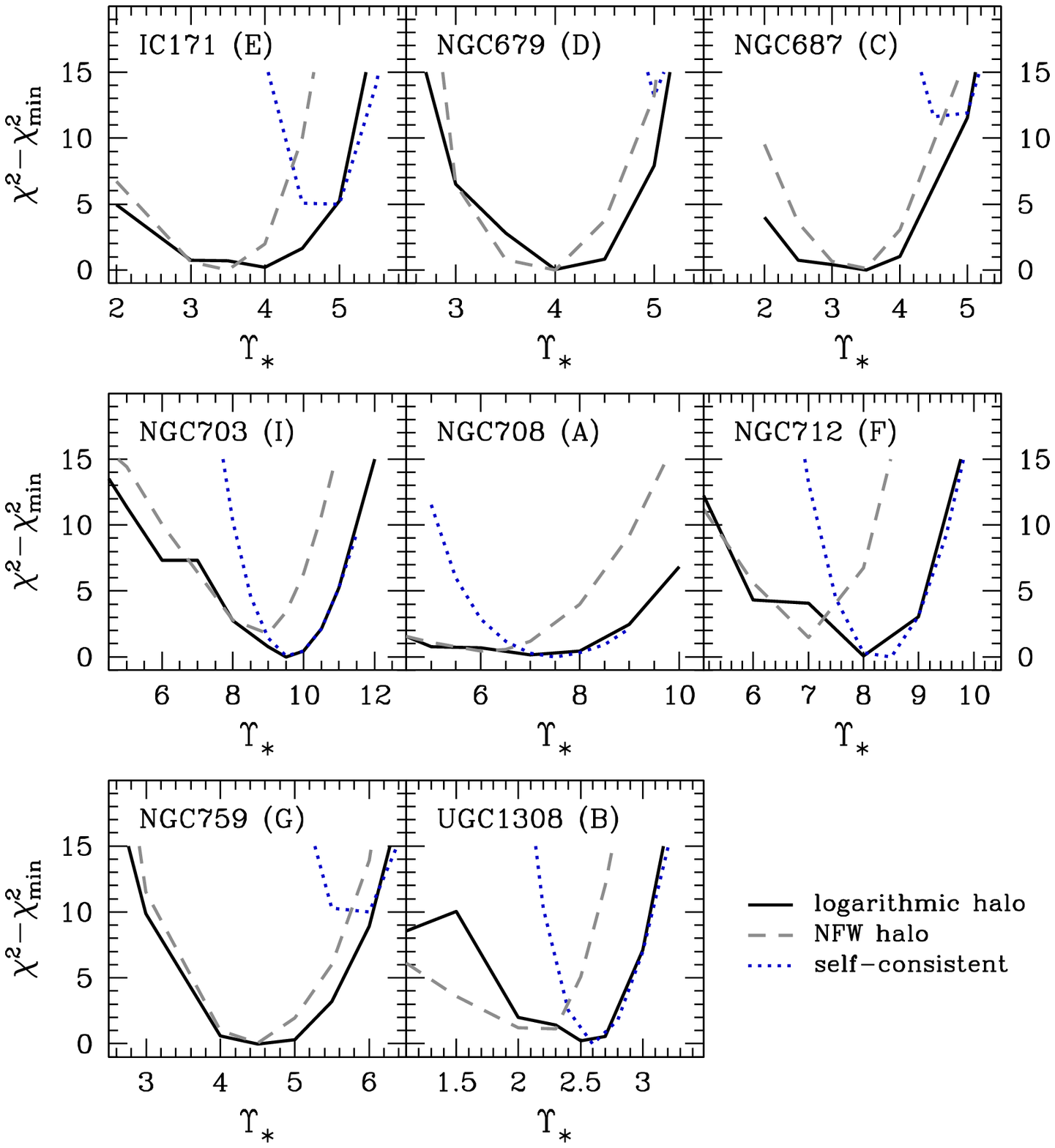} 
\caption{Goodness of the kinematic fit $\chi^2$
  (cf. Eq.~\ref{eq:entropy}) versus mass-to-light ratio
  $\mldyn$. Black solid lines are for fits including a
  logarithmic dark matter halo; grey and dashed lines are for NFW
  fits; the blue and dotted lines show models where all the mass
  follows the light ($\rhodm \equiv 0$ in Eq.~\ref{eq:rho}). For each
  galaxy the value $\chi^2_\mathrm{min}$ of the corresponding best-fit
  model is subtracted.}
\label{fig:chi} 
\end{figure*}

Fig.~\ref{fig:chi} shows the $\chi^2$ of Eq.~\ref{eq:entropy} against
the dynamical mass-to-light ratios $\mldyn$ for the Abell~262
galaxies. Three parameter fits including a logarithmic dark halo
component $\rhodm=\rho_\mathrm{LOG}$ (Eq.~\ref{eq:log}; { the three
  parameters are $\mldyn$, $r_{\rm c}$, and $V_{\rm c}$}) are
displayed by solid lines and we marginalized over the halo
parameters. Similarly, NFW halos are shown by the dashed lines and the
dotted curves are for models in which all the mass follows the light
($\rhodm \equiv 0$).
 
\begin{figure*}[t!] 
\centering
\includegraphics[bb=18 144 555 700, width=0.95\textwidth]{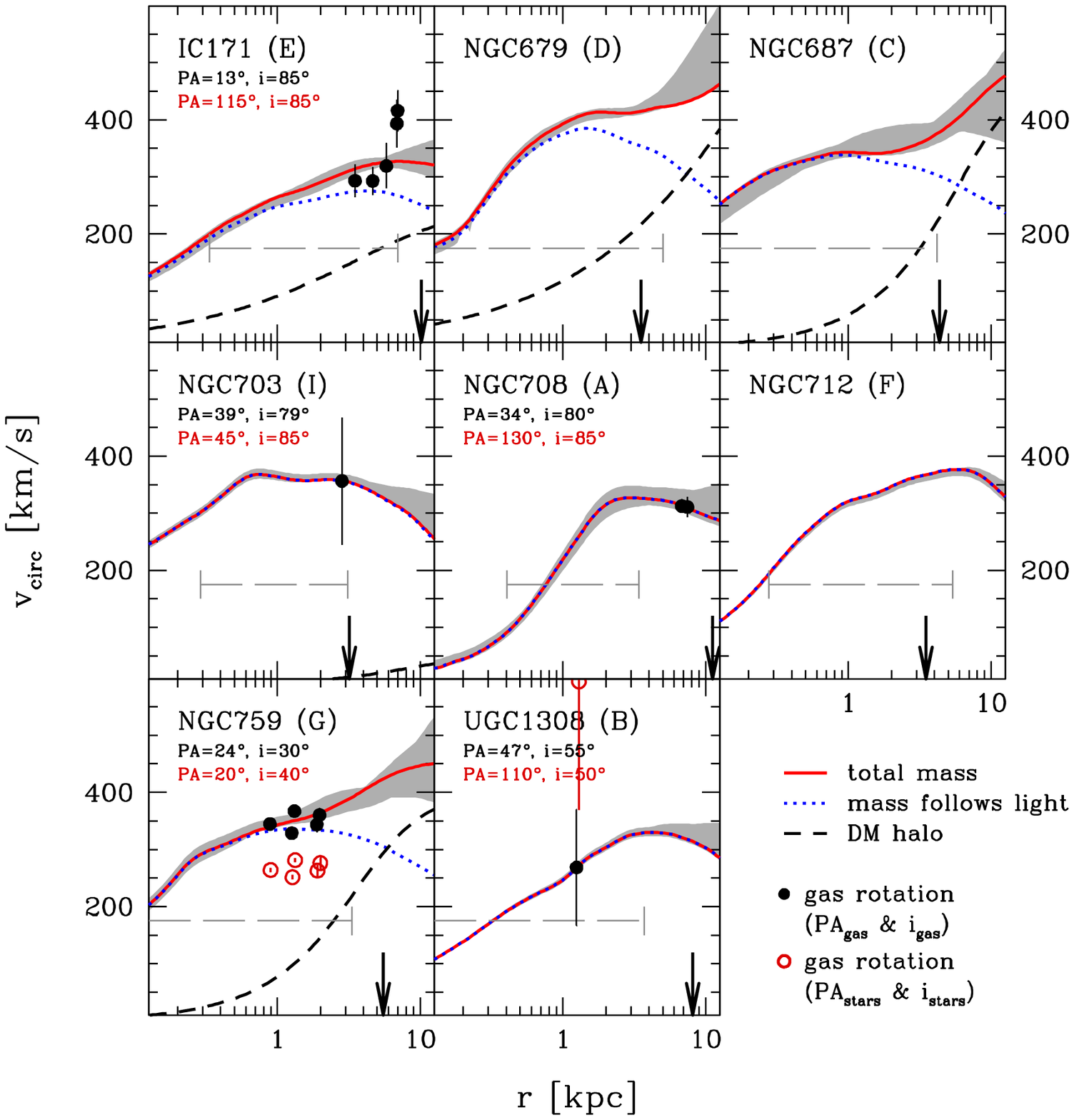} 
\caption{Rotation curves of the Abell~262 galaxies derived from
  dynamical modeling. Blue dotted lines show the contribution of {
    the mass following the light}; black dashed lines are for dark
  { halos}; red solid lines give the total mass (with the 68\%
  confidence region in grey).  Long-dashed lines delimit the region
  with kinematic observations and the effective radius $\reff$ of each
  galaxy is marked by an arrow.  Where detectable, we also show gas
  rotation measurements: red open circles correspond to the
  deprojected gas velocities assuming for the gaseous disk orientation
  the values of $\pastars$ and $\istars$ adopted for dynamical models
  (Table~\ref{tab:dynres}); black filled circles are obtained by
  varying $\pagas$ and $\igas$ of the gaseous disk to match the gas
  velocity to the circular velocity predicted by dynamical models. The
  best-fit $\pagas$ and $\igas$ are given. The central regions with
  significant pressure support ($\sigma_\mathrm{gas} > 60\,\kms$ ) are
  omitted.}
\label{fig:rotation}
\end{figure*}

The kinematics of NGC~679, NGC~687, and NGC~759 cannot be fit without
$\rhodm > 0$. The statistical significance is at least $3\,\sigma$ and
the dark halos dominate beyond $r\,\approx\,1.7\,\reff$, $1.4\,\reff$,
and $1.0\,\reff$, respectively (Fig.~\ref{fig:rotation}). These radii
are only slightly larger than the radius $r_\mathrm{max}$ of the
farthest kinematic data point.  For IC~171 the evidence for $\rhodm >
0$ is about $2\,\sigma$ and { the dark halo outweighs the mass
  following the light} beyond $r\,\approx\,2.9\,\reff$, which is
significantly on the far side of $r_\mathrm{max}$. For the remaining
objects (NGC~703, NGC~708, NGC~712, and UGC~1308) a second
mass-component $\rhodm$ does not improve the fit significantly. In
fact, the best-fitting models of NGC~708, NGC~712, and UGC~1308 {
  are obtained without any dark matter in a halo with a different mass
  distribution than the stars}, while the best-fit halo-mass fraction
inside $\reff$ in NGC~703 is $\dmfrac = 0.001$, its upper limit
$\dmfrac \le 0.06$.

In the Coma galaxy sample of \citet{Thomas2007a} the statistical
significance for dark { halos} is over $95\%$ for 8 (out of 17)
galaxies and { the halo-mass} takes over { the mass that follows
  the light} roughly between $\reff$ and $2 \, \reff$.  IC~171,
NGC~679, NGC~687, and NGC~759 have comparable dark halos. In Coma,
however, we found only one galaxy with $\dmfrac \approx 0$ (GMP~1990,
\citealt{Thomas2007a}), whereas the Abell~262 sample reveals 4 (out of
8) galaxies of this kind.
 
The evidence for a dark component aside { mass that follows light}
is not directly connected to the spatial extent of the kinematic data
(indicated by the dashed horizontal lines in
Fig.~\ref{fig:rotation}). For example, the presence of { halo mass}
in NGC~759 is highly significant, even though
$r_\mathrm{max}\,<\,\reff$. By contrast, our measurements reach
$r_\mathrm{max}\,\approx\,1.5 \, \reff$ in NGC~712 yet the mass
distribution follows the light almost exactly. Likewise, in the Coma
galaxy GMP~1990 mass follows light out to
$r_\mathrm{max}\,\approx\,3\,\reff$.  The kinematic evidence for extra
mass beyond the light is also not coupled to the degree of rotation or
flattening. For example, UGC~1308 is round and shows almost no
rotation, whereas NGC~712 is a moderately flattened, strong rotator
and the Coma galaxy GMP~1990 is a highly-flattened
($\epsilon_\mathrm{max}\,\approx\,0.6$) fast rotating system as well.
 
As already noted for the Coma galaxies, we can not discriminate
between NFW halos and logarithmic halos based on the quality of the
kinematic fits. There is only one Abell~262 galaxy where NFW halos fit
worse by $\Delta \chi^2 \approx 2$ (NGC~703), in all other galaxies
the difference is $\Delta \chi^2 \le 1$.

\subsection{Mass that follows the light}
\label{subsec:lummass} 
 
\begin{figure}[t!] 
\centering \includegraphics[width=0.45\textwidth]{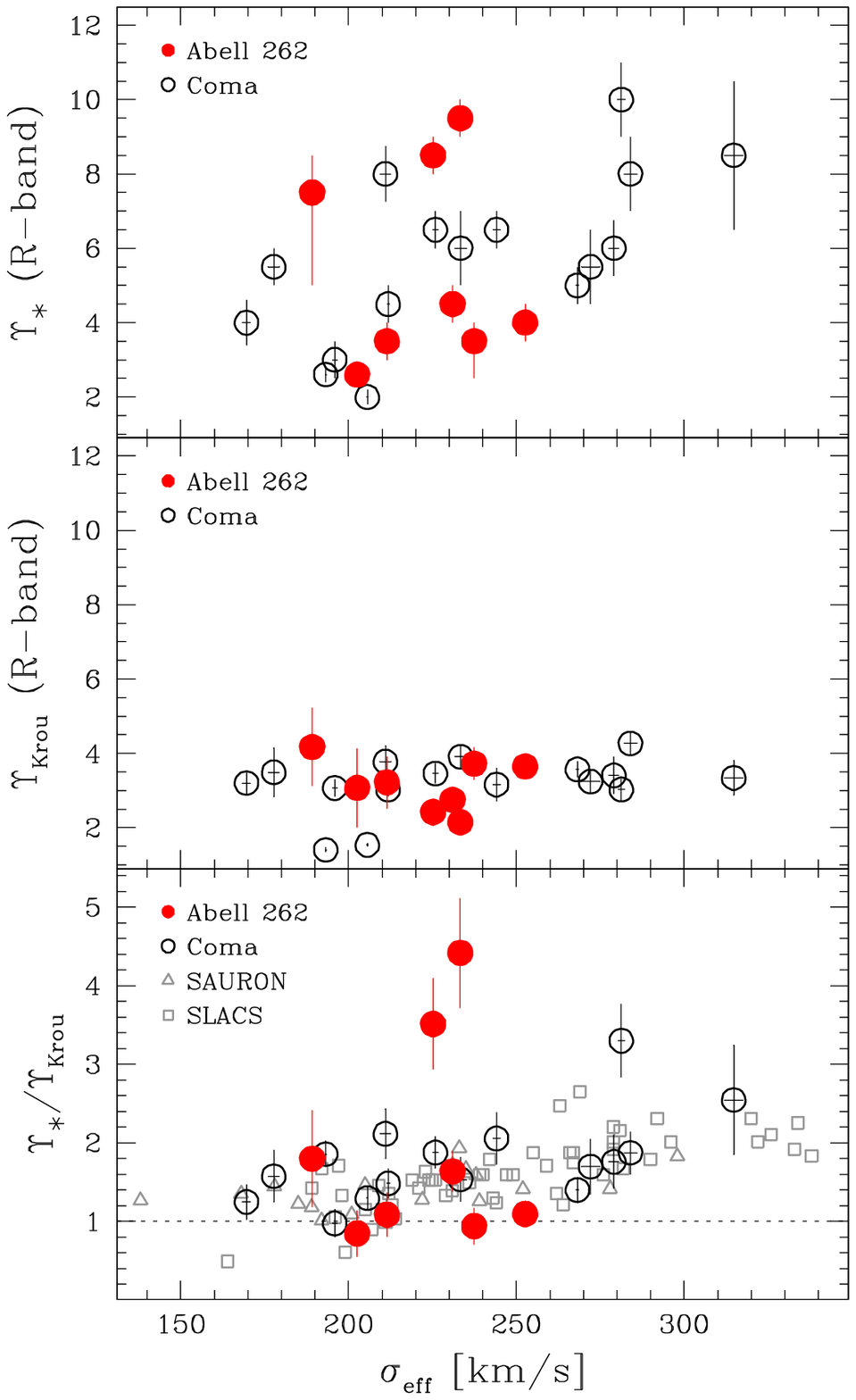}
\caption{Dynamical $\mldyn$ (upper panel), stellar-population
  $\kroupa$ (middle panel), and the ratio $\ratml$ (bottom panel) as a
  function of the effective velocity dispersion, $\sigeff$, for
  galaxies in Abell~262 (filled circles) and Coma Cluster \citep[open
    circles,][]{Thomas2011}. In the bottom panel, the results for the
  Abell~262 and Coma galaxies are also compared to SLACS galaxies with
  combined dynamical and lensing analysis \citep[open
    squares,][]{Treu2010} and SAURON galaxies with dynamical models
  { lacking a separate dark halo} \citep[open
    triangles,][]{Cappellari2006}.}
\label{fig:mlplot} 
\end{figure}

The two upper panels of Fig.~\ref{fig:mlplot} plot $\mldyn$ and the
corresponding SSP $\kroupa$ values as a function of $\sigeff$, i.e.,
the velocity dispersion averaged within $\reff$. The galaxies of
Abell~262 follow trends very similar to the galaxies in the Coma
cluster.  While the dynamically determined $\mldyn$ increase strongly
with $\sigeff$, the SSP models result in almost constant
$\kroupa$. This implies that the ratio $\ratml$ increases with
$\sigeff$, as shown explicitly in the bottom panel of
Fig.~\ref{fig:mlplot}.
 
Around $\sigeff\,\approx\,200\;\kms$ the distribution of $\ratml$ has
a sharp cutoff with almost no galaxy below $\ratml = 1$
(Fig.~\ref{fig:mlplot}, bottom panel). For
$\sigeff\,\gtrsim\,250\;\kms$ the lower bound of $\ratml$ increases
{ to} $\ratml \gtrsim 2$ at
$\sigeff\,\approx\,300\;\kms$. Similar trends are also observed in the
SAURON sample (\citealt{Cappellari2006}; cf. the triangles in the
bottom panel of Fig.~\ref{fig:mlplot}), in SLACS galaxies
(\citealt{Treu2010}; squares in Fig.~\ref{fig:mlplot}) { and,
  recently, in the ATLAS3d survey \citep{Cappellari2012}}.

\subsection{Dark halo densities}

\begin{figure}[t!] 
\centering
\includegraphics[bb=30 148 530 700, width=0.45\textwidth]{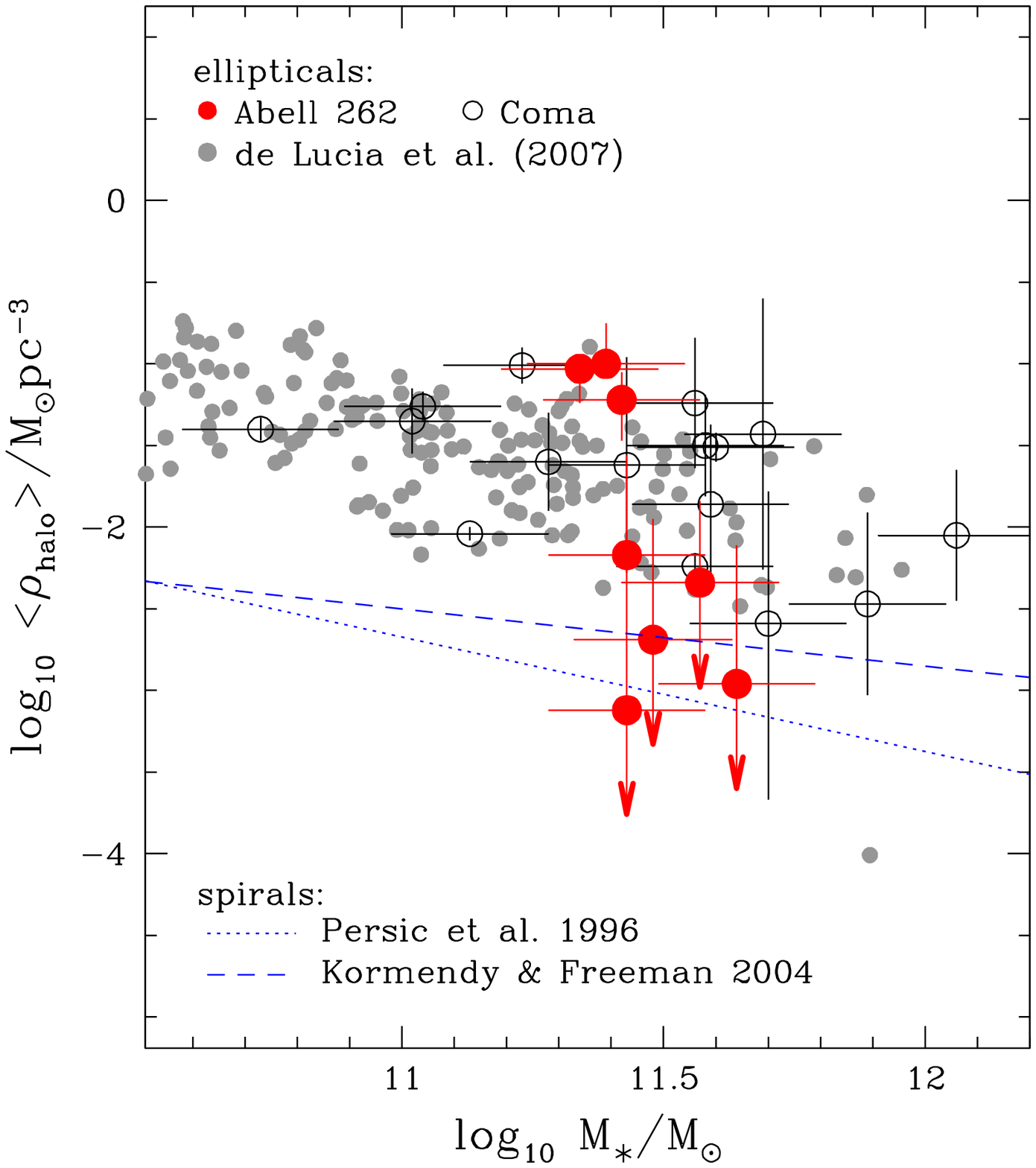} 
\caption{Average dark halo density from dynamical modeling,
  $\langle \rhodm \rangle$, inside two effective radii as a function
  of the mass that follows light, $\mstars$, for the galaxies in Abell 262 (red
  filled circles) and Coma Cluster \citep[open
    circles,][]{Thomas2009a}. The best-fitting halo densities among
    all models with a dark halo are plotted. Note that in some
  galaxies a better fit to the data was achieved without a separate
  halo component. Predictions from semi-analytic galaxy formation
  models \citep{Delucia2007} are indicated by the grey filled
  circles. The lines show spiral galaxy scaling relations from two
  different studies (blue dotted line: \citealt{Persic1996a,
    Persic1996b}; blue dashed line: \citealt{Kormendy2004}).}
\label{fig:rhodm_ms} 
\end{figure}

Fig.~\ref{fig:rhodm_ms} displays the average dark { halo} densities
$\langle \rhodm \rangle$ inside $2\,\reff$ of Coma and Abell~262
galaxies against the { dynamical mass that follows the light}
$\mstars = \mldyn \times L$. In contrast to the Coma galaxies not all
the Abell~262 galaxies have denser dark halos than spirals. For the
three Abell~262 galaxies closest to the spiral galaxy scaling
relations (NGC~703, NGC~708, and UGC~1308) we can only determine upper
limits for the amount of mass that is distributed unlike the light. 
In these cases, models 
without a halo fit the data equally well within the
uncertainties, or even better than models with halo.

{ Still, the majority of cluster early-types has $2-10$ times
  denser halos than local spirals, implying a $1.3-2.2$ times higher
  $(1+\zdm)$ (assuming
  $\langle\,\rho_\mathrm{DM}\,\rangle\,\sim\,(1+z_\mathrm{DM})^3$). Thus,
  if spirals typically formed at $z\,\approx\,1$, then cluster
  ellipticals assembled at $\zdm\,\approx\,1.6-3.4$.}

\subsection{Dark matter fractions}
\label{subsec:dmfrac}

Averaging over all Abell~262 galaxies we find that a fraction of
$\avdmfrac = 0.19$ of the total mass inside $\reff$ is { in a dark
  halo} distinct from the light.  The mean over the Coma galaxies is
$\avdmfrac = 0.23$ and similar fractions come from other dynamical
studies employing spherical models \citep[e.g.,][]{Gerhard2001,
  Tortora2009}. The Abell~262 and Coma samples indicate an
anti-correlation between $\ratml$ and $\avdmfrac$. { Galaxies where
  the dynamical mass following the light exceeds the Kroupa value by
  far ($\ratml > 3$) seem to lack matter following the halo
  distribution inside $\reff$ ($\avdmfrac = 0.004$). Not so in
  galaxies near the Kroupa limit ($\ratml < 1.4$), where the dark-halo
  mass fraction is at its maximum ($\avdmfrac = 0.30$)}.  The
remaining galaxies are intermediate in both their $\ratml$ and their
{ halo-mass} fractions ($\avdmfrac = 0.22$). We will return to this
anti-correlation in Sec.~\ref{sec:variable_imf}.

\section{Discussion}       
\label{sec:discussion}

{ Fig.~\ref{fig:mlplot} provides strong evidence for large central
  mass-to-light ratios $\mldyn$ in many early-type galaxies --
  preferably in those with large velocity dispersions $\sigeff$.
  However, in all gravity-based methods there is a fundamental
  degeneracy concerning the interpretation of mass-to-light ratios.
  Such methods can not uniquely discriminate between luminous and dark
  matter once they follow similar radial distributions. The
  distinction is always based on the assumption that the mass density
  profile of the dark matter differs from that of the luminous matter.
  Thus, if dark matter in massive ellipticals follows the light closer
  than that can be resolved by kinematic and/or lensing observations,
  then the dynamical mass associated to the light might not be stars
  only.

In fact, if the actual stellar mass-to-light ratio of a 
galaxy is $\mltrue$ then the stellar mass-density is
\begin{equation}
\label{eq:stardens} 
\rho_\mathrm{stars} \equiv \mltrue \times \nu, 
\end{equation}
while the remaining mass that follows the light,
\begin{equation}
\label{eq:dmlight} 
\rho_\mathrm{DM,\ast} = (\mldyn-\mltrue) \times \nu,
\end{equation}
is a component of dark matter that follows the light -- 
at least so closely that it is captured by $\mldyn$ 
rather than the nominal $\rhodm$. Rigorously speaking, 
any $0 \le \mltrue \le \mldyn$ is equally consistent 
with the observational data and the dynamically derived 
dark-matter distributions are correspondingly 
uncertain. In the following we will discuss the issue 
in detail.

\subsection{Maximum stellar mass and variable initial mass function}
\label{sec:variable_imf}

One extreme point of view is the assumption that the stellar masses in
early-type galaxies are maximal, i.e. $\mltrue = \mldyn$ (in analogy
to the maximum-disk interpretation of spiral galaxy rotation curves;
e.g., \citealt{vanAlbada1986}). The immediate consequence is that the
stellar IMF in early-type galaxies is not universal. While around
$\sigeff\,\approx\,200\;\kms$ there are some galaxies with $\mldyn
\approx \kroupa$ such galaxies are lacking at higher dispersions
$\sigeff$ (Fig.~\ref{fig:mlplot}). Recent attempts to try and measure
the stellar IMF directly from near-infrared observations point in the
same direction \citep{vanDokkum2010,vanDokkum2011,Conroy2012}. These
spectroscopically derived IMFs (Salpeter or steeper around
$\sigeff\,\approx\,300\;\kms$) correspond to $\ratml \approx 1.6$ or
larger (consistent with Fig.~\ref{fig:mlplot}) favoring the maximum
stellar mass interpretation and an IMF varying from Kroupa-like at low
velocity dispersions to Salpeter (or steeper) in the most massive
early-types.

As already noticed for the Coma galaxies \citep{Thomas2011}, we do not
see any correlation of $\ratml$ and the SSP ages, metallicities, and
$\alpha$-elements overabundances of the Abell~262 galaxies. Such
correlations could be expected if the IMF is variable, since a change
in the relative number of low and high-mass stars has some influence
on the chemical evolution of the galaxies
\citep[e.g.,][]{Graves2010}. However, the stellar IMF acts on the
metallicities and element abundances simultaneously with other variable
conditions, like the depth of the potential well and star-formation
timescale. Hence, the lack of significant stellar-population
differences between galaxies with high $\ratml$ on the one hand and
galaxies with low $\ratml$ on the other is inconclusive with respect
to the IMF.

An intriguing observation is the fact that the galaxies with the
highest $\ratml$ in the Abell~262 and Coma clusters ($\ratml > 3$)
have the lowest dark matter fractions ($\avdmfrac \approx 0$). In
contrast, galaxies with $\ratml < 1.4$ have the highest dark matter
fractions ($\avdmfrac = 0.30$). Such an anti-correlation between
$\ratml$ and $\dmfrac$ can easily be understood if the occasionally
large $\mldyn$ in some galaxies results from their dark matter density
following the light so closely that it becomes indistinguishable from
the stellar mass (at least in the inner parts at $r\la \reff$
considered here). The more that the actual dark matter of a galaxy
leaks into the mass component of our models that follows the light,
the less is expected to reside in its nominal halo component $\rhodm$,
resulting in the observed anti-correlation between $\dmfrac$
(measuring only the dark matter in the halo) and $\ratml$ (including
also the dark matter following the light). We are not aware of an
immediate physical explanation why a more `massive' IMF should occur
in galaxies with less dark matter.

\subsection{Cold dark matter and halo contraction}
\label{sec:contraction}

One option to further constrain the mass-decomposition of
gravity-based models is to incorporate predictions from cosmological
simulations that confine the maximum amount of dark matter that can be
plausibly attached to a galaxy of a given stellar
mass. \citet{Napolitano2010} found that the velocity dispersions of a
large sample of low-redshift early-type galaxies require high central
mass-to-light ratios for the stars (consistent with a Salpeter IMF on
average) when using such cosmological halos without baryon
contraction.  Adiabatic contraction \citep{Blumenthal1986}, instead,
is able to increase the amount of dark matter in the centers of
ellipticals and to lower the required stellar masses towards a Kroupa
IMF \citep[see also][]{Napolitano2011}. The recent study of
\citet{Cappellari2012}, with a not further specified halo-contraction
scenario, points against halo contraction to be sufficient to lower
the central stellar masses. Lensing studies go in the same direction
\citep{Auger2010} and numerical simulations generally indicate that
the classic adiabatic scenario overpredicts the actual contraction of
dark matter halos \citep[e.g.,][]{Gnedin2004}.

Therefore, while halo contraction could in principle reconcile the
observed central dynamical masses with a `light' stellar IMF, the
required amount of dark-matter contraction seems to be rather strong.
However, this constraint against a `light' IMF is not purely empirical
and we here briefly discuss the halo-assembly redshifts of the
Abell~262 and Coma galaxies that result in the context of a Kroupa IMF
and adiabatic contraction.

An immediate consequence of a universal Kroupa IMF is that some of the
mass that follows the light is actually dark matter and needs to be
combined with $\rhodm$, i.e.
\begin{equation} 
\label{eq:kroupadm} 
\rho_\mathrm{DM,Krou} = \rhodm + (\mldyn-\kroupa) \times \nu. 
\end{equation} 
The effect is to increase the dark matter fractions by about a factor
2, for example it is $\langle f_\mathrm{DM,Krou} \rangle = 0.47$ in
the Abell~262 sample and similarly in the Coma cluster
\citep{Thomas2011}.  Dark-matter fractions of $\approx\,50\%$ or more
are generally found and required in the context of the assumption of a
`light' IMF \citep[e.g.,][]{Napolitano2010,Barnabe2011}.

Based on $\rho_\mathrm{DM,Krou}$ we derived the dark-halo assembly
redshift $\zdm$ for the Abell~262 galaxies following
\citet{Thomas2011}. The basic assumption is that the (decontracted)
halo density scales with the mean density of the universe at the
assembly epoch, i.e.  $\langle \rho_\mathrm{DM} \rangle \sim
(1+z_\mathrm{DM})^3$. Fig.~\ref{fig:zform} compares the values of
$\zdm$ to the star-formation redshifts $\zstars$ calculated from the
stellar-population ages.
 
The derivation of $\zdm$ involves the adiabatic decontraction of the
dark matter density $\rho_\mathrm{DM,Krou}$.  In NGC~687 and UGC~1308
the dynamical $\mldyn$ turns out to be lower than $\kroupa$, such that
the dark halo to be decontracted for NGC~687 would be exactly cored
logarithmic. However, the adiabatic decontraction of cored logarithmic
halos is not possible (cf. \citealt{Thomas2011}) and NGC~687 has been
omitted from the analysis. The dynamical model of UGC~1308 formally
leaves no space for dark matter at all and the galaxy is not
considered furthermore as well.

For the Abell~262 galaxies in Fig.~\ref{fig:zform} we estimated the
uncertainty in $\zdm$ that 
comes from the scatter in the correlation between 
$\langle
\rho_\mathrm{DM} \rangle$ and $(1+z_\mathrm{DM})^3$ 
  derived from cosmological simulations (see Fig. 9 in
  \citealt{Thomas2009a})
Note that this does not include the uncertainties in
$\rhodm$ or any uncertainties related to the contraction scenario.
In the majority of Coma galaxies $\zdm \approx \zstars$ and the
assembly of these galaxies seems to have stopped before $\zdm \approx
1$. In four Coma and three Abell~262 galaxies the stars appear younger
than the halo, which indicates a secondary star-formation episode
after the main halo assembly. 

In one Coma galaxy (GMP~5568) $\zdm$ is significantly smaller than
$\zstars$. The $\reff$ of GMP~5568 is, however, about 3 times larger
than expected for its $B$-band luminosity, when compared to the rest
of the Coma sample. The average of its dark halo density is low
because of this change in the physical scale (\citealt{Thomas2011}).
In the Abell~262 cluster we found two galaxies with $\zdm < \zstars$
as well (IC~171 and NGC~708). Both galaxies have significant isophote
twist (Sec.~\ref{sec:gasresults}) and, in addition, IC~171 has boxy
outskirts (Fig.~\ref{fig:photometry}) while NGC~708 is a very slow
rotator (Fig.~\ref{fig:kinematics}). The two galaxies may therefore be
the remnants of gas-poor binary mergers. The virial theorem predicts
the average density to drop by a factor of 4 during such a merger
(assuming homology and equal-mass progenitors), whereas the stellar
$\zstars$ remains constant (assuming similar stellar-population ages
for the progenitors) due to the lack of gas to form new stars. Thus,
gas-less major mergers move galaxies downward in
Fig.~\ref{fig:zform}. In fact, the approximate decrease of the mean
density to a quarter of its original value implies assembly redshifts
of the hypothetical progenitors to be $\zdm \approx 1.7$ and $\zdm
\approx 1.9$ for IC~171 and NGC~708, respectively. Hence, if actually
the two galaxies have been recently undergone a major merger, then
their progenitors may well have originated from the vicinity of the
one-to-one relation in Fig.~\ref{fig:zform} where we find the majority
of early-type galaxies with homogeneously old stellar populations.

\begin{figure}[t!]  
\centering 
\includegraphics[bb=30 148 530 700, width=0.45\textwidth]{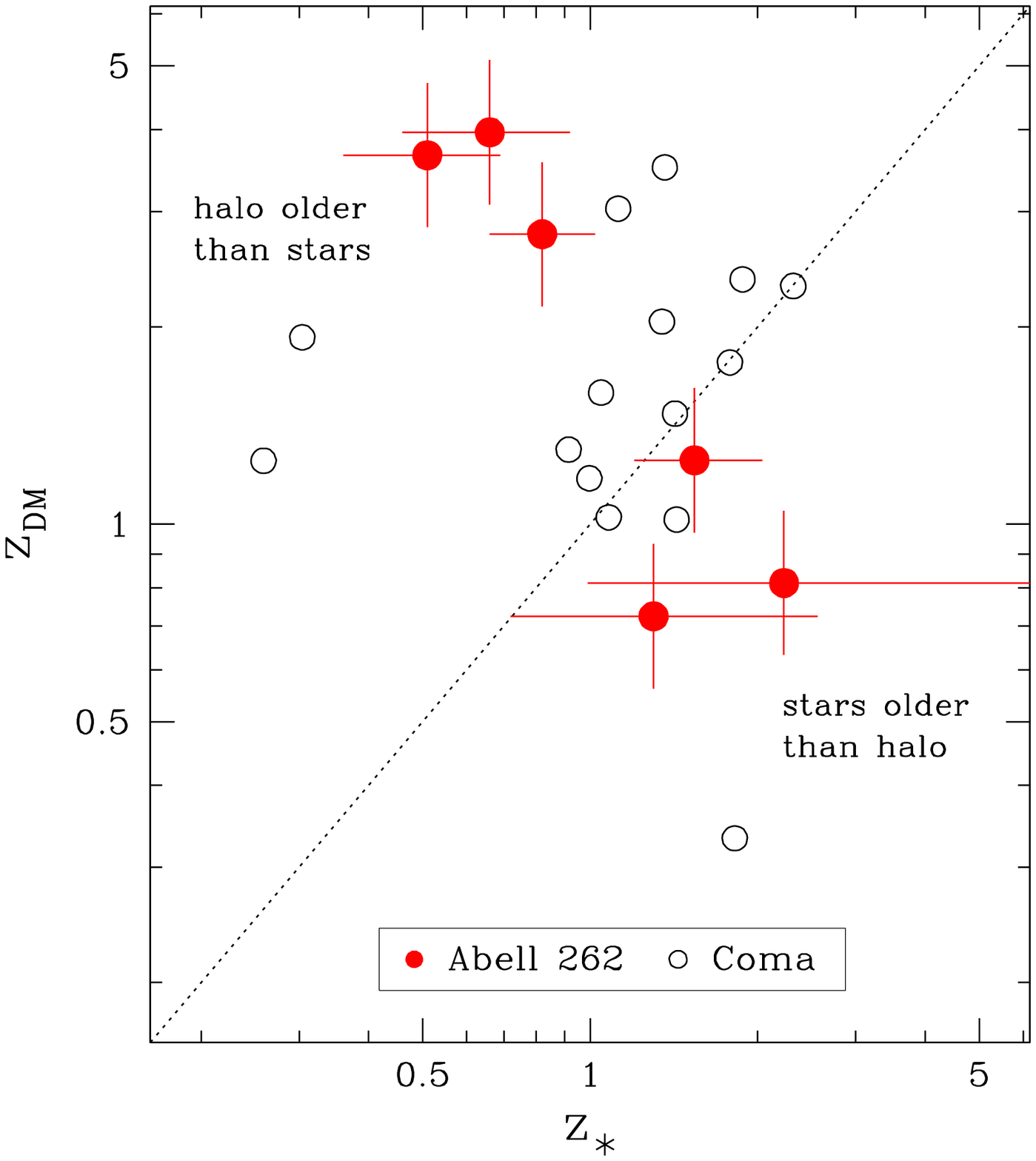}
\caption{Kroupa-IMF based halo-assembly redshifts $\zdm$ and
  star-formation redshifts $\zstars$ from stellar-population ages
  against each other for the galaxies in Abell~262 (filled circles)
  and Coma Cluster \citep[open circles,][]{Thomas2011}. The dotted
  line indicates where $\zdm = \zstars$. The Coma galaxy GMP~5568 is in
  the lower right of the diagram.}
\label{fig:zform} 
\end{figure} 

Without trying to overinterpret the result given the assumptions going
into Fig.~\ref{fig:zform} (i.e, SSP models, cosmological relations
between halo density and formation redshift, and halo contraction), it
seems that an universal Kroupa IMF gives a consistent picture for the
formation redshifts of our galaxies. 

\subsection{Violent relaxation}
\label{sec:relaxation}

Apart from baryonic contraction, the central dark-matter and stellar
mass profiles could also become similar as a result of violent
relaxation. This is usually invoked to explain the orbital structure
of elliptical galaxies formed via collapse or merging
\citep{LyndenBell1967, Nakamura2000}. Although many of its details are
not fully understood \citep[e.g.,][]{Arad2005}, some general
predictions can be made in the simplest cases of gas-rich (wet) and
gas-poor (dry) major mergers, but it is apparent that their general
evolution can be more complex.

In this paper, we have added data for some Abell~262 early-type
galaxies to those of our previously studied Coma galaxies. We derived
the redshifts when the stars and the dark matter halos formed and
compared the profiles of the luminous and dark matter components. Some
of this information could be interpreted as being consistent with
signatures of violent relaxation. In particular, there are two pieces
of evidence which are consistent with violent relaxation. The first is
the division of the ellipticals into the two merger types as shown in
Fig.~\ref{fig:zform} and explained by models. Indeed recent analyses
of the light distributions and dynamics of ellipticals support a
picture requiring violent relaxation \citep{Cox2006, Hopkins2009b,
  Hopkins2009a, Hoffman2010} with the conclusion that cuspy
ellipticals grew from the mergers of gas-rich spirals while core
ellipticals resulted from mergers of earlier gas-poor ellipticals, the
orbital and mass distributions of the merger remnants providing a
record of their progenitors (cf. \citealt{Benson2010} for a review).
The second comes from Figs.~\ref{fig:rotation} and
\ref{fig:mlplot}. If one makes the assumption that the stellar IMF is
fixed, then the stellar and luminous matter distributions are close to
being the same in several of the galaxies, which is what it would be
expected from violent relaxation in the merging which formed them.

Nevertheless, there are problems from some of the other galaxies. The
dark matter follows the light closely in both the Abell~262 galaxy
NGC~712 and Coma galaxy GMP 1990. They are fast flattened rotators and
show the least randomized orbital motions. Moreover, GMP 1990 is a
highly flattened system and, taking the model results at face value,
the assumption of a Kroupa IMF for this galaxy requires a highly
flattened dark matter component $\rho_\mathrm{DM,\ast}$. However, as
we have not tried to fit a continuous sequence of ever more flattened
halos to the kinematic data, it is not clear if a model where all the
mass exactly follows the light is the only way to explain the
observational data. In fact, the preference for a high $\mldyn$ in the
models could just reflect that the actual dark halo is flatter than
the (spherical) $\rhodm$, though still being rounder than the light
distribution -- on a level that the data are just better approximated
by increasing the mass that follows the light rather than to make the
spherical component $\rhodm$ more massive. The actual flattening of
the galaxy's potential might be somewhere in between that of the light
and $\rhodm$ (spherical).  Lensing galaxies do not show a significant
difference between the flattening of their total mass distributions
and that of their light, at least not the massive ones
\citep{Koopmans2006,Barnabe2011}. If this also holds for the galaxies
studied here, then either the mass that follows the light is indeed
stars, or, if not, then there has to be a dark matter component that
is as flattened as the stellar distribution
($\epsilon_\mathrm{max}\,\approx\,0.6$).

Admittedly, we are dealing with a small sample of imperfect
observations, but at present it appears that the best one can conclude
is that these data do not offer strong proof for violent relaxation
being a major mechanism to make sufficiently similar the distributions
of stars and dark matter.}

{ Further numerical investigations are required to test 
if a predominantly collisionless-accretion growth of 
the outer parts of massive early-types can lead to a 
similarisation of the mass-density profiles of stellar 
and dark matter. Simulations show that accreted stars 
have shallower radial mass profiles than in-situ formed 
stars \citep{Naab2009}. Since the dark-matter accretion 
is collisionless as well, the stellar and dark-matter 
profiles around $\reff$ (where accreted stars are 
expected to dominate) might be similar.}

Finally, we note that it seems unlikely that uncertainties in the
modeling process (due to limited kinematic observations and/or
symmetry assumptions) can explain the particular large mass-to-light
ratios of some early-type galaxies.  Mainly, because our total (i.e.,
luminous and dark) dynamical masses are consistent with completely
independent results from strong gravitational lensing and from our gas
rotation velocities being consistent with the circular velocities
derived from dynamical modeling.

\section{Summary}
\label{sec:conclusions} 

We presented new radially resolved spectroscopy of 8 early-type
galaxies in the Abell~262 cluster. We measured the stellar rotation,
velocity dispersion, and the $H_3$ and $H_4$ coefficients of the
line-of-sight velocity distribution along the major axis, minor axis
and an intermediate axis. In addition, we derived line index profiles
of Mg, Fe and H$\beta$. The ionized-gas velocity and velocity
dispersion were measured along different axes of 6 sample
galaxies. { Axisymmetric orbit-based dynamical models were used to
  derive (1) the mass-to-light ratio of all the mass that follows the
  light and (2) the dark halo parameters.}  The analysis of the
ionized-gas kinematics gave a valuable consistency check for the
circular velocity (and total mass distribution) predicted by dynamical
modeling.  Line-strength indices were analyzed by SSP models to derive
the galaxies' ages, metallicities, and $\alpha$-elements abundances.
 
This study of Abell~262 galaxies complements our earlier work on a
similar, though larger, sample of early-type galaxies in Coma. While
the Coma galaxies were selected to be mostly flattened, the Abell~262
galaxies regarded here appear predominantly round on the sky.
 
Three of the new Abell~262 galaxies show clear evidence for a mass
component that is unrelated to the light with a statistical
significance of at least $3\,\sigma$. In one further galaxy the
statistical evidence for dark matter is slightly lower (about
$2\,\sigma$). In four Abell~262 galaxies the addition of mass that is
distinct from the luminosity distribution does not improve the fit to
the observed kinematics significantly.

We find that { the dynamical mass-to-light ratios $\mldyn =
  \rholum/\nu$, where $\rholum$ is the density of all the mass that
  follows the light,} increase with the stellar velocity dispersion
$\sigeff$ (averaged inside $\reff$), while the corresponding
stellar-population values are almost constant. { Many galaxies
  around $\sigeff\,\approx\,200\;\kms$ have $\mldyn$ consistent with a
  Kroupa IMF. Around $\sigeff\,\approx\,300\;\kms$ the $\mldyn$ are
  typically 2 times larger than for a Kroupa IMF.}

This could reflect a change in the stellar IMF with
$\sigeff$. However, we also find an anti-correlation between the {
  average fraction of mass in the halo, $\dmfrac$, and the ratio
  $\ratml$. The galaxies with the largest $\ratml$ have the lowest
  halo-mass fractions inside $\reff$ and vice versa}. A possible
explanation for this finding is a dark matter distribution that
follows the light very closely in massive galaxies and contaminates
the measured $\mldyn$, while it is more distinct from the light in
lower-mass systems.

{ If there is indeed some ambiguity between stellar 
and dark mass,
   then the finding of $\mldyn > \kroupa$ does not 
necessarily indicate
   a `massive' IMF. In fact, galaxies where the 
dynamical mass that
   follows the light is in excess of a Kroupa stellar 
population do not
   differ in terms of their stellar population ages, 
metallicities and
   $\alpha$-elements abundances from galaxies where this 
is not the
   case. Taken at face value, our dynamical mass models 
are therefore
   as consistent with a universal IMF, as they are with 
a variable IMF.

   However, \citet{Conroy2012} have presented 
stellar-population models
   with a variable IMF that do show a correlation with
   $\alpha$-overabundance which can partly explain the 
large amount of
   mass that follows the light in massive early-type
   galaxies. Nevertheless, even these models produce 
stellar
   mass-to-light ratios that are smaller than the 
dynamical ones for
   the mass that follows the light. Therefore, a 
combination of both,
   IMF variation and some degeneracy between the mass 
profiles of
   stellar and dark matter could give a consistent 
explanation for our
   observations.  If the IMF indeed varies from galaxy 
to galaxy
   according to the average star-formation rate, as 
implicated by the
   correlation with $\alpha$-overabundance, then the 
assumption of a
   constant stellar mass-to-light ratio {\it inside} a 
galaxy should be
   relaxed in future dynamical and lensing models.

Finally, we have derived halo-assembly redshifts for the Abell~262 galaxies
by calibrating the average dark matter density inside $2\,\reff$ using
cosmological $N$-body simulations. As in Coma early-types, the typical
assembly redshift of the halos is around $\zdm\,\approx\,1-3$.
}

%

\acknowledgments
E.M.C. acknowledges the Max-Planck-Institut f\"ur extraterrestrische  
Physik for hospitality while this paper was in progress.  
E.M.C. receives support by Padua University through grants CPDA089220/08 
and CPDR095001/09 and by Italian Space Agency through grant ASI-INAF 
I/009/10/0. 
J.T. acknowledges the Padua University for hospitality while this paper 
was in progress. 
Support for Program number HST-GO-10884.0-A was provided by NASA 
through a grant from the Space Telescope Science Institute which is 
operated by the Association of Universities for Research in Astronomy, 
Incorporated, under NASA contract NAS5-26555. 
This work was in part supported by the Chinese National Science
Foundation (Grant No. 10821061) and the National Basic Research
Program of China (Grant No. 2007CB815406). We also gratefully
acknowledge the Chinese Academy of Sciences and Max-Planck-Institut
f\"{u}r Extraterrestrische Physik that partially supported this work.
The authors wish to acknowledge the anonymous referee who's comments 
substantially improved this paper.
 
\clearpage 
 
\bibliographystyle{apj} 
\bibliography{A262_astroph.bib}
 
\begin{rotate}  
\begin{deluxetable}{rrrrrrcrrrrrrc}  
\tablecolumns{14}  
\tabletypesize{\tiny}  
\tablewidth{0pt}  
\tablecaption{Photometric parameters of the sample galaxies. \label{tab:photometry}} 
\tablehead{  
\colhead{$a$} & \colhead{$\mu_R$} & \colhead{$e$} & \colhead{PA} &   
\colhead{$\Delta x_{\rm c}$} & \colhead{$\Delta y_{\rm c}$} & \colhead{Err.\tablenotemark{a}}&  
\colhead{$a_3/a$} & \colhead{$b_3/a$} &   
\colhead{$a_4/a$} & \colhead{$b_4/a$} &  
\colhead{$a_6/a$} & \colhead{$b_6/a$} & \colhead{Err.\tablenotemark{b}} \\  
\colhead{[arcsec]} & \colhead{[mag arcsec$^{-2}$]} & \colhead{} & \colhead{[$^\circ$]} &   
\colhead{[arcsec]} & \colhead{[arcsec]} & \colhead{[arcsec]} &  
\colhead{$\times 100$} & \colhead{$\times 100$} &  
\colhead{$\times 100$} & \colhead{$\times 100$} &  
\colhead{$\times 100$} & \colhead{$\times 100$} & \colhead{}}  
\startdata  
\cutinhead{IC 171}  
$ 0.202\pm0.000$ & $16.927\pm0.010$ & $0.000\pm0.000$ & $ 90.0\pm 0.0$ & $ 0.127$ & $ 0.594$ & 0.000 & $ 0.00$ & $ 0.00$ & $ 0.00$ & $ 0.00$ & $ 0.00$ & $ 0.00$ & 0.00\\  
$ 0.644\pm0.081$ & $17.005\pm0.014$ & $0.317\pm0.121$ & $ 66.7\pm16.1$ & $ 0.000$ & $ 0.000$ & 0.057 & $-2.21$ & $ 3.91$ & $ 2.94$ & $-1.95$ & $ 2.36$ & $-1.64$ & 1.25\\  
$ 1.104\pm0.010$ & $17.146\pm0.015$ & $0.182\pm0.010$ & $ 78.4\pm 1.9$ & $ 0.024$ & $ 0.006$ & 0.007 & $-0.94$ & $ 0.85$ & $ 0.55$ & $ 0.03$ & $-0.18$ & $-0.45$ & 0.41\\  
$ 1.482\pm0.007$ & $17.338\pm0.014$ & $0.186\pm0.006$ & $ 82.3\pm 1.1$ & $ 0.036$ & $ 0.012$ & 0.005 & $-0.67$ & $ 0.85$ & $ 0.27$ & $ 0.06$ & $-0.28$ & $-0.01$ & 0.29\\  
$ 1.797\pm0.006$ & $17.530\pm0.012$ & $0.170\pm0.004$ & $ 84.1\pm 0.7$ & $ 0.048$ & $ 0.006$ & 0.004 & $-0.23$ & $ 0.77$ & $-0.07$ & $-0.02$ & $-0.30$ & $-0.09$ & 0.17\\  
$ 2.118\pm0.006$ & $17.690\pm0.011$ & $0.176\pm0.003$ & $ 85.9\pm 0.7$ & $ 0.061$ & $ 0.000$ & 0.004 & $ 0.24$ & $ 0.75$ & $-0.21$ & $ 0.07$ & $-0.34$ & $-0.05$ & 0.12\\  
$ 2.455\pm0.007$ & $17.845\pm0.012$ & $0.186\pm0.003$ & $ 88.5\pm 0.6$ & $ 0.079$ & $-0.006$ & 0.005 & $ 0.51$ & $ 0.75$ & $-0.23$ & $ 0.13$ & $-0.12$ & $ 0.16$ & 0.15\\  
$ 2.805\pm0.009$ & $17.993\pm0.010$ & $0.187\pm0.004$ & $ 91.4\pm 0.7$ & $ 0.091$ & $-0.006$ & 0.006 & $-0.91$ & $-0.49$ & $ 0.14$ & $ 0.18$ & $-0.07$ & $ 0.40$ & 0.19\\
\enddata  
\tablenotetext{a}{Error on the center coordinates derived from the residual rms  
of the ellipse fit to the isophotes: Err$={\rm  
    rms_{fit}}/\sqrt{N}$ with $N\leq128$ the number of fitted points  
  of the isophote.}    
\tablenotetext{b}{Error of Fourier coefficients  
  defined as  
  Err$=\sqrt{\frac{\sum_{i=10}^{N/2}\,\left(a_i^2+b_i^2\right)}{N/2-10}}\times\frac{100}{a}$.}  
\end{deluxetable}  
\end{rotate}  
 
\clearpage 
           
\begin{deluxetable}{rrrrrrrr}	 			                  
\tablecolumns{7}	   	 			                  
\tabletypesize{\scriptsize}  
\tablewidth{0pt}  
\tablecaption{Stellar kinematics of the sample galaxies \label{tab:kinematics}}  
\tablehead{  
\colhead{$r$} & \colhead{$V$} & \colhead{$\sigma$} &   
\colhead{$H_3$} & \colhead{$H_4$} &   
\colhead{PA} & \colhead{Run}\\  
\colhead{[arcsec]} & \colhead{[$\kms$]} & \colhead{[$\kms$]} &   
\colhead{} & \colhead{} &   
\colhead{[$^\circ$]} & \colhead{}}  
\startdata  
\cutinhead{IC 171}  
$-11.71$ & $  -8.8\pm19.3$ & $130.4\pm24.0$ & $ 0.033\pm0.094$ & $ 0.068\pm0.114$ & $ 10$ & 2\\ 
$ -8.47$ & $  -3.5\pm14.2$ & $157.4\pm17.2$ & $ 0.036\pm0.059$ & $ 0.084\pm0.073$ & $ 10$ & 2\\ 
$ -5.27$ & $  16.9\pm12.8$ & $192.1\pm12.7$ & $-0.042\pm0.042$ & $ 0.056\pm0.043$ & $ 10$ & 2\\ 
$ -2.84$ & $  11.5\pm 8.7$ & $191.1\pm 7.2$ & $-0.006\pm0.023$ & $ 0.050\pm0.026$ & $ 10$ & 2\\ 
$ -1.08$ & $  31.1\pm 3.7$ & $178.9\pm 4.5$ & $-0.024\pm0.016$ & $ 0.039\pm0.018$ & $ 10$ & 2\\ 
$ -0.23$ & $   9.3\pm 3.6$ & $160.4\pm 4.1$ & $-0.019\pm0.018$ & $-0.010\pm0.021$ & $ 10$ & 2\\ 
$  0.38$ & $   0.6\pm 3.7$ & $157.7\pm 5.1$ & $-0.016\pm0.019$ & $ 0.059\pm0.023$ & $ 10$ & 2\\ 
$  0.98$ & $ -11.1\pm 8.0$ & $170.3\pm 4.9$ & $-0.024\pm0.019$ & $ 0.038\pm0.021$ & $ 10$ & 2\\
\enddata  
\end{deluxetable}  
  
\clearpage  
  
\begin{deluxetable}{rrrrr}	 			                  
\tablecolumns{5}	   	 			                  
\tabletypesize{\scriptsize}  
\tablewidth{0pt}  
\tablecaption{Ionized-gas kinematics of the sample galaxies \label{tab:gas}}  
\tablehead{  
\colhead{$r$} & \colhead{$V$} & \colhead{$\sigma$} & 
\colhead{PA} & \colhead{Run}\\  
\colhead{[arcsec]} & \colhead{[$\kms$]} & \colhead{[$\kms$]} &  
\colhead{[$^\circ$]} & \colhead{}}  
\startdata  
\cutinhead{IC 171}  
$ -7.34$ & $ 138.4\pm14.8$ & $  0.0\pm24.9$ & $  0$ & 4\\
$ -6.11$ & $ 146.2\pm15.4$ & $ 75.4\pm21.0$ & $  0$ & 4\\
$ -4.88$ & $ 123.1\pm23.5$ & $ 88.6\pm29.3$ & $  0$ & 4\\
$ -3.65$ & $  85.8\pm16.2$ & $ 78.6\pm21.6$ & $  0$ & 4\\
$ -2.42$ & $  54.8\pm13.7$ & $ 82.1\pm18.3$ & $  0$ & 4\\
$ -1.19$ & $  12.4\pm12.1$ & $ 93.4\pm16.0$ & $  0$ & 4\\
$  0.04$ & $ -16.9\pm32.8$ & $104.0\pm38.6$ & $  0$ & 4\\
$  1.27$ & $ -36.9\pm30.4$ & $ 81.8\pm38.5$ & $  0$ & 4\\
\enddata  
\end{deluxetable}  
  
\clearpage  
  
\begin{deluxetable}{rcccccrr}  
\tablecolumns{8}  
\tabletypesize{\scriptsize}  
\tablewidth{0pt}  
\tablecaption{Line-strength indices of the sample galaxies  \label{tab:indices}}  
\tablehead{  
\colhead{$r$} & \colhead{\Hb} & \colhead{[MgFe]} &   
\colhead{\Fe} & \colhead{\Mgb} & \colhead{\Mgd} &   
\colhead{PA} & \colhead{Run}\\  
\colhead{[arcsec]} & \colhead{[\AA]} & \colhead{[\AA]} &   
\colhead{[\AA]} & \colhead{[\AA]} & \colhead{[mag]} &  
\colhead{[$^\circ$]} & \colhead{}}  
\startdata  
\cutinhead{IC 171}  
$-11.71$ & $2.571\pm0.518$ & $3.004\pm0.562$ & $2.086\pm0.566$ & $4.326\pm0.445$ & $0.264\pm0.014$ & $ 10$ & 2\\ 
$ -8.47$ & $1.410\pm0.294$ & $3.287\pm0.376$ & $2.603\pm0.352$ & $4.151\pm0.388$ & $0.247\pm0.010$ & $ 10$ & 2\\ 
$ -5.27$ & $1.754\pm0.223$ & $3.394\pm0.281$ & $2.635\pm0.263$ & $4.371\pm0.289$ & $0.256\pm0.007$ & $ 10$ & 2\\ 
$ -2.84$ & $1.804\pm0.176$ & $3.386\pm0.157$ & $2.632\pm0.149$ & $4.357\pm0.159$ & $0.268\pm0.004$ & $ 10$ & 2\\ 
$ -1.08$ & $1.792\pm0.145$ & $3.367\pm0.119$ & $2.634\pm0.111$ & $4.303\pm0.123$ & $0.266\pm0.003$ & $ 10$ & 2\\ 
$ -0.23$ & $2.322\pm0.136$ & $2.908\pm0.130$ & $2.157\pm0.126$ & $3.920\pm0.122$ & $0.250\pm0.003$ & $ 10$ & 2\\ 
$  0.38$ & $2.585\pm0.148$ & $3.056\pm0.168$ & $2.404\pm0.167$ & $3.884\pm0.157$ & $0.269\pm0.004$ & $ 10$ & 2\\ 
$  0.98$ & $2.409\pm0.259$ & $3.342\pm0.196$ & $2.770\pm0.195$ & $4.031\pm0.191$ & $0.242\pm0.005$ & $ 10$ & 2\\
\enddata  
\end{deluxetable}

\end{document}